\documentclass[fleqn,usenatbib]{mnras}
\usepackage{newtxtext,newtxmath}
\usepackage[T1]{fontenc}
\DeclareRobustCommand{\VAN}[3]{#2}
\let\VANthebibliography\thebibliography
\def\thebibliography{\DeclareRobustCommand{\VAN}[3]{##3}\VANthebibliography}

\usepackage{epsfig}
\usepackage{mathtools}
\usepackage{cuted}
\usepackage[normalem]{ulem}

\title[Surface changes on Comet 67P]{$\mathbf{CO_2}$--driven surface changes in the Hapi region on Comet 67P/Churyumov--Gerasimenko}

\author[B. J. R. Davidsson et al.]{Bj\"{o}rn J. R. Davidsson,$^{1}$\thanks{E-mail: bjorn.davidsson@jpl.nasa.gov}
F. Peter Schloerb,$^{2}$
Sonia Fornasier,$^{3,4}$
Nilda Oklay,$^{5}$\newauthor
Pedro J. Guti\'errez,$^{6}$
Bonnie J. Buratti,$^{7}$
Artur B. Chmielewski,$^{8}$
Samuel Gulkis,$^{9}$\newauthor
Mark D. Hofstadter,$^{10}$
H. Uwe Keller,$^{11,12}$
Holger Sierks,$^{13}$
Carsten G\"{u}ttler,$^{13}$\newauthor
Michael K\"{u}ppers,$^{14}$
Hans Rickman,$^{15,16}$
Mathieu Choukroun,$^{7}$
Seungwon Lee,$^{17}$\newauthor
Emmanuel Lellouch,$^{3}$
Anthony Lethuillier,$^{11}$,
Vania Da Deppo,$^{18}$
Olivier Groussin,$^{19}$\newauthor
Ekkehard K\"{u}hrt,$^{20}$
Nicolas Thomas,$^{21}$
Cecilia Tubiana,$^{22}$
M. Ramy El--Maarry,$^{23}$\newauthor
Fiorangela La Forgia,$^{24}$
Stefano Mottola$^{12}$ and
Maurizio Pajola$^{25}$\\
\\
Affiliations are listed at the end of the paper.
}

\date{Accepted 2022 September 05. Received 2022 August 08; in original form 2022 May 04}

\pubyear{2022}

\begin{document}
\label{firstpage}
\pagerange{\pageref{firstpage}--\pageref{lastpage}}
\maketitle


\begin{abstract}
Between 2014 December 31 and 2015 March 17, the OSIRIS cameras on \emph{Rosetta} documented the growth of a $140\,\mathrm{m}$ wide and $0.5\,\mathrm{m}$ deep 
depression in the Hapi region on Comet 67P/Churyumov--Gerasimenko. This shallow pit is one of several that later formed elsewhere on the comet, all in smooth terrain 
that primarily is the result of airfall of coma particles. We have compiled observations of this region in Hapi by the microwave instrument MIRO on \emph{Rosetta}, acquired during 
October and November 2014. We use thermophysical and radiative transfer models in order to reproduce the MIRO observations. This allows us to place constraints on 
the thermal inertia, diffusivity, chemical composition, stratification, extinction coefficients, and scattering properties of the surface material, and how they evolved during 
the months prior to pit formation. The results are placed in context through long--term comet nucleus evolution modelling. We propose that: 1) MIRO observes 
signatures that are consistent with a solid--state greenhouse effect in airfall material; 2) $\mathrm{CO_2}$ ice is sufficiently close to the surface to have a 
measurable effect on MIRO antenna temperatures, and likely is responsible for the pit formation in Hapi observed by OSIRIS; 3) the pressure at the $\mathrm{CO_2}$ 
sublimation front is sufficiently strong to expel dust and water ice outwards, and to compress comet material inwards, thereby causing the near--surface compaction 
observed by CONSERT, SESAME, and groundbased radar, manifested as the `consolidated terrain' texture observed by OSIRIS.
\end{abstract}


\section{Introduction} \label{sec_intro}

Near--nucleus operations of the ESA \emph{Rosetta/Philae} spacecraft \citep{glassmeieretal07} at Comet 67P/Churyumov--Gerasimenko (hereafter, 67P) 
began 2014 August 6 and ended 2016 September 30 \citep{tayloretal17}. The OSIRIS \citep{kelleretal07} Narrow Angle Camera (NAC) and Wide Angle Camera (WAC) revealed a geologically diverse landscape 
with two major types of morphological units: \emph{consolidated terrain} that constitutes topographically complex structures, and \emph{smooth terrain} dominated by 
$\stackrel{<}{_{\sim}}1\,\mathrm{cm}$--sized chunks \citep{mottolaetal15,pajolaetal17b} that form vast plains \citep[e.g.][]{sierksetal15, thomasetal15a, elmaarryetal15} 
that closely follow equipotential surfaces \citep[i.e., slopes with respect to the local gravity field are small, often $<5^{\circ}$;][]{sierksetal15, augeretal15a, pajolaetal19}. 
The southern hemisphere, that is strongly illuminated near perihelion \citep{kelleretal15}, consists primarily of exposed consolidated terrain \citep{elmaarryetal16}. 
The southern hemisphere is a source of large coma particles that rain down on the northern hemisphere (that experiences polar night near perihelion) as airfall 
\citep{thomasetal15b, kelleretal15, kelleretal17, huetal17, davidssonetal21}. Airfall thereby contributes to the formation of smooth terrain, often on top of partially exposed 
consolidated terrain \citep{thomasetal15a, thomasetal15b, elmaarryetal15}.

In May 2015, three months pre--perihelion, the large ($0.8\,\mathrm{km^2}$) smooth terrain at Imhotep \citep[for region names and definitions, see][]{thomasetal18} started to display morphological changes 
in the form of several roundish shallow features that grew and merged over the following months \citep{groussinetal15b}. For brevity, we occasionally use the more informal 'pits' for such shallow depressions. 
The escarpments that constituted the rims of these pits had heights on the order of one metre and moved with speeds at $0.2$--$0.3\,\mathrm{m\,h^{-1}}$ \citep{groussinetal15b}. The largest single 
feature grew to a diameter of $220\,\mathrm{m}$, and these morphological changes eventually affected 40 per cent of the surface area of the Imhotep smooth terrain \citep{groussinetal15b}. 
Similar phenomena were later observed in several smooth terrains on different parts of the comet \citep{elmaarryetal17, huetal17, birchetal19,bouquetyetal22}. 

Thus far, the characterisation of expanding pits and moving escarpments has relied exclusively on visual images and spectrophotometry. 
This provides snapshots of the pit morphology that can be used to establish a timeline of how the pit size and shape evolved, and it places some 
constraints on composition. However, this information is not sufficient in order to understand why the pits form or what mechanisms are responsible for 
their evolution. We also do not know the physical properties of the near--surface material (such as temperature, porosity, and thermal inertia), and we 
do not know which volatiles are present, at what depths they are encountered, or what vapour pressures they are capable of reaching. Furthermore, it is 
necessary to know the thermal history of the location in question. This requires calculating the amount of energy that is available at any 
given moment to drive changes, and to understand the previous evolution that has led up to the current conditions. Only with such detailed information available it is possible to develop a pit 
formation and evolution scenario that is quantitative and not merely qualitative, that is physically realistic and consistent with observational data, and that 
explains why the phenomenon starts and ends at given points in time. We here take the first step of expanding the database of physical and chemical properties at 
pit formation sites by using observations made by \emph{Rosetta}/MIRO \citep[Microwave Instrument for \emph{Rosetta} Orbiter;][]{gulkisetal07}, and apply the 
state--of--the--art comet nucleus thermophysics model \textsc{nimbus} \citep[Numerical Icy Minor Body evolUtion Simulator;][]{davidsson21} in order to analyse the MIRO data and 
to provide the contextual information necessary to develop a quantitative understanding of the pit formation phenomenon. 

This work is important because we still have a poor understanding of comet activity \citep[a post--\emph{Rosetta} analysis of the state of affairs is made by][]{kellerkuhrt20}. 
Pit growth is one of the most dramatic expressions of comet activity observed by \emph{Rosetta}. Reaching an understanding of how and why these pits form and grow 
is therefore needed in order to better understand comet activity itself. One of the four main measurement goals of the \emph{Rosetta} mission was 
`Study of the development of cometary activity and the processes in the surface layer of the nucleus and inner coma (dust/gas interaction)' \citep{schwehmandschulz99}. 
If the problem of pit growth could be solved, substantial progress could be made to understand comet activity and fulfil a key \emph{Rosetta} science goal. 

As stated previously, Imhotep was not the only smooth terrain in which pits were formed. In this paper, we focus on one particular set of shallow pits in the Hapi region, 
that formed around 2014 December 31 and grew until growth stopped sometime between 2015 February 28 and March 17. 
Hapi is a smooth terrain located on the northern hemisphere on the neck between the two lobes of the comet \citep{sierksetal15}. Its properties have been 
described in detail by \citet{pajolaetal19}. We use a sequence of OSIRIS images to reconstruct the temporal evolution of the depressions, as well as OSIRIS spectrophotometry 
to obtain constraints on the composition of the material within, and around, the pits. We use observations by MIRO acquired in October and November 2014, 
that provides the thermal emission of the surface material during the months leading up to pit formation. A combination of thermophysical and radiative transfer 
models are employed in order to analyse the MIRO microwave data and to place constraints on the thermal inertia, diffusivity, chemical composition, stratification, extinction coefficients, 
and scattering properties of the near--surface material. From this analysis we infer that significant changes in chemical stratification and 
physical properties took place prior to pit formation. We use thermophysical models of the nucleus from the May 2012 aphelion to mid--March 2015 to place 
our findings into a broader context. This allows us to place important constraints on the properties of airfall material (by demonstrating the presence of 
a solid--state greenhouse effect), and on the mechanism responsible for morphological changes in smooth terrain (by demonstrating that the MIRO observations 
and the timing of pit formation are consistent with $\mathrm{CO_2}$--driven activity). Furthermore, we demonstrate that such activity may be responsible for 
the observed near--surface compaction of cometary material, and ultimately for the formation of consolidated terrain.

Section~\ref{sec_obs} describes the observational data, specifically, OSIRIS images and spectrophotometry in section~\ref{sec_obs_osiris}, and MIRO data in section~\ref{sec_obs_miro}. 
Section~\ref{sec_models} summarises our models: illumination conditions (section~\ref{sec_models_illum}); a relatively simple (section~\ref{sec_models_basic}) and a relatively 
advanced (section~\ref{sec_models_NIMBUS}) nucleus thermophysical model; and a radiative transfer model (section~\ref{sec_models_THEMIS}). Our results are 
described in section~\ref{sec_results}, focusing on the MIRO October 2014 data (section~\ref{sec_results_Oct}), the MIRO November 2014 data (section~\ref{sec_results_Nov}), 
and the contextual simulations (section~\ref{sec_results_context}). These results are discussed in section~\ref{sec_discuss} and our conclusions are summarised in section~\ref{sec_conclusions}.

\section{Observational data} \label{sec_obs}

\subsection{OSIRIS observations} \label{sec_obs_osiris}

\begin{figure*}
\centering
\begin{tabular}{cc}
\scalebox{0.45}{\includegraphics{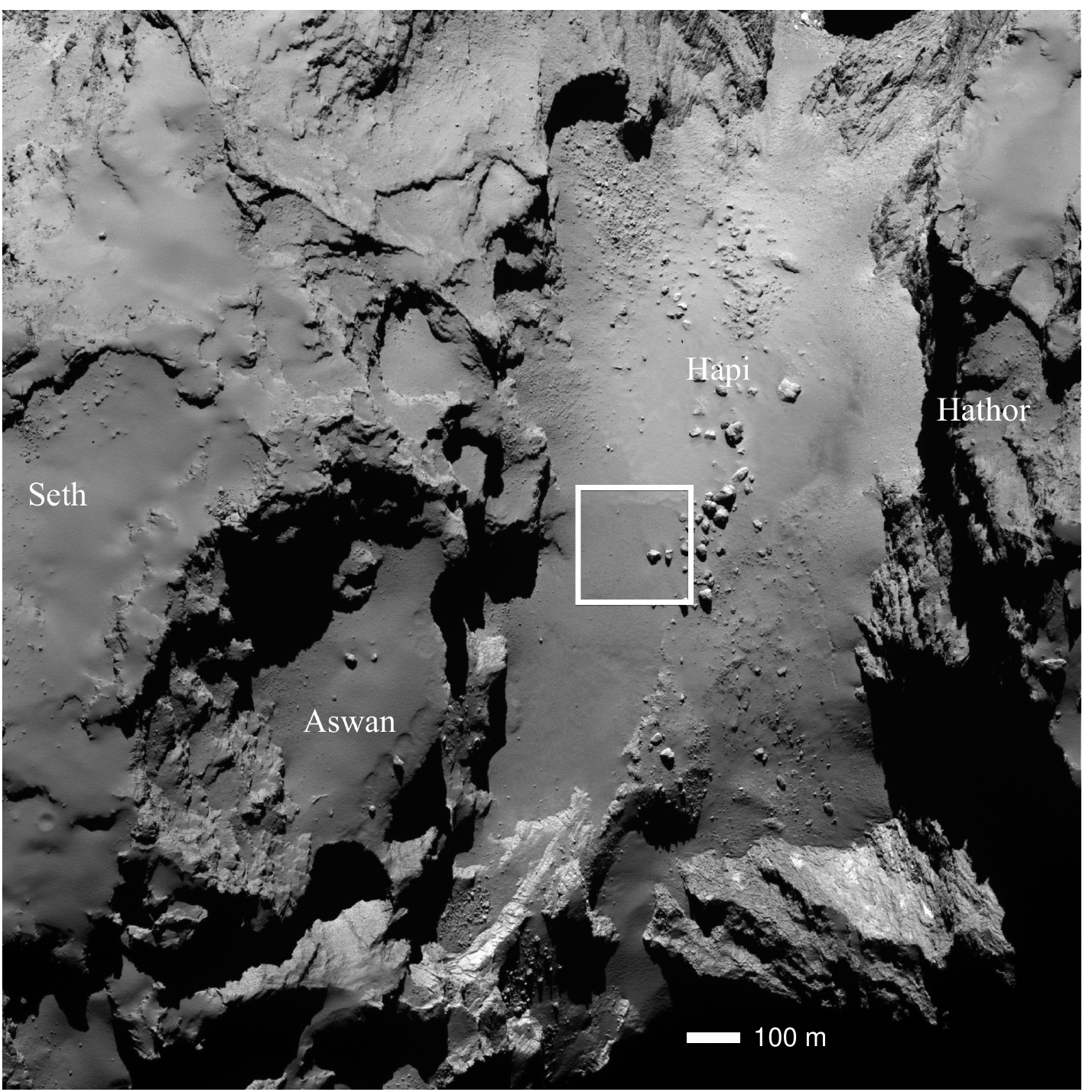}} & \scalebox{0.45}{\includegraphics{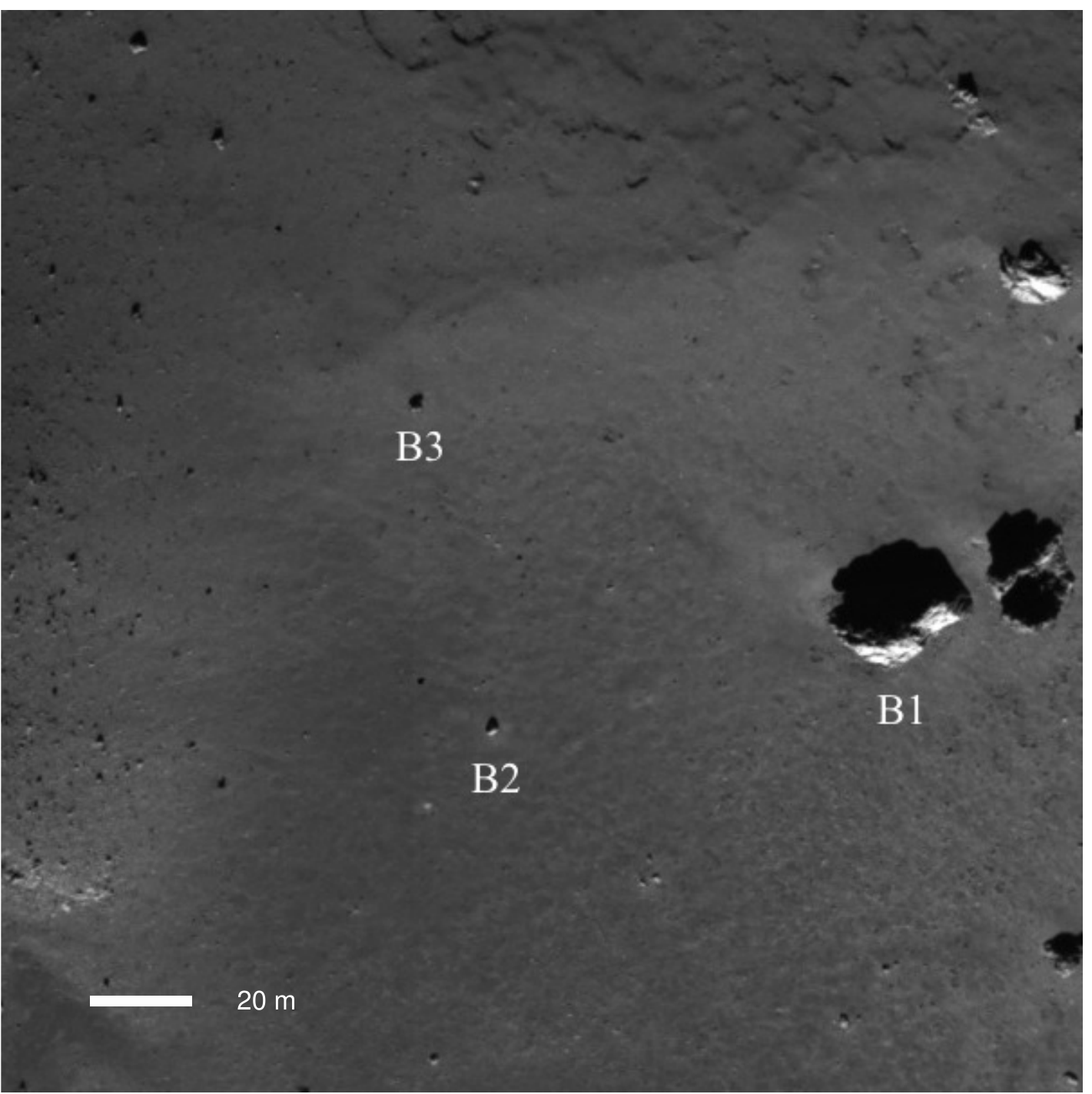}}\\
\scalebox{0.45}{\includegraphics{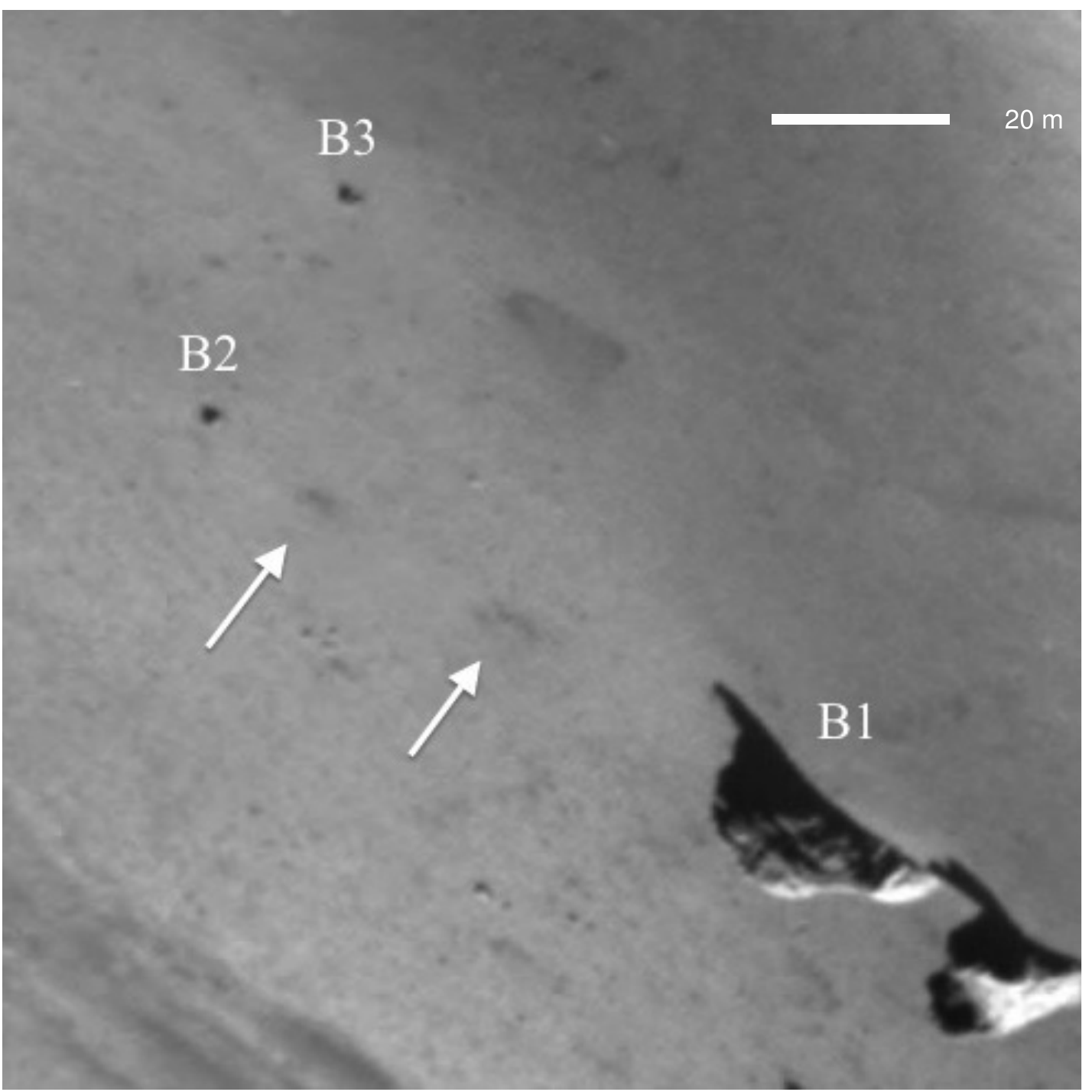}} & \scalebox{0.45}{\includegraphics{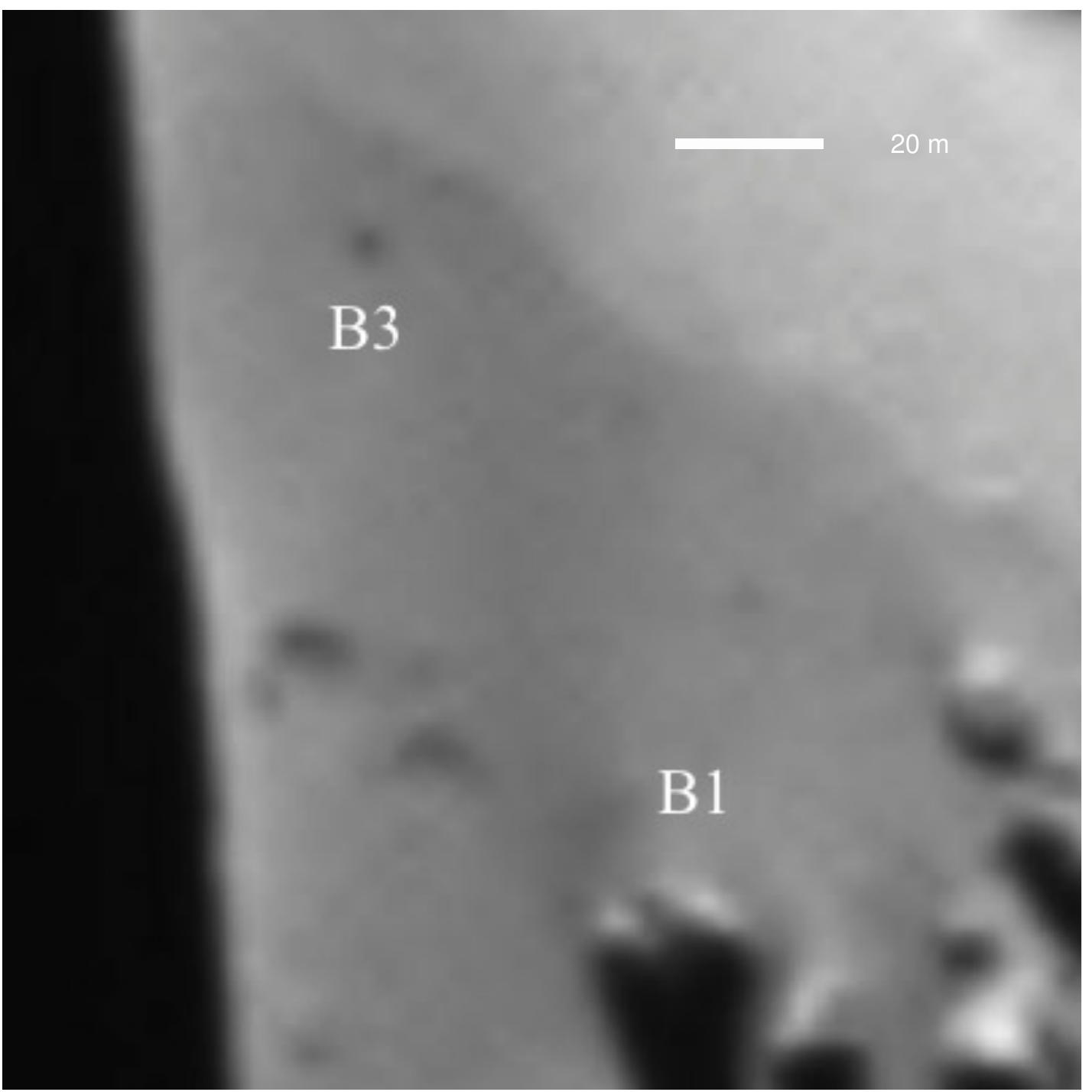}}\\
\end{tabular}
     \caption{\emph{Upper left:} NAC context image of Hapi and surroundings taken on 2014 August 30. The large lobe is to the left and the 
small lobe is to the right (outside the image). The distance from \emph{Rosetta} to the comet surface was approximately 
$55.5\,\mathrm{km}$ and the resolution was $0.98\,\mathrm{m\,px^{-1}}$. The square marks the area in which 
the currently discussed changes took place, and is the approximate field of view in the upper right panel. Image 
MTP006/n20140830t034253546id20f22.img. \emph{Upper right:} This NAC image of a small portion of the Hapi region was acquired on 2014 December 10, prior to 
any detectable change (see the upper left panel for context). The distance from \emph{Rosetta} to the comet surface was approximately 
$19.9\,\mathrm{km}$ and the resolution was $0.35\,\mathrm{m\,px^{-1}}$. The boulder B1 measures $27.4\pm 0.4\,\mathrm{m}$ 
across. Boulders B2 and B3 are $1.8\pm 0.4\,\mathrm{m}$ and $2.1\pm 0.4\,\mathrm{m}$ across, respectively, 
and the projected distance between the two is $65.5\pm 0.4\,\mathrm{m}$. Image MTP010/n20141210t062855791id20f22.img. \emph{Lower left:} This NAC image, taken on 2014 December 30, 
shows the first known indications of changes (two dark features marked with arrows).  The distance from \emph{Rosetta} to the comet surface was approximately 
$27.8\,\mathrm{km}$ and the resolution was $0.49\,\mathrm{m\,px^{-1}}$. The surface is seen at a high 
emergence angle that distorts the perspective. The left feature is $4.9\pm 0.5\,\mathrm{m}$ across, and the 
right feature is $7.7\pm 0.5\,\mathrm{m}$ across. Image MTP011/n20141230t081300834id20f22.img. \emph{Lower right:} This WAC image was taken on 2015 January 11.  
The distance from \emph{Rosetta} to the comet surface was approximately $27.1\,\mathrm{km}$ and the resolution was $2.57\,\mathrm{m\,px^{-1}}$. 
The features have grown with respect to the lower left panel. The left feature is $12.9\pm 2.6\,\mathrm{m}$ across, and the 
right feature is $13.1\pm 2.6\,\mathrm{m}$ across. Image MTP011/w20150111t125858091id20f13.img.}
     \label{fig_hapi01}
\end{figure*}

\subsubsection{OSIRIS imaging} \label{sec_obs_osiris_images}

The images shown in this section are available on the ESA Planetary Science Archive (PSA\footnote{\url{https://www.cosmos.esa.int/web/psa/rosetta}}), as well as on the 
NASA Planetary Data System (PDS\footnote{\url{https://pds-smallbodies.astro.umd.edu/data_sb/missions/rosetta/index.shtml}}). 
Figure~\ref{fig_hapi01} (upper left), shows Hapi and its surroundings -- Seth on the large lobe to the left, and the steep Hathor cliff 
of the small lobe to the right. Figure~\ref{fig_hapi01} (upper right), shows a $213\times 213\,\mathrm{m}$ part of Hapi, at a location marked by a square in 
the upper left panel. Here, Hapi is dominated by material with a particle size smaller than the $0.35\,\mathrm{m\,px^{-1}}$ 
spatial resolution, forming a rather flat plain. The plain is covered with boulders with sizes ranging from the 
resolution limit to several tens of meters. Three boulders have been labelled for reference purposes (B1--B3). Note the 
ridge above B3, running parallel to the B1--B2 line.

\begin{figure*}
\centering
\begin{tabular}{cc}
\scalebox{0.45}{\includegraphics{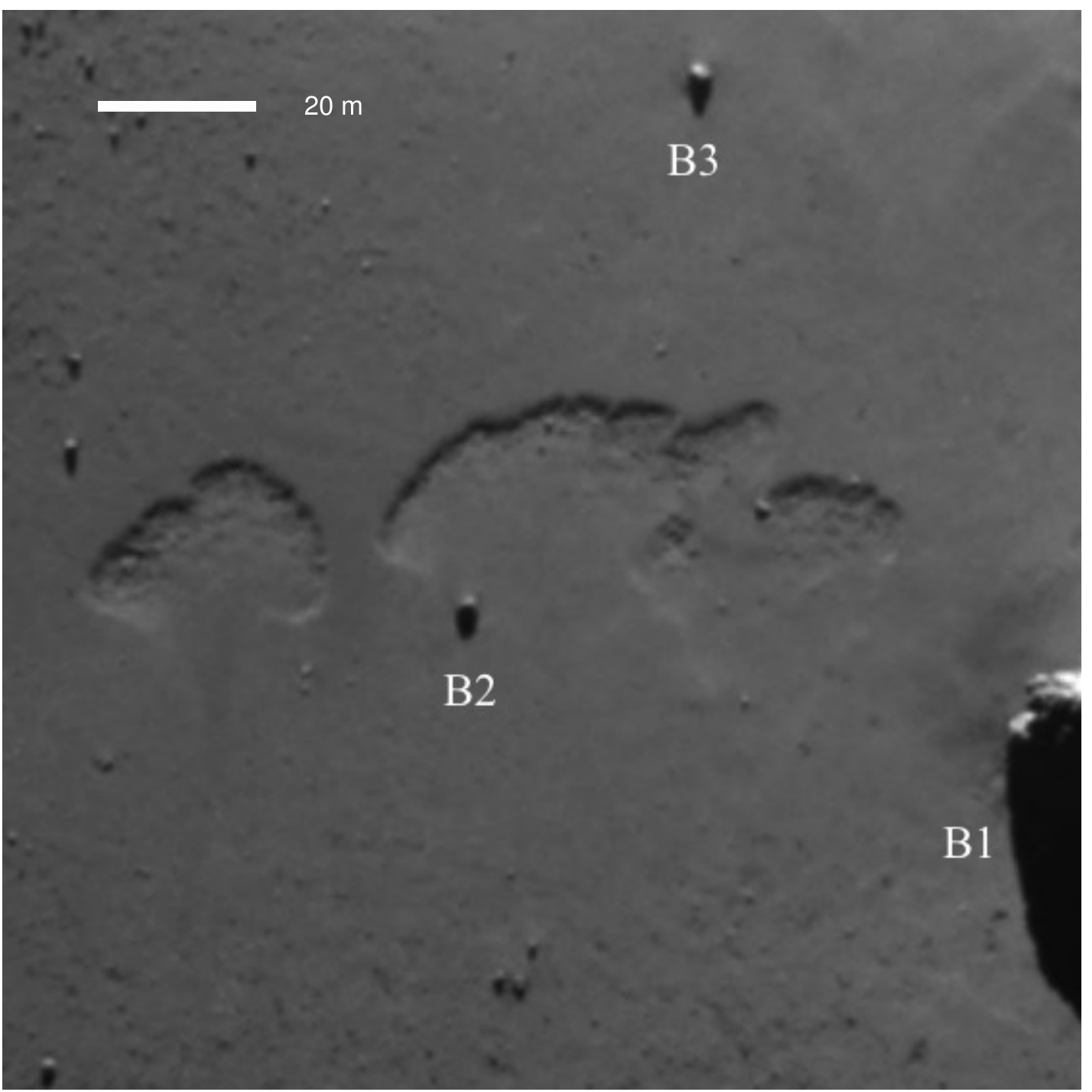}} & \scalebox{0.45}{\includegraphics{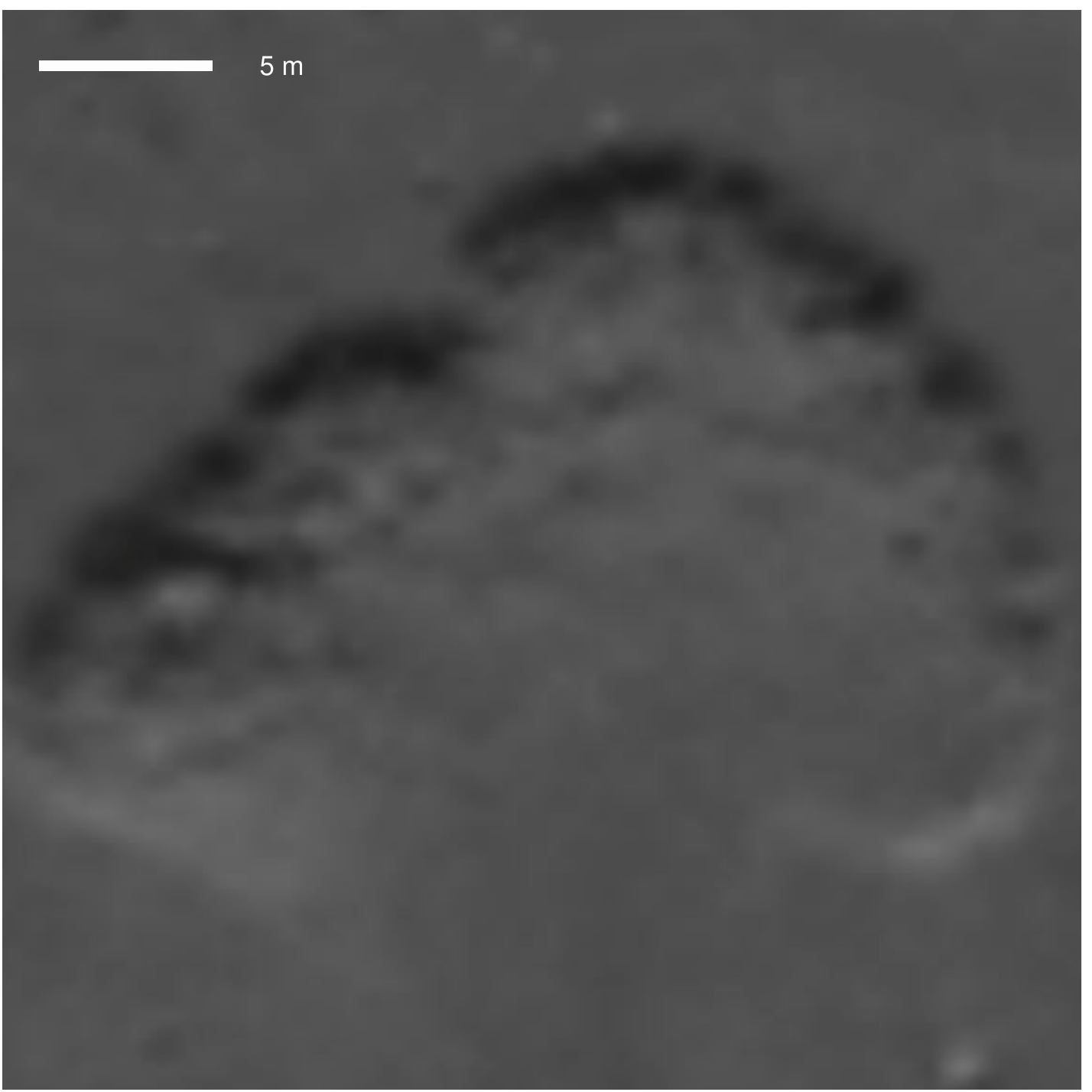}}\\
\scalebox{0.45}{\includegraphics{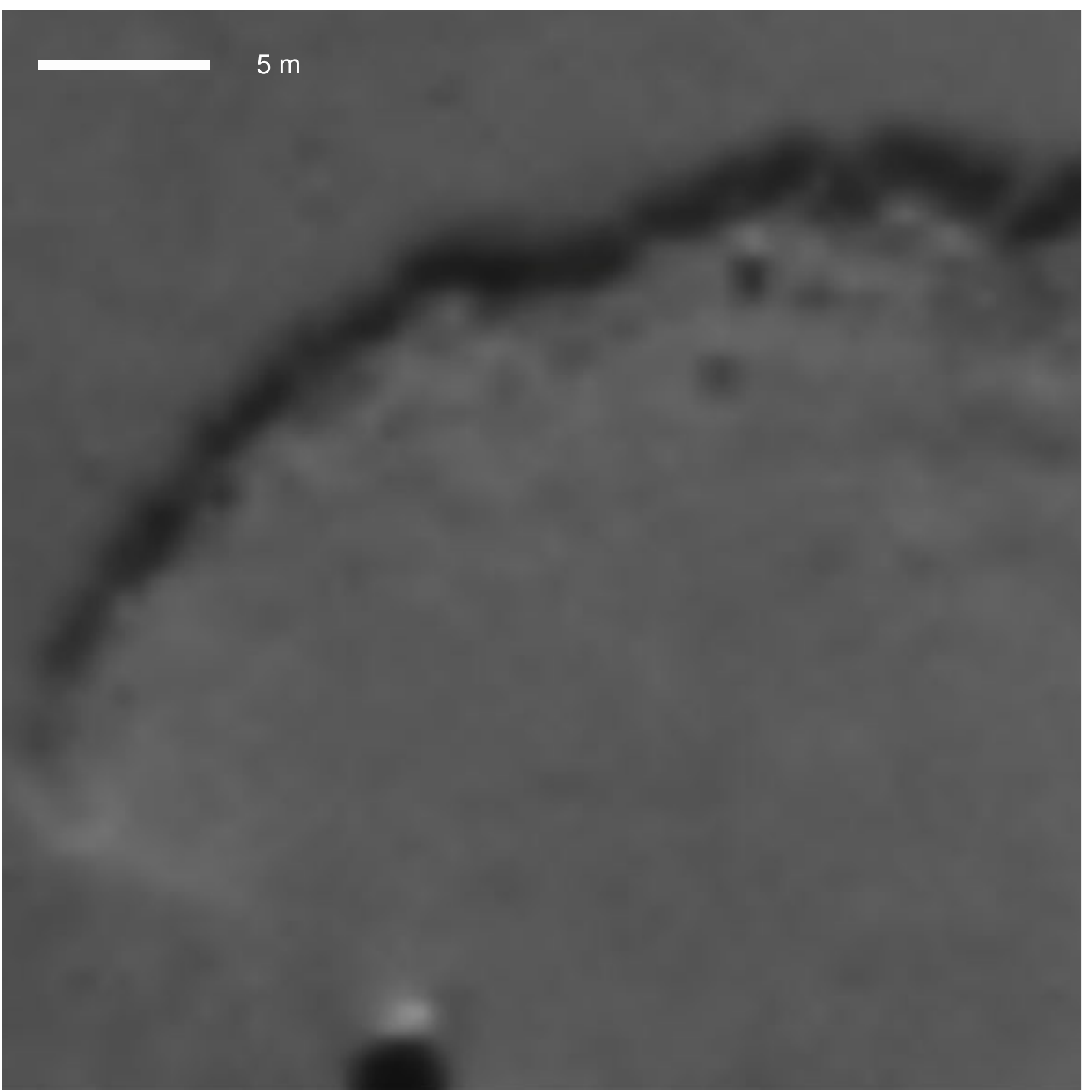}} & \scalebox{0.45}{\includegraphics{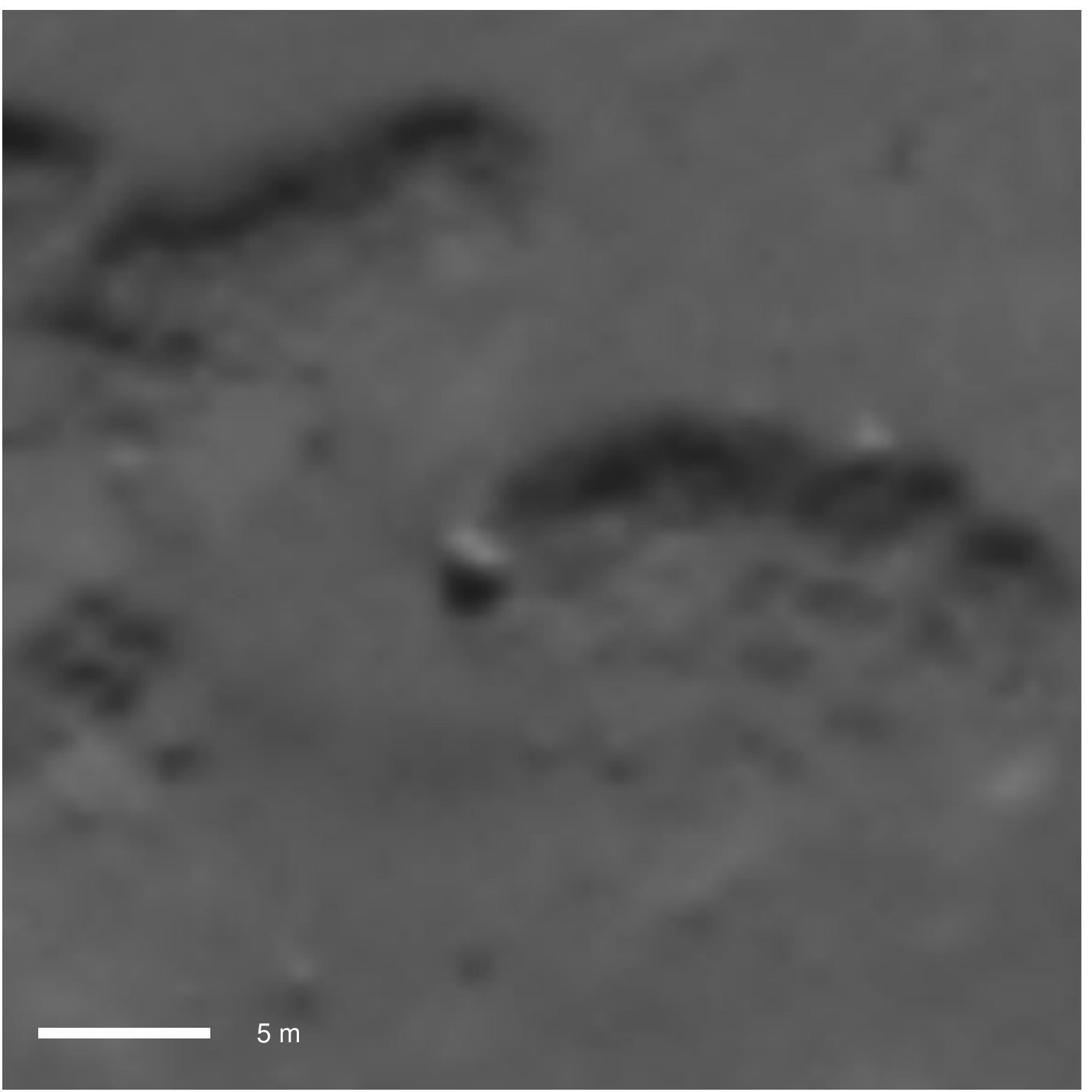}}\\
\end{tabular}
     \caption{\emph{Upper left:} This is a portion of a NAC image taken on 2015 January 22, and shows two shallow and flat--bottomed depressions that were 
not present six weeks earlier (Fig.~\ref{fig_hapi01}).  The distance from \emph{Rosetta} to the comet surface was approximately 
$27.3\,\mathrm{km}$ and the resolution was $0.49\,\mathrm{m\,px^{-1}}$. The projected distance between boulders B1 and B2 is 
$73.1\pm 0.4\,\mathrm{m}$. Image MTP012/n20150122t223400384id20f22.img. \emph{Upper right}: Close--up of the left feature in the upper 
left panel, on 2015 January 22. \emph{Lower left:} Close--up of the left rim of the right feature in the upper left panel, on 2015 January 22. 
\emph{Lower right:} Close--up of the right rim of the right feature in the upper left panel, on 2015 January 22.}
     \label{fig_hapi02}
\end{figure*}

Figure~\ref{fig_hapi01} (upper right), was acquired on 2014 December 10, prior to any detectable change. The first known 
indications of changes are from 2014 December 30, where Fig.~\ref{fig_hapi01} (lower left) shows two crescent--shaped 
dark features between boulders B1 and B2. The two features, indicated by the arrows, are $4.9\pm 0.5\,\mathrm{m}$ and $7.7\pm 0.5\,\mathrm{m}$ 
across, respectively. During the following 12.10 days (Fig.~\ref{fig_hapi01}, lower right), the features grew in size to $12.9\pm 2.6\,\mathrm{m}$ and 
$13.1\pm 2.6\,\mathrm{m}$. If assuming circular expansion, that corresponds to average radial propagation velocities of 
$0.33\pm 0.13\,\mathrm{m\,\mathrm{d^{-1}}}$ ($13.7\pm 5.4\,\mathrm{mm\,h^{-1}}$) and $0.22\pm 0.13\,\mathrm{m\,\mathrm{d^{-1}}}$ 
($9.4\pm 5.4\,\mathrm{mm\,h^{-1}}$).

The upper left image in Fig.~\ref{fig_hapi02} was taken on 2015 January 22, or 11.40 days after the lower right image in Fig.~\ref{fig_hapi01}. 
The two features have now merged into a single shallow depression located between boulders B2 and B3, and a new feature has appeared to the 
left of boulder B2. The large depression to the right has a length of $\sim 63\,\mathrm{m}$ and a width of $\sim 22\,\mathrm{m}$. 
The escarpment is located $30.9\,\mathrm{m}$ from B2 in the direction towards B3, and is $42.6\,\mathrm{m}$ from B3 along 
the same line. The features appeared near the line between B1 and B2, thus the escarpment has moved about 22 meters 
in 23.5 days, yielding an average propagation speed of approximately $0.9\,\mathrm{m\,\mathrm{d^{-1}}}$. 
About $28\,\mathrm{m}$ right of B2 there is a rim with a slope facing right, and $\sim 27\,\mathrm{m}$ farther away is a 
second rim with a slope facing left. If this quasi--circular structure is interpreted as the expanded $13.1\,\mathrm{m}$ feature 
in Fig.~\ref{fig_hapi01} (lower right), those rims propagated with a speed of about $0.6\,\mathrm{m\,\mathrm{d^{-1}}}$. These 
speeds are 2--4 times higher than those measured for the period 2014 December 30 to 2015 January 11, suggesting that 
the propagation speed may have accelerated.

At the time the image in Fig.~\ref{fig_hapi02} (upper left) was taken, the solar incidence angle was $\sim 68^{\circ}$ (for facet F\#1, see section~\ref{sec_models_illum}). 
Because the horizontally oriented rims in that image cast shadows into the depressions, we can infer that the 
rims have slopes steeper than $\sim 68^{\circ}$. Based on the length of these shadows being 1--$2\,\mathrm{m}$, the depth of the 
depressions can be estimated as $0.5\pm 0.1\,\mathrm{m}$.

The isolated depression to the left is roughly triangular in shape with dimensions $\sim 29\,\mathrm{m}$ by  $\sim 20\,\mathrm{m}$. 
The surface area is approximately $290\,\mathrm{m^2}$, and the volume affected by the change is about $140\,\mathrm{m^3}$, corresponding to nearly 80 metric 
tons of material if assuming that the density is identical to the nucleus bulk density $\rho_{\rm bulk}=535\,\mathrm{kg\,m^{-3}}$ 
\citep{preuskeretal15, jordaetal16}.

The lower rim is diffuse and featureless. The rest of the rim, seen magnified in Fig.~\ref{fig_hapi02} (upper right), 
is continuous and seems to consist of a number of weakly curved segments, each being a few meters in size. If mass wasting 
takes place at the steep rims, this material is too small to be resolved. Compared to the immediate surroundings, the bottom of the 
depression has a larger degree of resolved roughness, at least in the upper half of the depression. 

\begin{figure*}
\centering
\begin{tabular}{cc}
\scalebox{0.45}{\includegraphics{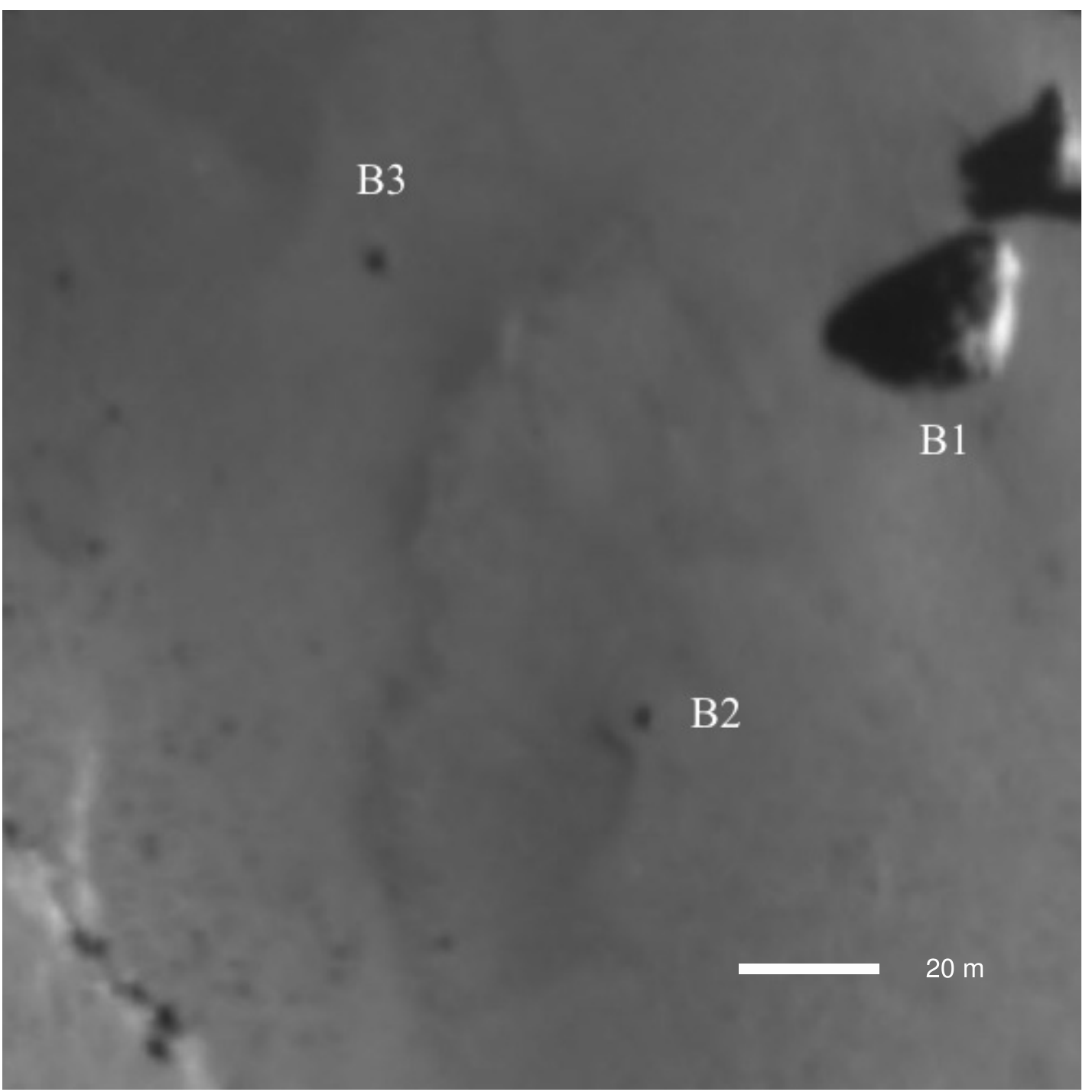}} & \scalebox{0.45}{\includegraphics{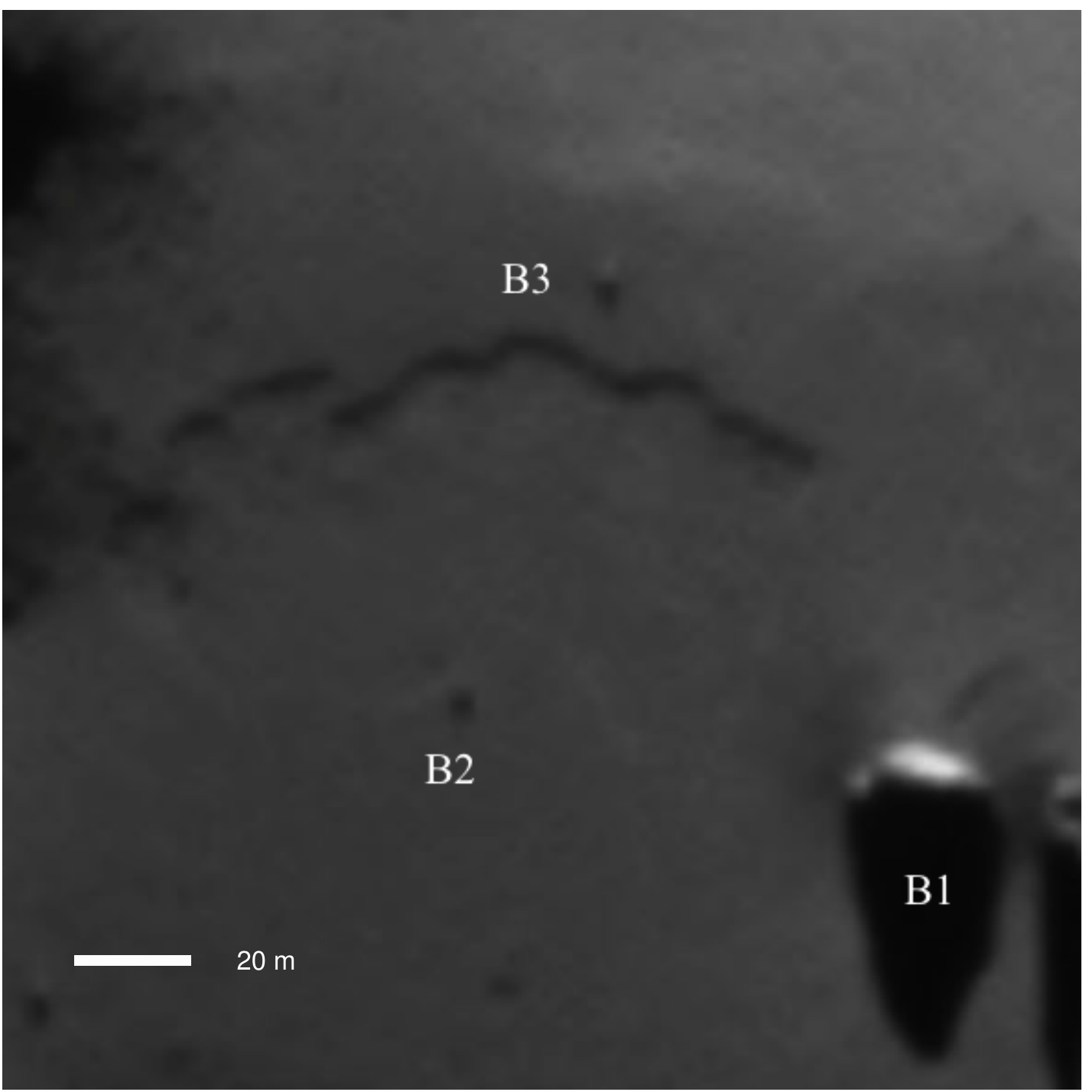}}\\
\scalebox{0.45}{\includegraphics{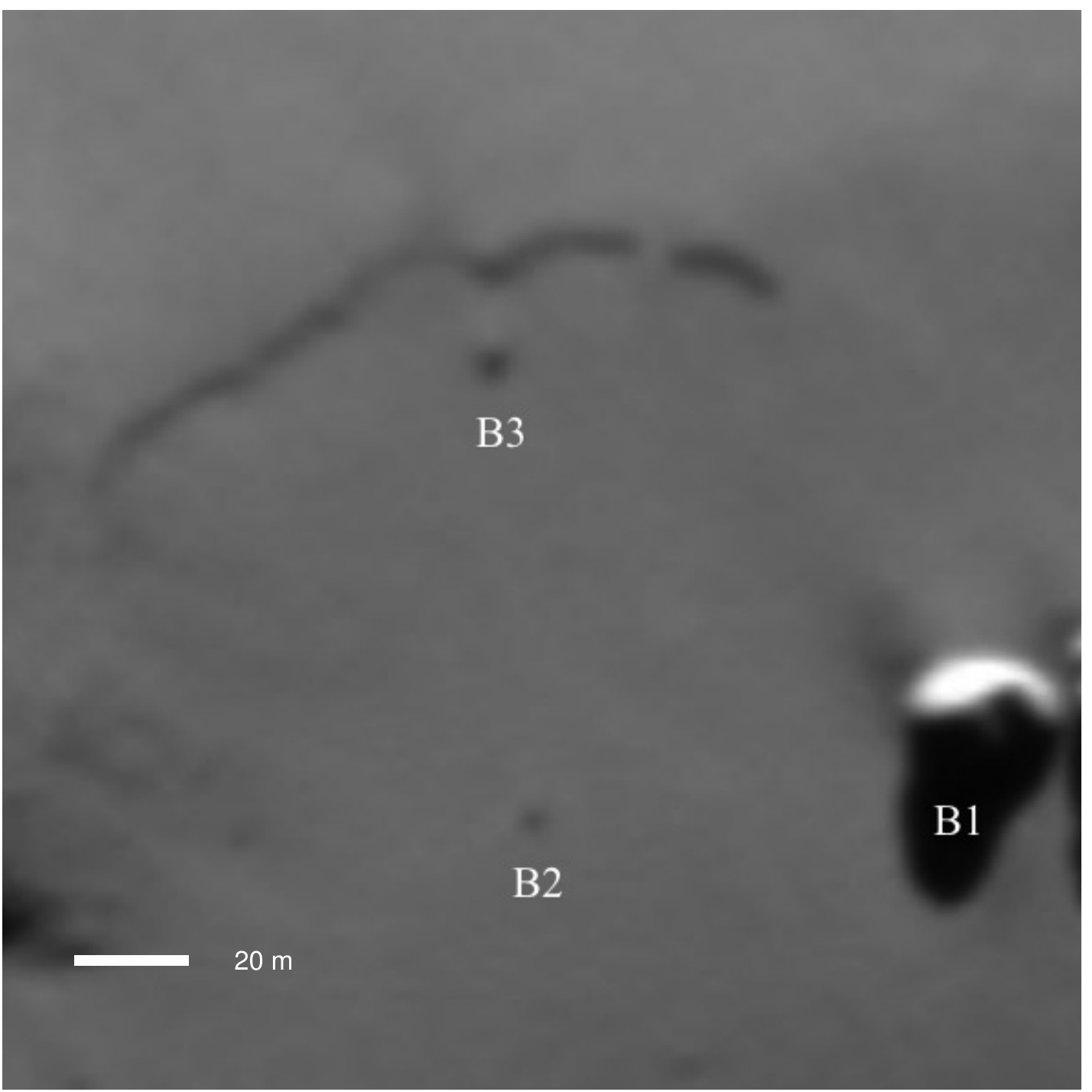}} & \scalebox{0.45}{\includegraphics{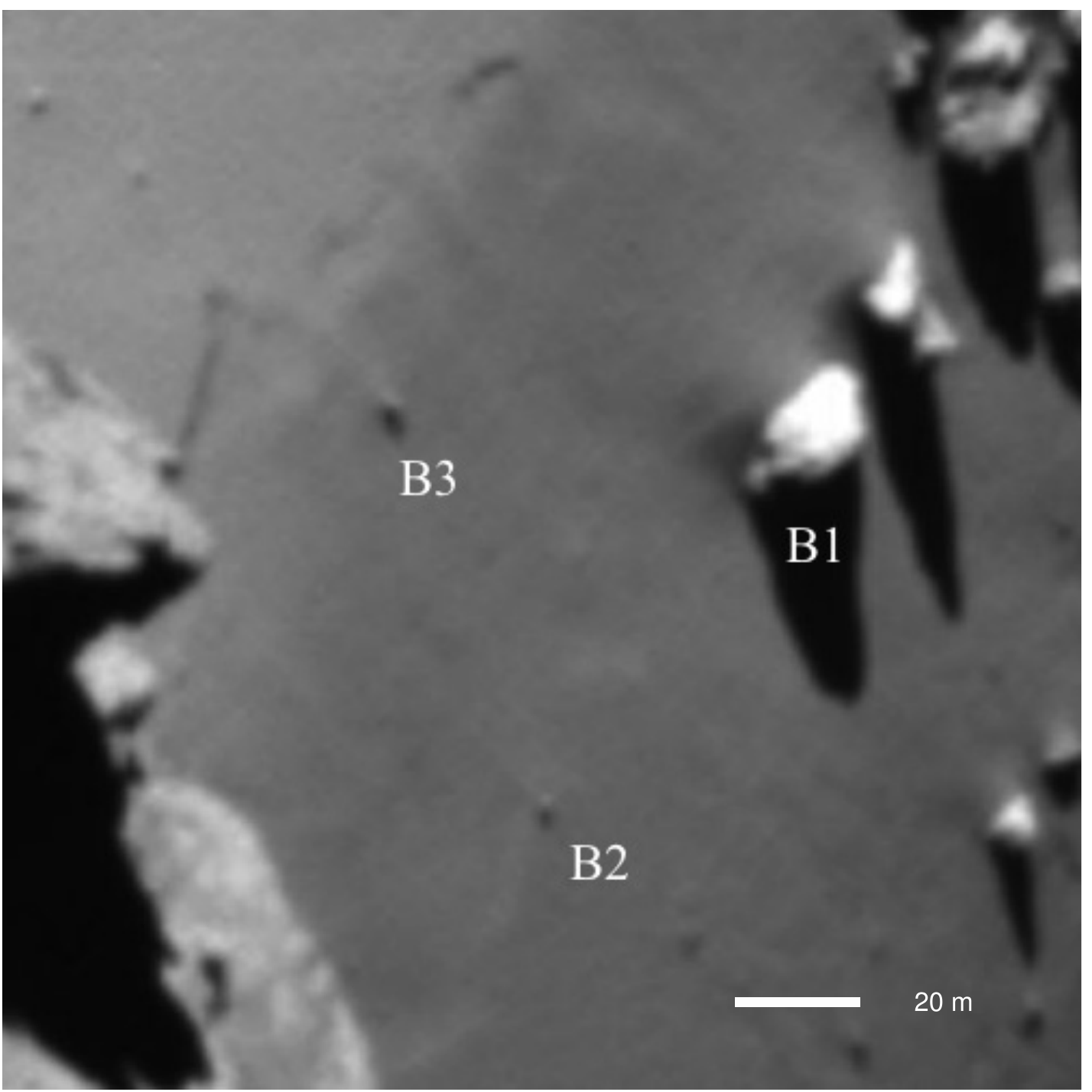}}\\
\end{tabular}
     \caption{\emph{Upper left:} On 2015 February 5, the two depressions have merged into a single one. The distance from \emph{Rosetta} to the 
comet surface was approximately $58.9\,\mathrm{km}$ and the resolution was $1.05\,\mathrm{m\,px^{-1}}$. Image MTP012/n20150205t013716672id20f41.img. 
\emph{Upper right:} This NAC image was taken on 2015 February 9. The distance from \emph{Rosetta} to the 
comet surface was approximately $106.1\,\mathrm{km}$ and the resolution was $1.89\,\mathrm{m\,px^{-1}}$. MTP012/n20150209t123142699id20f22.img. 
\emph{Lower left:} This NAC image was taken on 2015 February 28. The distance from \emph{Rosetta} to the 
comet surface was approximately $108.3\,\mathrm{km}$ and the resolution was $1.93\,\mathrm{m\,px^{-1}}$. Image MTP013/n20150228t044349351id20f22.img. 
\emph{Lower right:} This NAC image was taken on 2015 March 17. The distance from \emph{Rosetta} to the 
comet surface was approximately $77.2\,\mathrm{km}$ and the resolution was $1.37\,\mathrm{m\,px^{-1}}$. Image MTP014/n20150317t061250371id20f22.img.}
     \label{fig_hapi03}
\end{figure*}

The left half of the large depression to the right in Fig.~\ref{fig_hapi02} (upper left), is seen magnified in Fig.~\ref{fig_hapi02} (lower left). 
A $\sim 5\,\mathrm{m}$ wide region, tracing the curved escarpment, appears rougher at resolved size scales than the 
material below. The smoother material may form a tongue--shaped feature (just right of B2 in 
Fig.~\ref{fig_hapi02}), upper left panel, though the contrast is poor.

Figure~\ref{fig_hapi02} (lower right) is a close--up of the right part of the right depression in Fig.~\ref{fig_hapi02} (upper left). 
The rim lacks shadows in two places, suggesting that the slope locally is less than $\sim 68^{\circ}$. Just below the rim, the floor of the 
depression is hummocky in appearance, with a half--dozen relatively bright structures visible. Just right of the central boulder in Fig.~\ref{fig_hapi02} (lower right), 
a number of concavities and hills are seen on the floor of the depression.

The upper left image in Fig.~\ref{fig_hapi03} taken on 2015 February 5, or $13.12\,\mathrm{days}$ after the upper left image in Fig.~\ref{fig_hapi02}, shows that 
the two depressions have merged, having a common escarpment. The width of the bridge between the depressions in Fig.~\ref{fig_hapi02} (upper left) was 
$7.4\,\mathrm{m}$, suggesting a propagation velocity of $\geq 0.28\,\mathrm{m\,d^{-1}}$ if both depressions grew at similar speeds. The width of 
the depression, as measured from the B1--B2 line towards the escarpment near B2, varies between 30--$45\,\mathrm{m}$, suggesting an 
average propagation velocity of 0.8--$1.2\,\mathrm{m\,d^{-1}}$ for the escarpment. The distance between boulders B2 and B3 is $75.1\pm 2.1\,\mathrm{m}$ 
in the image plane. Since 2015 January 22, the distance from the escarpment to boulder B3 decreased from $\sim 43\,\mathrm{m}$ to 
$\sim 21\,\mathrm{m}$, suggesting a propagation velocity of $\sim 1.7\,\mathrm{m\,d^{-1}}$. Considering that the propagation speed 
from Fig.~\ref{fig_hapi01} (lower right) to Fig.~\ref{fig_hapi02} (upper left) was estimated as $0.9\,\mathrm{m\,d^{-1}}$, again suggests 
acceleration. At this stage, the full length of the depression was $\sim 140\,\mathrm{m}$.

Figure~\ref{fig_hapi03} (upper right) was taken on 2015 February 9, $4.45$ days after Fig.~\ref{fig_hapi03} (upper left). The escarpment is now 
$13\pm 4\,\mathrm{m}$ from B3, suggesting a propagation speed of $1.8\pm 0.9\,\mathrm{m\,d^{-1}}$, which is 
similar to the speed during the previous two weeks. The irregular shape of the escarpment suggests that there are some 
differences in propagation speed in different places. Although the resolution is comparatively poor, there is some indication 
of a rougher region just behind the escarpment, compared to much earlier locations. 

\begin{figure*}
\centering
\scalebox{0.3}{\includegraphics{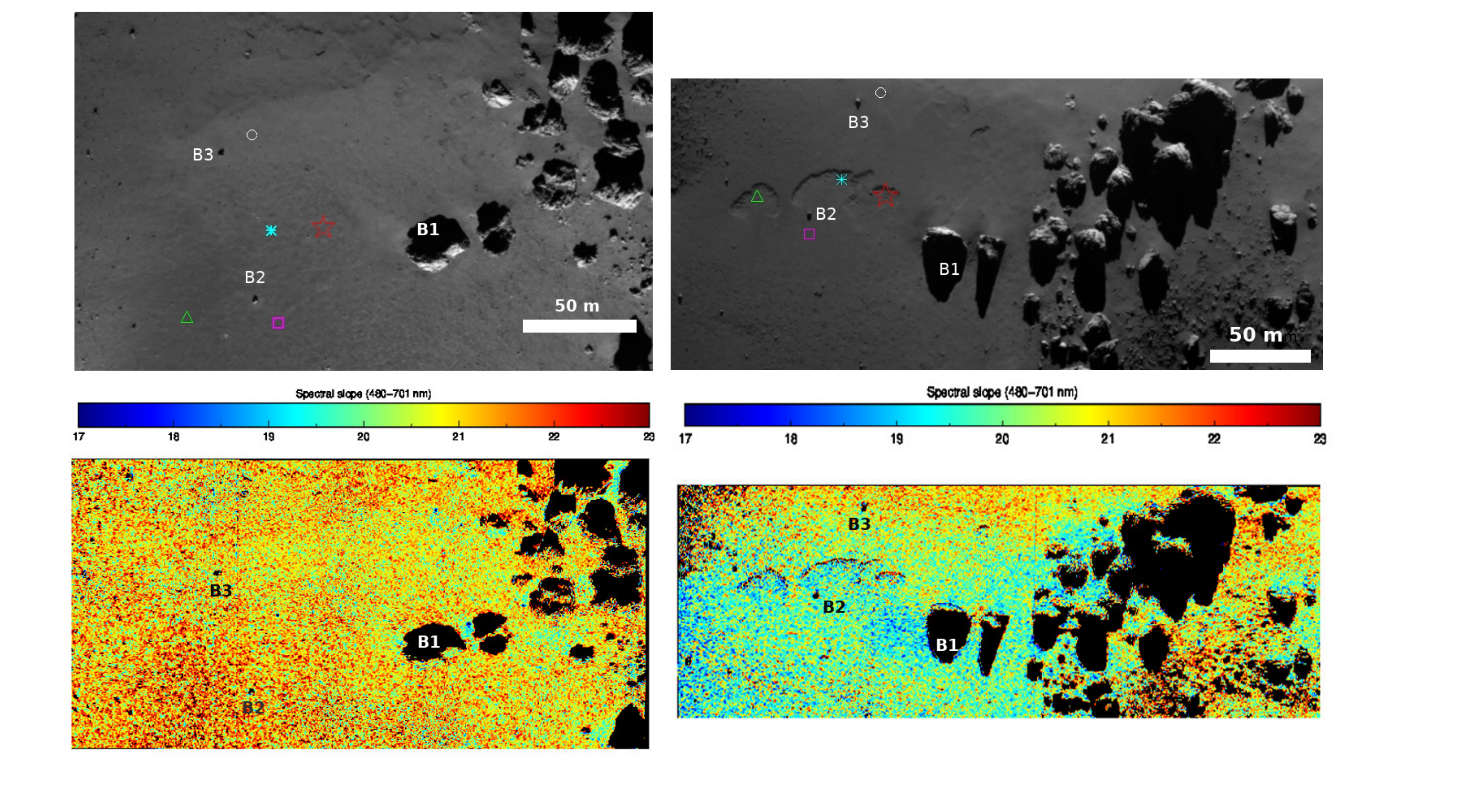}}
     \caption{Images and associated spectral slope $S$ map, evaluated in the $480$--$701\,\mathrm{nm}$ range and normalised at $480\,\mathrm{nm}$, for data acquired on 2014 December 10 
(left; MTP010/n20141210t062911447id4bf24.img and MTP010/n20141210t062919757id4bf27.img) and on 2015 January 22 (right; MTP012/n20150122t223416529id4bf24.img and 
MTP012/n20150122t223425034id4bf27.img). In both figures the symbols represent ROIs 
near and inside the pits, before and after their formation, for which we tabulate $S$ in Table~\ref{tab:slopes}.}
     \label{fig_hapi_fornasier}
\end{figure*}

The next image of this region was taken 23.13 days after Fig.~\ref{fig_hapi03} (upper right) on 2015 February 28. As seen in Fig.~\ref{fig_hapi03}  (lower left), 
the escarpment passed underneath boulder B3, and continued $\sim 15\,\mathrm{m}$ beyond it. The projected distance between B2 and B3 is 
$\sim 83\,\mathrm{m}$. The distance from boulder B2 to the escarpment increased by 
$\sim 34\,\mathrm{m}$ with respect to Fig.~\ref{fig_hapi03} (upper right), implying an escarpment propagation velocity of 
$\sim 1.5\,\mathrm{m\,d^{-1}}$. This is similar to the speed measured since 2015 January 22, implying quasi--constant progression for about a month.

The last image of this region before perihelion with $\sim 1\,\mathrm{m\,px^{-1}}$ resolution was taken on 2015 March 17, and 
is shown in Fig.~\ref{fig_hapi03} (lower right). During these 17.06 days, the escarpment moved very little, suggesting that it came 
to a halt at the beginning of March. The total size of the region affected by this phenomenon is about $100\,\mathrm{m}$ wide and about $140\,\mathrm{m}$ long. 
For simplicity, this region is hereafter referred to as `Hapi~D' (D for depression). Thus, a total area of $\sim 14,000\,\mathrm{m^2}$ was crossed by the 
escarpment, and considering the $0.5\,\mathrm{m}$ depth, the volume affected was $\sim 7,000\,\mathrm{m^3}$.  This corresponds to $\sim 3,700$ metric tons of mass.

\begin{table}
         \begin{center}          
        \begin{tabular}{|lll|} \hline
ROI & Spec. slope  & Spec. slope  \\
   & $S\,\mathrm{[\%\,(100\,nm)^{-1}]}$ & $S\,\mathrm{[\%\,(100\,nm)^{-1}]}$\\
    & Dec. 2014 & Jan. 2015  \\ \hline
Circle          & 21.2$\pm$0.2 &  21.2$\pm$0.2  \\
Red star        & 21.7$\pm$0.2 &  20.7$\pm$0.3    \\
Cyan asterisk   & 21.5$\pm$0.2 &  19.8$\pm$0.2  \\
Green triangle  & 21.5$\pm$0.2 &  19.0$\pm$0.2 \\
Magenta square  & 21.3$\pm$0.3 &  19.7$\pm$0.2  \\ \hline
        \end{tabular}
         \caption{Spectral slope $S$, computed in the $480$--$701\,\mathrm{nm}$ wavelength range and normalised at $480\,\mathrm{nm}$, for the 5 ROIs selected around 
and inside the pits (see Fig.~\ref{fig_hapi_fornasier}), selected at the same position in the images acquired before (December 2014) and during (January 2015) their formation.}
\label{tab:slopes}
\end{center}
 \end{table}

\subsubsection{OSIRIS spectrophotometry} \label{sec_obs_osiris_spectra}

Multispectral analysis of Hapi, together with other active areas, shows sub--units within the region \citep{oklayetal16}. While the region was interpreted as 
covered with well--mixed icy and non--icy materials, Hapi~D is known to have had lower spectral slopes than the areas beyond (towards B3 in Fig.~\ref{fig_hapi02}, upper left) 
and compared to the large boulders (towards Hathor in Fig.~\ref{fig_hapi01}, upper left). The high--resolution views of Hapi~D on 2014 December 10 (Fig.~\ref{fig_hapi01}, upper right) 
acquired before visible changes, and that on 2015 January 22 (Fig.~\ref{fig_hapi02}, upper left) obtained when pit formation was well underway, are available in different 
camera filters. Specifically, the filters F24 ($480.7\,\mathrm{nm}$), F23 ($535.7\,\mathrm{nm}$), F22 ($649.2\,\mathrm{nm}$), and 
F27 ($701.2\,\mathrm{nm}$), are common to the two dates. This allows for the detection of potential exposures of volatiles at the surface 
\citep{pommeroletal15,oklayetal16,oklay_2016b,fornasieretal16}, as well as spatial and temporal variability in ice abundance. We calculated the 
spectral slope for the two filters with the largest wavelength difference,
\begin{equation} \label{eq:slopes}
S=\frac{(R_{701.2}-R_{480.7})\cdot 10^4}{(701.2-480.7)R_{480.7}}. 
\end{equation}
Here, $S$ is in units $\mathrm{\%\,(100~nm)^{-1}}$ and $R$ is the radiance factor at the indicated wavelengths (i.~e., the observed radiance in units 
$\mathrm{W\,m^{-2}\,ster^{-1}\,nm^{-1}}$ divided by $F_{\lambda}/(r_{\rm h}^2\upi)$, where $r_{\rm h}$ is the heliocentric distance and $F_{\lambda}$ with 
units $\mathrm{W\,m^{-2}\,nm^{-1}}$ is the monochromatic solar irradiance at the central wavelength of the filter at $1\,\mathrm{au}$). These radiance factor 
images are produced using the OSIRIS standard pipeline, including corrections for geometric distortions, following the reduction steps described in \citet{tubianaetal15b} 
and \citet{fornasieretal15,fornasieretal19}. The two data sets were  acquired at high and very similar phase angles ($92^{\circ}$ on 2014 December 10, and $93^{\circ}$ on 2015 January 22), 
thus the phase reddening effect, observed on Comet 67P \citep{fornasieretal15,fornasieretal16} should be negligible.

The reliable determination of the subtle variations in spectral slope is delicate, and the results are susceptible to artefacts. To assess the reliability 
of the results, two independent groups analysed the data with two different methods. In the first approach, the sequences were coregistered using the F22 NAC filter 
as reference with dedicated python scripts \citep{fornasieretal19}. To improve the quality of the coregistration to sub--pixel level, the full images were cropped and we 
coregistered only the region of interest around the pit location.  Considering that the shape model does not take into account the morphological surface changes, 
we did not apply a Lommel--Seeliger disk function correction, as normally done for the 67P spectrophotometry \citep{fornasieretal17}, in order 
to avoid biases in the illumination corrections. However, considering that $S$ is normalised at a given wavelength ($480.7\,\mathrm{nm}$), this should not be an issue, because 
the same disk function appears at the numerator and denominator of equation~(\ref{eq:slopes}), and thus any normalisation applied would cancel out. The slope $S$ obtained with the first method 
is shown in Fig.~\ref{fig_hapi_fornasier}.

A comparison of the spectral slope at Hapi~D before and during the appearance of pits reveals some local colour changes. Globally, for the area shown in Fig.~\ref{fig_hapi_fornasier}, the 
spectral slope decreases from $21.0\,\mathrm{\%\,(100\,nm)^{-1}}$ to $20.2\,\mathrm{\%\,(100\,nm)^{-1}}$. A substantial fraction of the region, including the pits, is 
spectrally bluer in January 2015 compared to December 2014. This cannot be explained by phase reddening effects, but points to a local removal of dust caused by the 
cometary activity. In fact, seasonal spectral slope variations have been reported for 67P, with progressively bluer colours as the level of activity increases when approaching 
perihelion \citep{fornasieretal16,fornasieretal17}. This seasonal variation in colours has been attributed to the progressive thinning of the dust coating with increasing activity, with 
relatively bluer colour and thus smaller spectral slope values associated to the exposure of the underlying layers richer in volatiles. Simultaneous VIRTIS and OSIRIS observations have 
indeed confirmed that a smaller spectral slope in the visible range is associated with absorption bands in the near--infrared region due to $\mathrm{H_2O}$ or $\mathrm{CO_2}$ ice 
\citep{baruccietal16,filacchioneetal16,deshapryiaetal17}. Similar correlations are seen in Comets 9P/Tempel~1 and 103P/Hartley~2 \citep{oklay_2016b}.

We also investigated the spectral slope in specific Regions Of Interest or ROI (measuring 3$\times$3 pixels and shown in Fig.~\ref{fig_hapi_fornasier}) at the floor of the pits and 
surroundings before and during their formation. The $S$ values for the ROIs are reported in Table.~\ref{tab:slopes}. We notice that the floor of the pits, represented by the red star, 
cyan asterisk, and green triangle, are spectrally less red than surroundings in January 2015 data, notably compared to the ROI represented by the circle that appears unchanged 
both in morphology and colours in the two selected datasets. The spectral slope inside the pits decreased by about $\Delta S=1.0$--$1.7\,\mathrm{\%\,(100\,nm)^{-1}}$ from December 10, 2014 
to January 22, 2015, or a relative 7 per cent change in only 40 days, compared to the 1 per cent error bars on spectral slope. The slope at the magenta square becomes smaller despite not being located within a pit, 
suggesting that resurfacing may take place without resulting in detectable morphological changes. This is an important point, that we will return to in section~\ref{sec_obs_miro_footprint}.

\begin{figure}
\centering
\scalebox{0.3}{\includegraphics{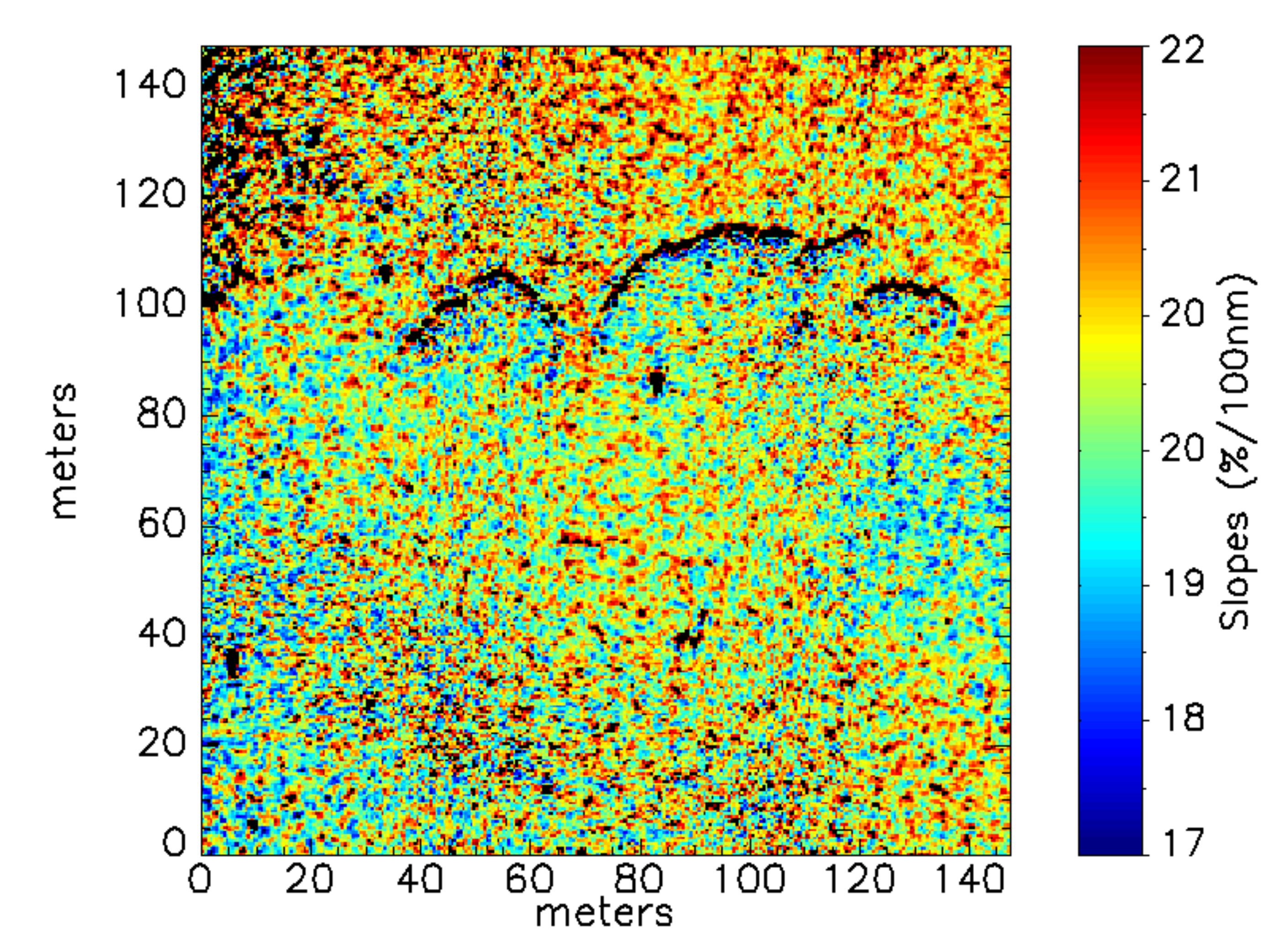}}
     \caption{Spectral slope $S$ map, calculated in an alternative way than was used in Fig.~\ref{fig_hapi_fornasier} (see text), evaluated in the $480$--$701\,\mathrm{nm}$ range and 
normalised at $480\,\mathrm{nm}$. The spectral slopes in the close vicinity of the depressions are about $2\,\mathrm{\%\,(100~nm)^{-1}}$ 
lower than the average of their surroundings. Images MTP012/n20150122t223416529id4bf24.img  and MTP012/n20150122T223425034id4bf27.img.}
     \label{fig_hapi_oklay}
\end{figure}

The 2015 January 22 data were also analysed using a second, alternative, approach. Here, the images taken with different filters were coregistered to a reference image 
(F23 at $535.7\,\mathrm{nm}$) in sub--pixel accuracy using Integrated Software for Imagers and Spectrometers~\citep[USGS \textsc{isis3} software\footnote{\url{http://isis.astrogeology.usgs.gov/index.html}},][]{anderson2004}. 
In this way, the colour artefacts introduced due to rotation of the comet and the motion of the spacecraft are eliminated. Every step of this procedure can be found in \citet{oklayetal16}. The spectral slope $S$ 
was calculated according to equation~(\ref{eq:slopes}) and is shown in Fig.~\ref{fig_hapi_oklay}. The two methods give consistent results -- the depressions are spectrally different from their surroundings. 
In the close vicinity of the depressions in the direction of B1 and B2 there are small areas with mean spectral slopes of $17.60\,\mathrm{\%\,(100~nm)^{-1}}$, which is lower than their surrounding 
with a value of $20.25\,\mathrm{\%\,(100~nm)^{-1}}$. 

Similar feature formation associated with sub--surface water--ice exposures were observed later in various places on the comet including the Hapi region \citep{birchetal19}. 
While the spectral slope definition in the scarps study of \citet{birchetal19} is slightly different than ours (the F41 filter at $882.1\,\mathrm{nm}$ was used instead of the F27 filter at 
$701.2\,\mathrm{nm}$), both studies find about $2\,\mathrm{\%\,(100\,nm)^{-1}}$ lower spectral slopes in the areas close to the depressions. The escarpments in the Imhotep 
region reported in \citet{groussinetal15} are different from those in Hapi. The Imhotep escarpments contained bright material that had almost neutral spectra, indicating those were 
rich in water ice. While the Hapi depressions expose local small areas of bright and presumably water--ice--rich material, those are not as prominent as seen in the escarpments 
described by \citet{groussinetal15}. In conclusion, the pit formation in Hapi~D exposed material that was somewhat richer in ice than the undisturbed surface, but not by much. 
At the time of pit formation, the top $\sim 0.5\,\mathrm{m}$ appears to have been already largely devolatilised.

\subsection{MIRO observations} \label{sec_obs_miro}

\begin{figure*}
\centering
\begin{tabular}{cc}
\scalebox{0.35}{\includegraphics{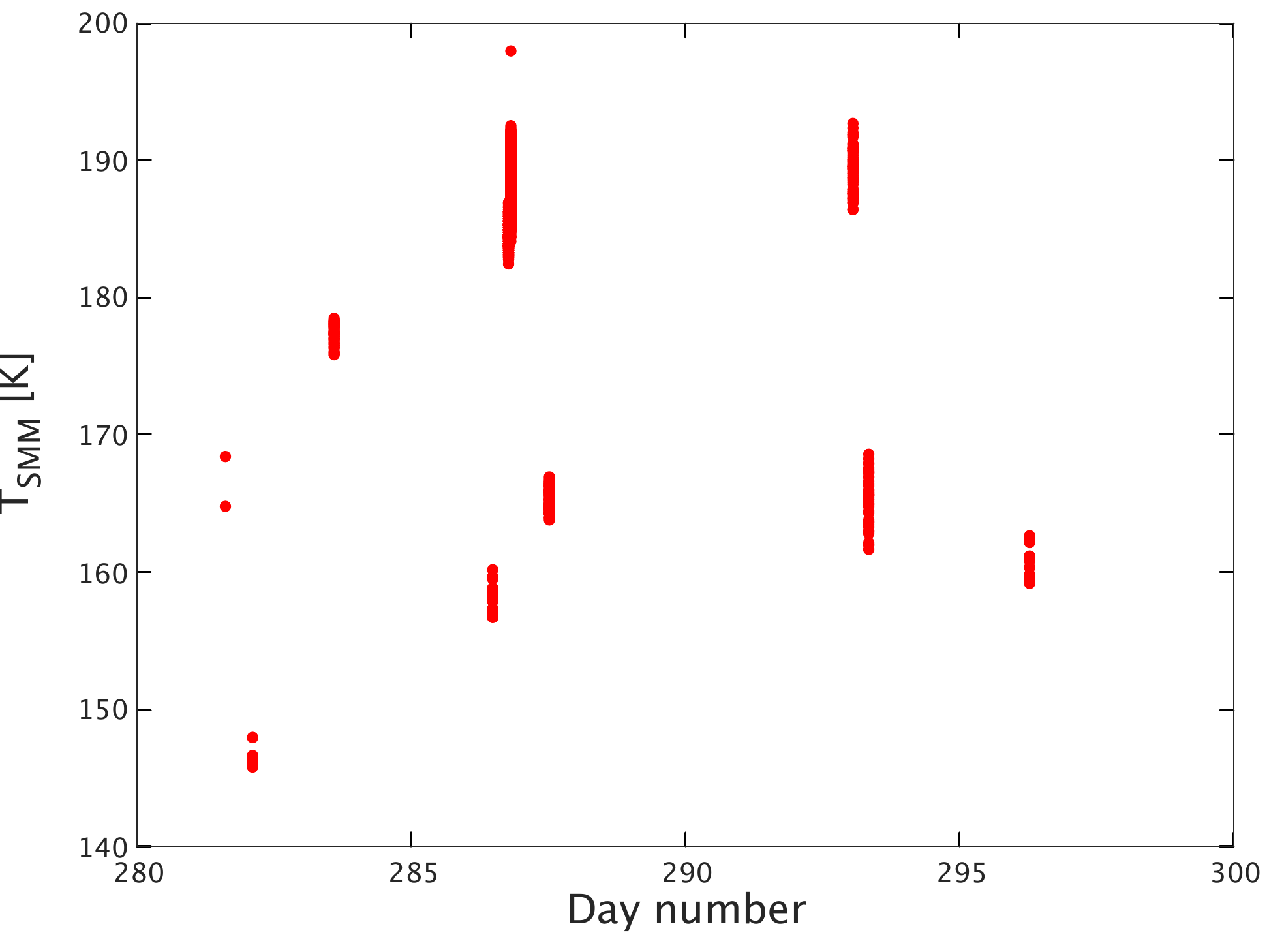}} & \scalebox{0.35}{\includegraphics{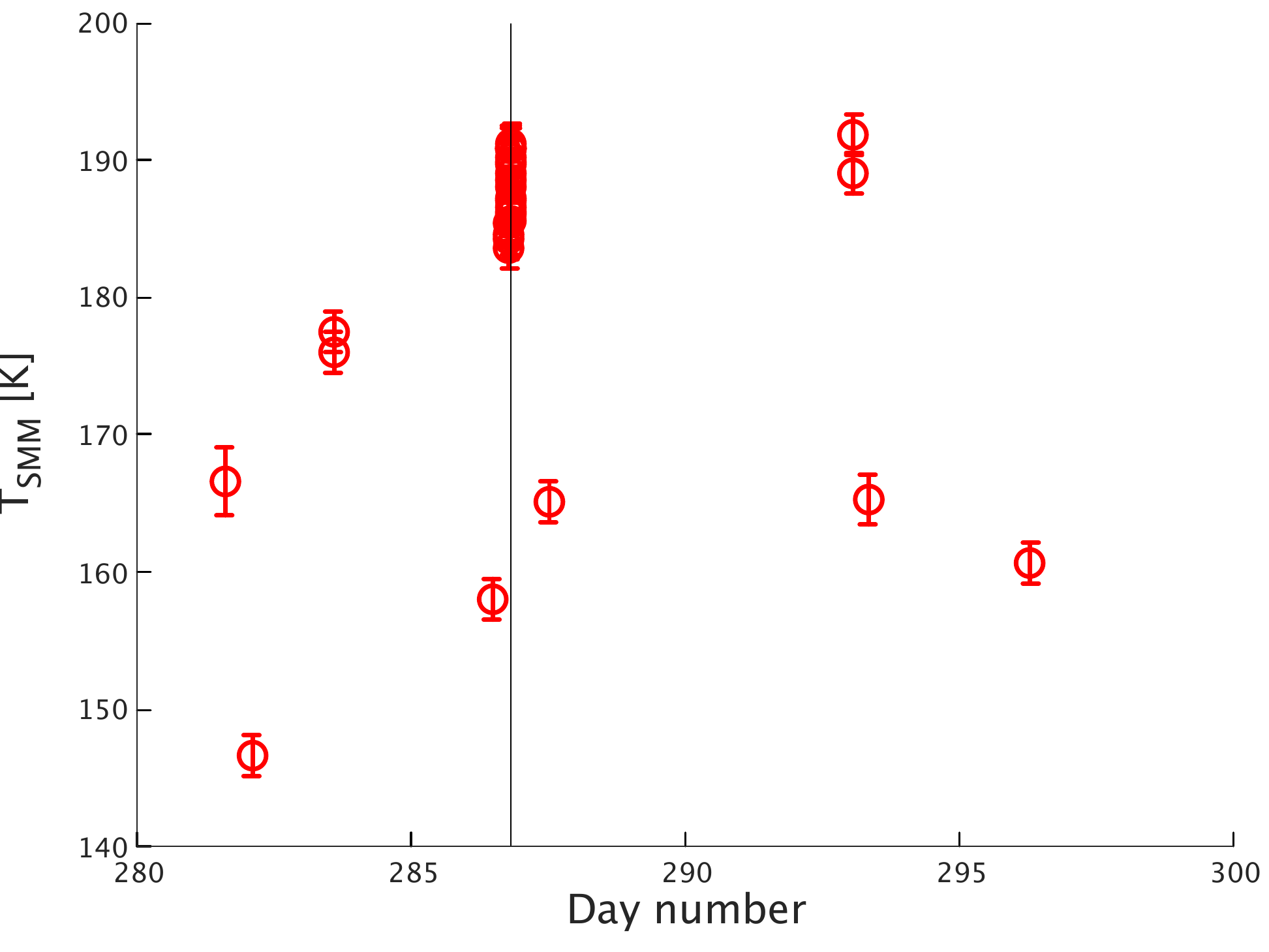}}\\
\scalebox{0.35}{\includegraphics{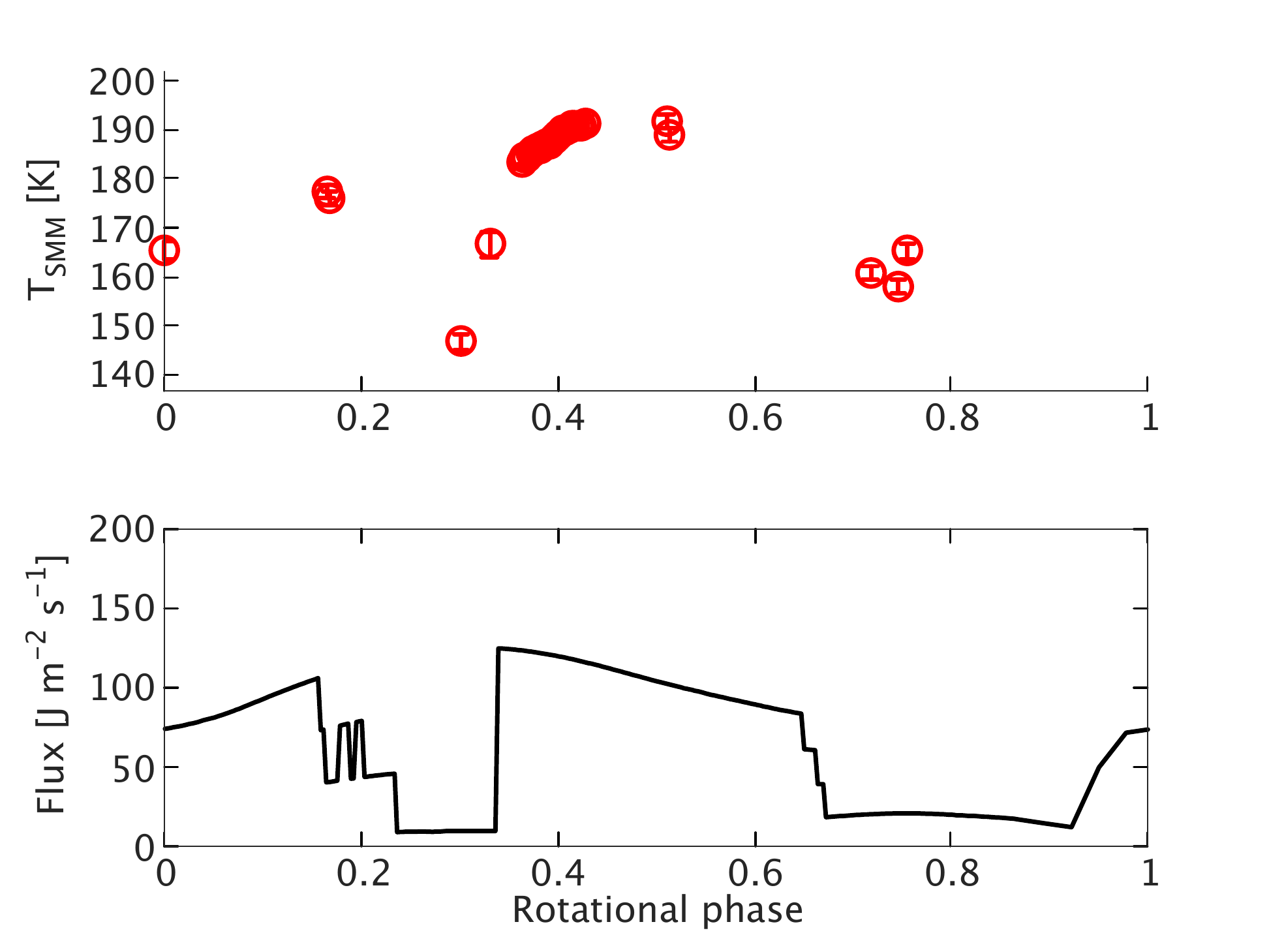}} & \scalebox{0.35}{\includegraphics{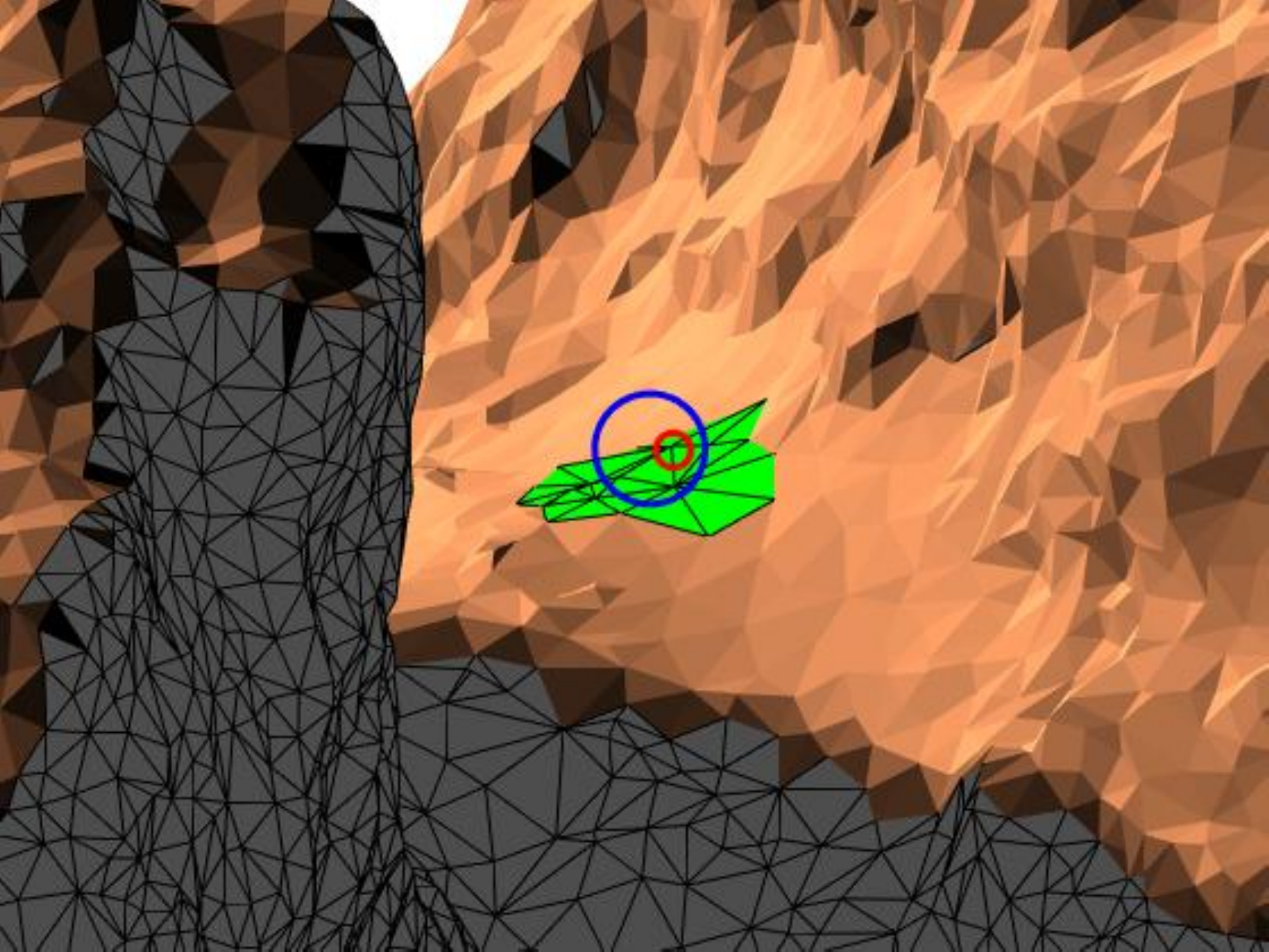}}\\
\end{tabular}
     \caption{\emph{Upper left:} Raw sub--millimetre 1--second continuum MIRO antenna temperature observations of Hapi~D in October 2014, on the original time--line, assuming a beam efficiency (see text) of 
$B_{\rm e}=0.96$. \emph{Upper right:} Binning the SMM data into $2.4\,\mathrm{min}$--wide bins (error bars show the standard deviation for the data within a given bin), on the original time--line. \emph{Middle left:} The binned 
SMM data is here time--shifted to a common master period. \emph{Lower left:} The time--shifted data is here compared to the flux (direct solar illumination plus infrared self--heating) as 
described in section~\ref{sec_models_illum}. \emph{Lower right:} The nucleus of 67P as seen from the \emph{Rosetta} spacecraft on 2014 October 13, at $19:21\,\mathrm{UTC}$, during the continuous stare. 
The intensity of the copper colour scales with the local solar flux, grey areas are in shadow, the green field shows Hapi~D, and the red (SMM) and blue (MM) rings are the MIRO FWHM beam footprints.}
     \label{fig_hapi04}
\end{figure*}

Observations by MIRO have previously been analysed and discussed by, e.~g., \citet{gulkisetal15}, \citet{schloerbetal15}, \citet{choukrounetal15},
\citet{leeetal15}, \citet{biveretal19}, \citet{marshalletal18}, and \citet{rezacetal19,rezacetal21}. We here consider thermal emission from 67P observed by MIRO in two broadband continuum 
channels centred at the wavelengths $\lambda=0.533\,\mathrm{mm}$ and $\lambda=1.594\,\mathrm{mm}$ \citep{gulkisetal07, schloerbetal15}, 
referred to as the sub--millimetre (SMM) and millimetre (MM) channels, respectively. The measured antenna temperatures have been averaged over $1\,\mathrm{s}$ intervals 
and stored at NASA's Planetary Data System (PDS) with a wealth of ancillary information, including the time of observation and the Cheops--system spherical coordinates \citep{preuskeretal15} 
of the interception point of the SMM and MM beam centres with the nucleus surface. Time is measured in `day numbers' with $d_{\rm n}=1$ occurring 2014 January 1 at $00:00\,\mathrm{UTC}$ and 
incremented by unity every $24\,\mathrm{h}$.

The archived antenna temperatures $T_{\rm SMM}''$ and $T_{\rm MM}''$ need to be corrected for two instrumental effects, in order to obtain values relevant for the main beams: spillover and 
beam efficiency \citep{frerkingetal20}. Spillover refers to losses due to incomplete interception of radiated power at the optical components, and is 1.5 per cent at the secondary mirror and 2.5 per cent 
at the primary mirror, amounting to a total of 4 per cent in both 
channels. Spillover--corrected antenna temperatures are therefore $T_{\rm SMM}'=T_{\rm SMM}''/0.96$ and $T_{\rm MM}'=T_{\rm MM}''/0.96$. Beam efficiency refers to 
additional losses caused by ${\rm \mu m}$--scale roughness on optical surfaces and optics misalignment (coma effects). By integrating over the laboratory--measured beam 
patterns out to $100\arcmin$ \citep[far beyond the main beams with Full Width at Half Maximum, FWHM, of $7.5\arcmin$ at SMM and $23.8\arcmin$ at MM;][]{schloerbetal15}, the beam efficiency is found to 
be nearly complete at MM, but $B_{\rm e}=0.96\pm 0.02$ at SMM. Therefore, we apply finally calibrated antenna temperatures $T_{\rm MM}=T_{\rm MM}'$ and $T_{\rm SMM}=T_{\rm SMM}'/B_{\rm e}$. 
We nominally use $B_{\rm e}=0.96$, but occasionally apply other values in the $0.94\leq B_{\rm e}\leq 0.98$ range (sometimes $B_{\rm e}=1$ for comparison), as indicated in the text. When the nucleus is close enough to enter the antenna side lobes (at $1^{\circ}$--$6^{\circ}$ from the beam centre, or at $\stackrel{<}{_{\sim}} 20$--$100\,\mathrm{km}$ from the nucleus), 
small reductions of $T_{\rm MM}$ and $T_{\rm SMM}$ take place because of the extra peripheral signal, that slightly counter the effect of spillover and beam efficiency. We estimate that these are sufficiently small to 
be absorbed by the uncertainties assigned to the antenna temperatures and $B_{\rm e}$. 
Calibrated antenna temperatures are here generically referred to as $T_{\rm A}$ if we do not need to distinguish SMM and MM.

The archive was searched for suitable observations as follows. The $3.1\cdot 10^6$--facet SHAP5 version 1.5 shape model of 67P \citep{jordaetal16} was read into 
the \textsc{meshlab}\footnote{Visual Computing Lab -- ISTI -- CNR, http://meshlab.sourceforge.net/} 
tool that allows the user to visualise the geometry of the nucleus. The region corresponding to the 2015 March~17 extension of the depression (Fig.~\ref{fig_hapi03}, lower right) was identified visually 
and marked using the Z--painting tool of \textsc{meshlab}. Searching the shape model data files for marked facets showed that the region is located between longitudes $6^{\circ}$--$44^{\circ}\,\mathrm{E}$, 
latitudes $44^{\circ}$--$71^{\circ}\,\mathrm{N}$, at a distance $0.53$--$0.58\,\mathrm{km}$ from the nucleus core. The MIRO database was searched for entries having simultaneous SMM and MM 
observations with beam centres within the specified longitude, latitude, and radial ranges, excluding observations at emergence angles $e\geq 80^{\circ}$. This was done for October and November 2014, 
the months prior to the onset of surface changes when \emph{Rosetta} was closest to the nucleus.

\begin{figure*}
\centering
\begin{tabular}{cc}
\scalebox{0.45}{\includegraphics{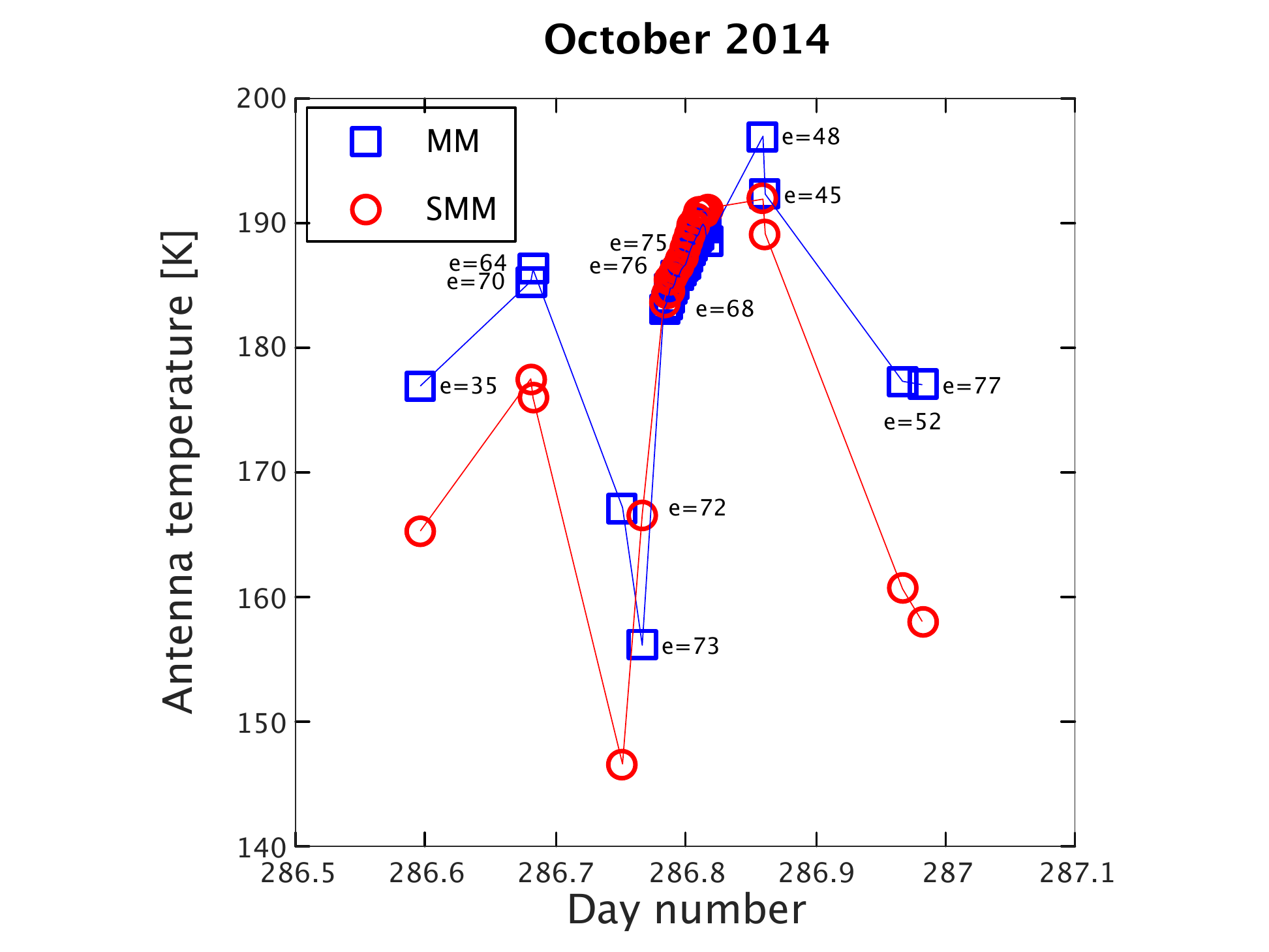}} & \scalebox{0.45}{\includegraphics{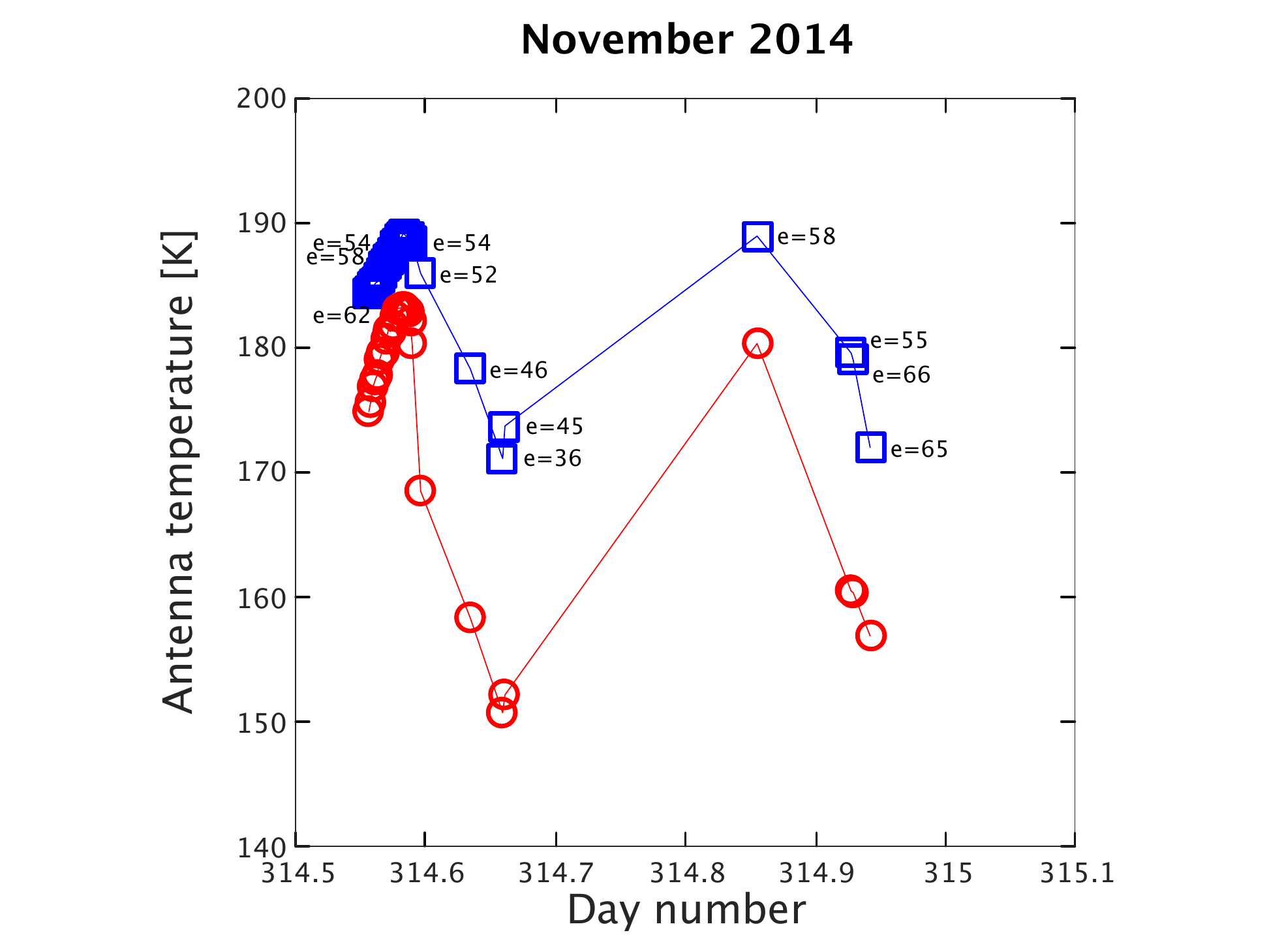}}\\
\end{tabular}
     \caption{The MM and SMM (with $B_{\rm e}=0.96$) binned 1--second continuum antenna temperatures, shifted to the October 2014 (left) and November 2014 (right) 
master periods. Both panels show the beam centre emergence angle $e$ at the time of observation.}
     \label{fig_hapi05}
\end{figure*}

We first discuss the retrieved observations for October (section~\ref{sec_obs_miro_oct}) and November (section.~\ref{sec_obs_miro_nov}). 
We then discuss the spatial resolution of these observations in relation to the size of the area of interest (section~\ref{sec_obs_miro_footprint}).

\subsubsection{MIRO observations in October 2014} \label{sec_obs_miro_oct}

For October 2014, a total of 2,712 observations were found, as shown in the upper left panel of Fig.~\ref{fig_hapi04}. All occurred during the 15--day period between October 8--23. 
Ideally, the region should have been observed continuously during a nucleus revolution to 
allow for a comparison between calculated and observed diurnal temperature variations. However, the longest continuous stare lasted $48\,\mathrm{min}$, and the other data 
points were acquired when the MIRO beams passed Hapi~D briefly and serendipitously during scanning. The dispersions of antenna temperature during these 
crossings are roughly $2$--$8\,\mathrm{K}$, exemplifying the sensitivity to the exact pointing in this region. An empirical diurnal temperature curve was created by sampling 
the available data in $2.4\,\mathrm{min}$--wide bins (upper right panel of Fig.~\ref{fig_hapi04}), roughly corresponding to a nucleus angular rotation of 
$1^{\circ}$, and time--shifting those bins onto a common nucleus master period starting 2014 October 13 at $14:18:12\,\mathrm{UTC}$ ($d_{\rm n}=286.596$). The error bars of the bins are the standard deviation of the temperatures 
within each bin, ranging between $1.5$--$2.4\,\mathrm{K}$. These are larger than the $\sim 1\,\mathrm{K}$ absolute calibration error of each data point \citep{schloerbetal15}, reflecting the fact 
that each bin consists of numerous data points with some temperature dispersion. It was decided to apply a flat $\pm 2.5\,\mathrm{K}$ error bar for the purpose of assessing 
goodness--of--fit with respect to synthetic temperature curve (section~\ref{sec_models_THEMIS_basics_Q}). The temperature increase due to a reduced heliocentric distance during the 
period of observation is smaller than $2.5\,\mathrm{K}$. The time--shifted curve, consisting of 31 bins, is shown in the middle left 
panel of Fig.~\ref{fig_hapi04}, above the calculated illumination conditions at this location during the master period (see section~\ref{sec_models_illum}). The rise and fall of 
the antenna temperature with rotational phase correlates with changes of the incident flux, as expected. As a precaution, the observing geometry at the time of each 
bin was visualised as seen from \emph{Rosetta}, including the illumination and shadowing conditions on the nucleus, as well as the location of Hapi~D 
with respect to the MM and SMM FWHM beams (one example is shown in the lower right panel of Fig.~\ref{fig_hapi04}). If the beams intercepts foreground or background 
terrain in addition to the region of interest (potentially having strong temperature differences), the corresponding bins would be removed for both MM and SMM.  A single bin had to be removed, situated at 
the very end of the time--shifted curve. At the time of observation, Hapi~D was in darkness and observed just above the fully illuminated small lobe in the foreground that 
intercepted parts of the beams. The antenna temperature of the deleted bin is some $\sim 10\,\mathrm{K}$ warmer than other observations acquired at similar rotational phases, 
consistent with the suspicion that the warm small lobe is influencing the measurements.

\subsubsection{MIRO observations in November 2014} \label{sec_obs_miro_nov}

A similar 1--$\mathrm{s}$ continuum database search for November 2014 resulted in 4,936  observations concentrated in a 
20--day period between November 10--30. These data contained a $55\,\mathrm{min}$ stare at Hapi~D plus several substantially 
shorter glimpses. These were binned and time--shifted to create a single diurnal temperature curve, using a master period 
starting on 2014 November 10 at $13:21:01\,\mathrm{UTC}$ ($d_{\rm n}=314.5563$). The temperature increase due to a reduced 
heliocentric distance during the period of observation is smaller than the $\pm 2.5\,\mathrm{K}$ uncertainties. The curve consisted of 36 bins. However, four 
of those had to be removed. In one case, Hapi~D was in darkness and both MIRO beams contained foreground terrain on 
the small lobe that was in full illumination. In the other three cases, Hapi~D was illuminated but the MIRO MM beam 
contained foreground terrain on the large lobe that was in darkness. In all cases, the substantial temperature difference 
between interfering terrain and Hapi~D caused significant anomalies that could not be tolerated.  

Figure~\ref{fig_hapi05} shows the MM and SMM antenna temperature curves plotted in the same diagram, with October and November shown 
side by side for comparison (using the nominal $B_{\rm e}=0.96$ for SMM). Two main differences between the October and November data sets are discernible. Both concern \emph{the first peak and dip} of the curves. 
At these rotational phases, the maximum and minimum incident fluxes are $\sim 100\,\mathrm{J\,m^{-2}\,s^{-1}}$ and $\sim 10\,\mathrm{J\,m^{-2}\,s^{-1}}$, 
respectively, which is true for both months. Despite the similarity in illumination conditions it is seen that: 1) the MM amplitude is reduced from $\sim 30\,\mathrm{K}$ 
in October to $\sim 15\,\mathrm{K}$ in November, but no corresponding change is seen at SMM; 2) the SMM first peak and dip are warmer by some $5$--$8\,\mathrm{K}$ 
in November compared to October, and the MM curve is somewhat warmer as well. These systematic changes from one month to the other, under similar 
illumination conditions, suggest a significant change in the physical conditions of the surface material between October and November.

\begin{figure*}
\centering
\begin{tabular}{cc}
\scalebox{0.35}{\includegraphics{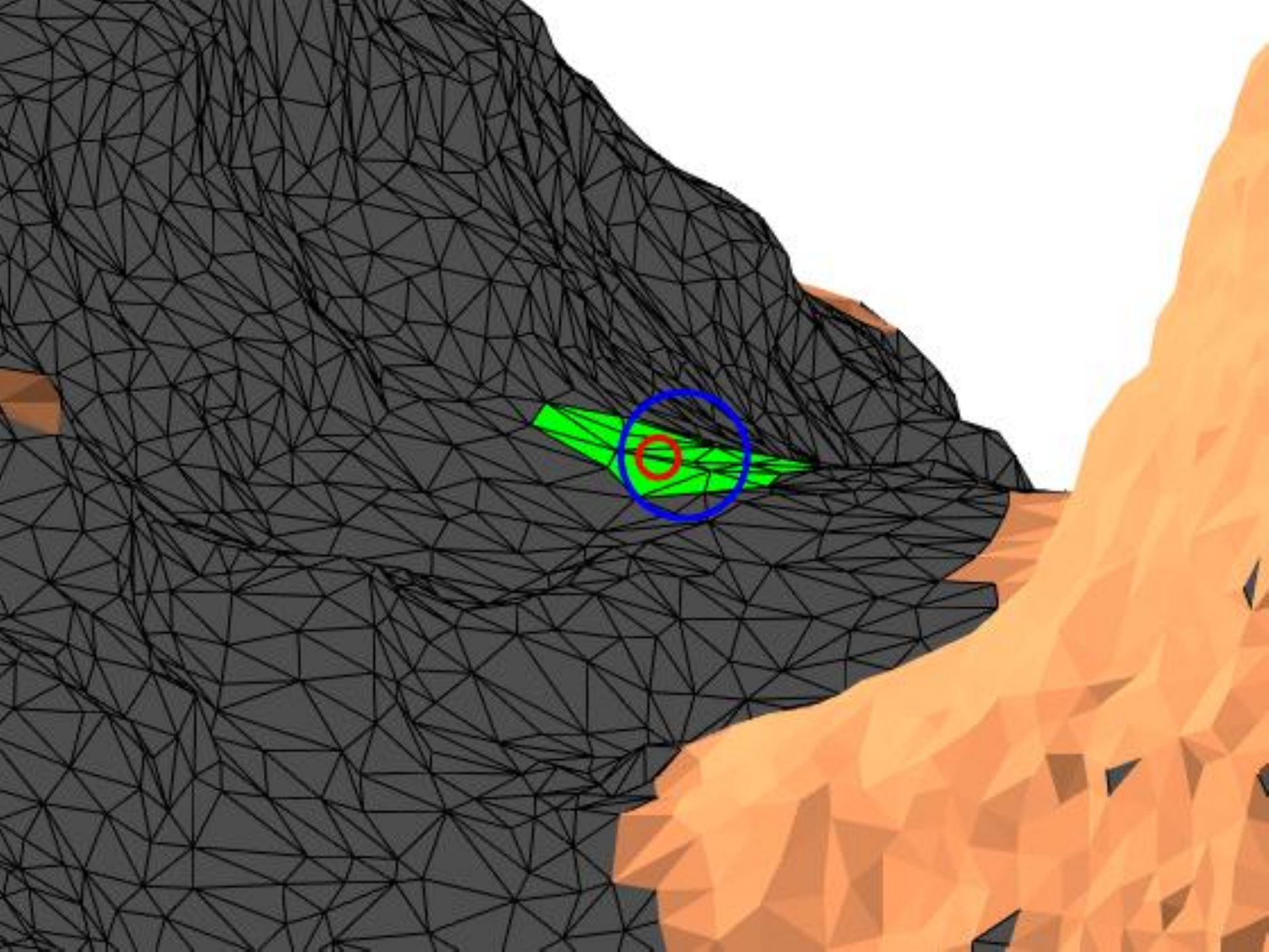}} & \scalebox{0.35}{\includegraphics{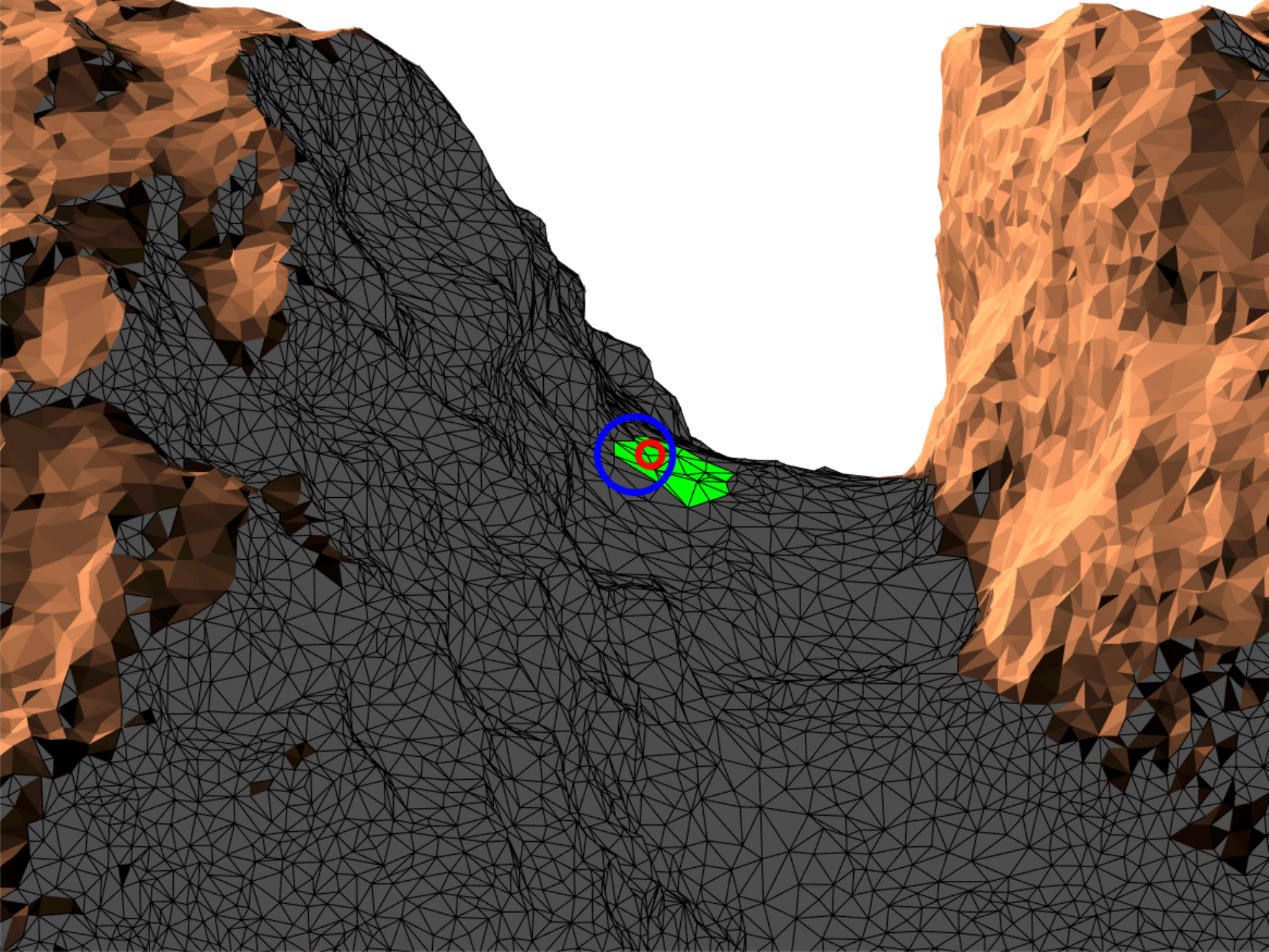}}\\
\end{tabular}
     \caption{\emph{Left:} The viewing geometry on 2014 October 13, at 11:09~UTC, when the emergence angle was $e=77^{\circ}$ and the MIRO beam footprints were the 
largest for the selected October observations: $37$--by--$171\,\mathrm{m}$ in the SMM (red) and $119$--by--$544\,\mathrm{m}$ in MM (blue), when projected onto the nucleus. 
\emph{Right:} The viewing geometry on 2014 November 18, at 04:15~UTC, when the emergence angle was $e=65^{\circ}$ and the MIRO beam footprints were the 
largest for the selected November observations ($71$--by--$166\,\mathrm{m}$ in the SMM and $226$--by--$527\,\mathrm{m}$ in MM). The SMM and MM FWHM footprints are 
shown as red and blue circles, respectively.}
     \label{fig_hapi05a}
\end{figure*}

\subsubsection{MIRO beam footprint sizes} \label{sec_obs_miro_footprint}

As stated previously, the Hapi~D pits covered a region that measured $100\,\mathrm{m}\times 140\,\mathrm{m}$. Pit formation was associated with 
a reduction of the spectral slope, presumably because ice--free material was ejected, thus exposing material containing small amounts of water ice. However, 
Figs.~\ref{fig_hapi_fornasier} and \ref{fig_hapi_oklay} show that similar spectral slope changes took place in a larger region extending for at least another $100\,\mathrm{m}$ 
below block B2. It is reasonable to assume that the mechanism that ejected dust at the pits also was active below Hapi~D, albeit causing too subtle morphological 
changes to be easily recognised at the available resolution. The larger region that experienced spectral slope changes roughly coincides with the white square in the upper left panel of Fig.~\ref{fig_hapi01}, with 
approximate dimensions $200\,\mathrm{m}\times 200\,\mathrm{m}$. The green regions in Figs.~\ref{fig_hapi04} (lower right) and \ref{fig_hapi05a} show shape 
model facets with centres located within Hapi~D, but because the facets are rather large\footnote{The graphical rendering uses a degraded shape 
model with $5\cdot 10^4$ facets.}, they extend somewhat beyond and cover an area similar in 
size to the white square. MIRO receives radiation from an extended area of the nucleus, primarily from within the FWHM beams 
of $7.5\arcmin$ at SMM and $23.8\arcmin$ at MM (amounting to 76 per cent of the collected power). We here compare the sizes of these footprints with the area of interest.

When the October observations were acquired, \emph{Rosetta} was $9.2$--$18.2\,\mathrm{km}$ from the nucleus centre. 
At such distances, the SMM footprint is $20$--$40\,\mathrm{m}$ across on perpendicular surfaces. Considering the spacecraft--Hapi~D distances and 
the emergence angles at the time of observations, the long--axes of the resulting elliptic footprints are larger, 24--$171\,\mathrm{m}$, due to the cosine--effect. 
However, the short--axes are still $20$--$40\,\mathrm{m}$. Note that 22 out of 30 bins have footprint long--axes smaller than $140\,\mathrm{m}$ (the largest extension of Hapi~D). 
Observations were selected so that SMM and MM beam centres are placed within Hapi~D, and visual inspection shows that the SMM FWHM falls entirely 
within the green field. This is exemplified by the red circle in Fig.~\ref{fig_hapi04} (lower right), that projects as $35$--by--$123\,\mathrm{m}$ on the nucleus. 
The confinement to the green field is also illustrated by the largest October 2014 SMM footprint ($37$--by--$171\,\mathrm{m}$ and corresponding to the last bin in Fig.~\ref{fig_hapi05}, 
left, near $d_{\rm n}=287$), shown in Fig.~\ref{fig_hapi05a} (left). This bin is merely $\sim 3\,\mathrm{K}$ cooler than the second last bin, at nearly the same rotational phase but having a 
much smaller $20$--by--$32\,\mathrm{m}$ footprint. This indicates that footprint size effects the antenna temperature rather little. 

During the November observations, \emph{Rosetta} was $29.6$--$41.7\,\mathrm{km}$ from the comet. The SMM footprints are then $65$--$91\,\mathrm{m}$ on 
perpendicular surfaces.  The long--axes of the elliptic footprints during slanted observations ranged $79$--$166\,\mathrm{m}$, but only three of 32 bins had 
footprints in excess of $140\,\mathrm{m}$. The median long--axis decreased from $129\,\mathrm{m}$ in October to $115\,\mathrm{m}$ in 
November, due to smaller $e$--values. The viewing geometry for the largest November 2014 SMM footprint ($71$--by--$166\,\mathrm{m}$ and corresponding to the last 
bin in Fig.~\ref{fig_hapi05}, right, near $d_{\rm n}=314.95$) is shown in Fig.~\ref{fig_hapi05a} (right). 

Clearly, the SMM beam is dominated by emission emanating from the pit--forming Hapi~D region in both October and November, with small contributions from the 
immediate surroundings. Judging from the similarity in spectral slope evolution, which suggests an enhanced capability of ejecting material to space, these surroundings 
likely had thermal properties comparable to those of Hapi~D itself. It will therefore be assumed that the SMM observations sampled the special conditions that led to pit formation.

The situation is more complex at the MM wavelength, because of the $\sim 3$ times larger MM footprint, which made presence of undesirable terrain within the MM beam unavoidable. 
The green area typically fills 25--50 per cent of the MM FWHM, as illustrated by the blue circles in Figs.~\ref{fig_hapi04} and \ref{fig_hapi05a}. In case the physical properties of the 
top few centimetres of Hapi~D surface material are drastically different from those of the surrounding smooth terrain, the dominating signal from the surroundings could distort 
signatures that are unique to Hapi~D.

We approach this problem in two ways. First, we independently search for physical conditions (temperature and composition versus depth and time) that reproduce 
the observed antenna temperature, at both MM and SMM, for a given month. If those solutions are unique and turn out to be identical for MM and SMM, we consider this evidence of 
insignificant distortion of the MM signal. That investigation is presented in section~\ref{sec_results}. Second, we investigate whether a substantial dislocation of the MIRO pointing 
results in a significant change of the measured antenna temperature. If this is not the case, then the thermophysical properties in the vicinity of 
Hapi~D are similar to those in the pit forming area, and distortion effects should be small. 

For this reason, a nearby region $\sim 100\,\mathrm{m}$ to the northwest of Hapi~D was selected, called `Hapi~C' (C for control unit). It consists of 
smooth terrain (seen in the lower half of the upper right image of Fig.~\ref{fig_hapi01}) that is visually indistinguishable from that of Hapi~D. Because both terrains sit on a relatively flat portion of 
the nucleus their illumination conditions are similar. SMM data for Hapi~C, located at longitudes $18^{\circ}\,\mathrm{W}$--$27^{\circ}\,\mathrm{E}$, 
latitudes $61^{\circ}$--$78^{\circ}\,\mathrm{N}$, and $0.49$--$0.53\,\mathrm{km}$ from the nucleus centre, were binned 
and time--shifted to the same November 2014 master period as the Hapi~D data. The Hapi~C curve has only four bins. 
The two data sets overlap temporally at points where Hapi~D has high, intermediate, and low antenna temperatures, i.e., covering a 
wide range of conditions. At these points, the Hapi~C $T_{\rm SMM}$ are located within the error bars of the Hapi~D $T_{\rm SMM}$. The strong similarity between 
the neighbouring regions suggests that Hapi~D is similar to its surroundings, at least in the more shallow surface layer sampled by the SMM. 
The Hapi~C MM curve is also very similar to the Hapi~D MM curve. Although the SMM observations have sufficiently 
high resolution to not mix signals from Hapi~C and D, this is not necessarily the case at MM. However, even if the MM 
observations of Hapi~C might contain peripheral signal from Hapi~D, that contribution is very small due to the 
Gaussian shape of the instrument sensitivity profile that peaks at the beam centre. If Hapi~D had been drastically 
different from its surroundings (at the slightly larger depth sensed at MM wavelengths), the antenna temperatures 
ought to have changed measurably when the beam centre moved from Hapi~D to Hapi~C. The lack of such a change again 
suggests that the two regions have very similar physical properties. It is therefore likely that the MM observations indeed 
are representative of Hapi~D despite the relatively large footprint that includes surrounding terrain. We recall that 
the spectral slope change in Hapi~C (represented by the magenta square in Fig.~\ref{fig_hapi_fornasier} and in Table~\ref{tab:slopes}) 
also was very similar to that in Hapi~D. This further strengthens the notion that both regions experienced the same thermophysical and 
spectral evolution.

\section{Numerical models} \label{sec_models}

Section~\ref{sec_models_illum} describes the calculations performed to obtain illumination conditions for Hapi~D. These are fed to 
two nucleus thermophysical models of different levels of complexity, \textsc{btm} (section~\ref{sec_models_basic}) and 
\textsc{nimbus} (section~\ref{sec_models_NIMBUS}). The resulting physical nucleus temperature, as function of time and depth, are 
fed to the radiative transfer equation solver \textsc{themis} described in section~\ref{sec_models_THEMIS}, that calculates the emitted radiance and 
converts those to synthetic antenna temperatures that can be directly compared with MIRO observations. Note that 
we first describe our nominal approach, which takes advantage of the global flatness of Hapi~D to calculate temperatures for a 
representative location within that region, instead of performing point--by--point evaluations of temperature within the MIRO beams. 
This approach is later motivated in section~\ref{sec_models_THEMIS_basics_test}.

\subsection{Illumination conditions} \label{sec_models_illum}

In order to calculate the illumination conditions at Hapi~D we proceed as follows. The facets at Hapi~D on the $3.1\cdot 10^6$--facet 
SHAP5 version 1.5 shape model had already been identified (section~\ref{sec_obs_miro}). The average outward surface normal vector 
was calculated (because the region is rather flat, the normal vectors of individual facets deviate at most a few degrees from the average). 
Because the ~million--facet shape model is too large to practically carry out some of the calculations to follow, \textsc{meshlab} was then used 
to degrade the SHAP5 version 1.5 shape model to one with $5\cdot 10^4$ facets, and the facets corresponding 
to Hapi~D on that model were identified. Among those, a centrally placed facet with a surface normal close to the average Hapi~D normal 
vector of the high--resolution model was found (aligned to within $1.6^{\circ}$). We took this facet, here called F\#1, to represent Hapi~D.

An algorithm developed by \citet{davidssonandrickman14} was used to identify the 2,698 facets on the degraded shape model that 
are visible from F\#1 (here called `the terrain'). We emphasise that this method carefully avoids including facets that formally 
are along a line--of--sight, but are located behind foreground topography. Terrain facets are capable of shadowing F\#1 by intercepting the line 
between F\#1 and the Sun. They are also capable of illuminating F\#1 with parts of their scattered visual and emitted infrared radiation (this process 
is referred to as `self--heating', meaning one facet of the nucleus radiatively heats another facet). We also identified the 14,672 facets that are visible from at 
least one terrain facet (here called `the surroundings'). These are facets capable of shadowing the terrain. We used the model by \citet{davidssonandrickman14} 
in order to calculate the total flux at F\#1 (direct solar and diffuse self--heating by scattered visual and emitted infrared radiation from the terrain) 
at specific nucleus rotational phases, throughout one 67P orbit around the Sun \citep[applying the nucleus spin axis determined by][]{jordaetal16}. 

Specifically, we calculated all view factors of the terrain facets with respect to F\#1, and the approximate temperatures of terrain facets by balancing 
local direct solar illumination with thermal reradiation, assuming zero albedo. The temperatures were set to zero if a terrain facet was shadowed by 
the surroundings. This temperature distribution across the terrain was then used to calculate the self--heating flux onto F\#1 at any given rotational phase and 
orbital position. The direct solar flux was added, unless the Sun was located behind nucleus topography, as seen from F\#1. 

This is a simplification with respect to the 
nominal model of \citet{davidssonandrickman14}, that evaluates temperatures based on \emph{all mutual exchanges of radiation between facets}, and 
additionally accounts for heat conduction (either along each facet surface normal, or in full 3D). The simplifications meant substantial savings in calculation time, 
while still providing reasonably accurate illumination conditions at F\#1. By ignoring self--heating at terrain facets themselves, the local fluxes are off by typically 
$\sim 10$--$20$ per cent, corresponding to $\stackrel{<}{_{\sim}} 10\mathrm{K}$. By setting the albedo of terrain facets to zero, we artificially remove scattered 
visual radiation (with respect to the real surface) but increase the thermal emission by the corresponding amount. As seen from F\#1, it still receives the same amount 
of energy it would have done for a realistic albedo (assuming that scattering is Lambertian, as is the case for thermal emission). By ignoring heat conduction effects, 
terrain facets do not experience the modest thermal lag of the real nucleus surface. By setting the temperature of shadowed terrain facets to zero, we somewhat reduce 
the self--heating flux at F\#1 (but it would have been worse to allow those facets to illuminate F\#1 as if having been fully exposed to sunlight). In reality, shadowed regions 
would have temperatures of $\sim 130\,\mathrm{K}$, compared to the $\sim 215\,\mathrm{K}$ of the surrounding cliff walls (if illuminated) that provide most self heating. 
To evaluate the error in the flux onto F\#1, introduced by assuming $0\,\mathrm{K}$ for shadowed terrain facets, we made test simulations with shadows at $130\,\mathrm{K}$ 
for the two master periods. In October, the total flux increased by 6.2 (mean) and 3.7 (median) per cent. The corresponding numbers for November were 7.4 and 3.9 per cent. 
In terms of physical temperature, that corresponds to a 2.2--$2.5\,\mathrm{K}$ increase if assuming $\Gamma=0$, but smaller values for our actual modelling, that includes 
a non--zero thermal inertia. The antenna temperatures would be affected even less, therefore we consider our assumptions acceptable.

To further test the effect of uncertainties in the calculated self--heating, it was reduced by half during the October master period and the model re--run.  The resulting reduction of the MM antenna 
temperature was at most $3.8\,\mathrm{K}$, and on average it was $1.6\,\mathrm{K}$.  For the SMM the maximum antenna temperature drop was $5.0\,\mathrm{K}$, and 
the average was $2.0\,\mathrm{K}$. It therefore seems that the self--heating would have to be off by a factor $\sim 2$ in order to give errors in the antenna 
temperatures that start to approach (and locally exceed) the $\pm 2.5\,\mathrm{K}$ error margins applied for the binned data. To have errors in the self--heating flux 
of this magnitude, the temperatures would have to be systematically off by more than $30\,\mathrm{K}$ (e. g., a drop from $200\,\mathrm{K}$ to $168\,\mathrm{K}$ 
corresponds to a 50\% lower flux). Systematic errors in temperature of that magnitude are not likely.

We performed illumination condition calculations for F\#1 every $10^{\circ}$ of nucleus rotational phase for a full $360^{\circ}$ nucleus rotation, at every 12$^{\rm th}$ nucleus rotation, throughout the orbit. 
Because illumination conditions change very slowly with time, an accurately evaluated illumination sequence was therefore copied to the next 11 rotations, 
before the actual conditions were calculated anew (in order to obtain a continuous time sequence). The master periods were evaluated accurately with a higher resolution 
($1^{\circ}$ in nucleus rotation angle) compared to the $10^{\circ}$ applied elsewhere. The calculation of the nucleus rotational phase accounted for the changes in nucleus rotation period caused by 
outgassing torques. We used a table of nucleus rotation periods throughout the mission as determined by ESA in weekly internal \emph{Rosetta}--team communications, 
assembled by Dr. H. U. Keller (private communication), of which some have been published \citep{kelleretal15b}. We validated our calculations of the nucleus rotational phase 
by generating synthetic views of the nucleus (including shadows caused by topography) as seen from \emph{Rosetta}, that were cross--compared with actual OSIRIS images. 

Figure~\ref{fig_hapi04} (lower left) exemplifies the evaluated flux for F\#1 at Hapi~D during the 2014 October 13 master period. As seen, the daytime illumination 
of Hapi~D is interrupted by shadowing (caused by the small lobe) during a $\sim 2\,\mathrm{h}$ period. According to our calculations, the onset of sudden full 
illumination at F\#1 took place just prior to the MIRO observations corresponding to the fifth bin in Fig.~\ref{fig_hapi05} (left). Because of the finite sizes of the SMM and MM 
footprints, the transition is in reality not instantaneous, and it took place at different times in the two channels. At the time of the fifth bin, calculations show that the SMM FWHM 
only viewed illuminated terrain. Indeed, the SMM channel has registered a substantial antenna temperature increase with respect to the previous bin. However, at this point, a 
substantial fraction of the MM FWHM was still in shadow, and the corresponding antenna temperature bin is the coldest in the sample. During analysis, several variants of 
partial illumination were tested, but as it turned out, it was very difficult to match the observed low MM antenna temperature. Therefore, the nominal approach for analysing 
the MM data was to extend the period of shadowing by an additional $\sim 15\,\mathrm{min}$, so that the fifth bin was still in full darkness. In section~\ref{sec_results_Oct_NIMBUS}, we 
find that models including $\mathrm{CO_2}$ reaches the low $T_{\rm MM}$ more readily than models only containing $\mathrm{H_2O}$. If partial illumination of the MM beam at the dip 
had been included, the need for such additional cooling would only have become stronger. The solar flux file generated as described in section~\ref{sec_models_illum} was passed to both 
thermophysical models (described in the following).

\subsection{The basic thermophysical model \textsc{btm}} \label{sec_models_basic}

The first thermophysical model applied in this paper is relatively simple. It considers a homogeneous, porous medium with 
fixed (i.e., temperature--independent) heat conductivity and heat capacity. The Hertz factor (correcting the heat conductivity for porosity) 
is used to give the medium a desired thermal inertia $\Gamma$. The temperature (as function of depth and time) is obtained by solving the 1D energy
 conservation equation (accounting for heat conduction). The boundary condition at the upper surface balances the absorbed radiation (the flux in section~\ref{sec_models_illum} 
corrected for albedo), the thermal emission to space, the conductive heat flow to/from the surface and the interior, and energy consumption due to sublimation 
of surface water ice. Sublimation only takes place from a part of the available surface area, given by the volumetric ice fraction $f_{\rm i}$ of the medium. 
The refractory and icy patches are assumed to be sufficiently small and well--mixed to be isothermal. 

Accounting for the possibility of having thermally isolated patches of hot dust and cold ice would only be important if the ice coverage is 
large, but in Hapi it is $\ll 5$ per cent, except in small zones near shadows \citep{desanctisetal15}. The model does not consider sub--surface sublimation, 
condensation, or vapour transport, nor stratification or erosion.

We refer to this as the Basic Thermophysical Model, or the \textsc{btm}. The governing equations have been provided 
elsewhere \citep[see section.~2.3,][]{davidssonetal21} and are not repeated here. The physical properties taken into consideration 
are admittedly simple: this is intentional. If observational data can be fitted by such a simple model, it means that the 
effects of real and significant deviations from the model limitations are not detectable. This, by itself, disqualifies any claim that higher--order 
physics is necessary in order to explain the observations. However, if no \textsc{btm} fits the data convincingly, or only does so for unphysical parameter values, there is a real need to 
introduce a more elaborate description of the physical environment in the upper layer of the comet nucleus. By carefully scrutinising if,  how, and 
when \textsc{btm} fails, important information is provided that can be used to better understand what additional physical processes 
need to be introduced. For this reason, we make quite some effort in describing `failed' solutions, because we believe they 
are illuminating in the process of better understanding the cometary near--surface region. 

The parameters used to run the \textsc{btm} are summarised in Table~\ref{tab_BTM}. We solve the 1D energy conservation equation using 
the Finite Element Method. We always resolved the diurnal skin depth by $\sim 6$ equidistant grid cells. With the parameters in Table~\ref{tab_BTM}, this corresponded 
to grid cell thicknesses of $(1$--$7.7)\cdot 10^{-3}\,\mathrm{m}$ for a thermal inertia $\Gamma=30$--$230\,\mathrm{MKS}$. We always 
applied 3,400 grid cells, thus modelling the upper $3.4$--$26.2\,\mathrm{m}$ of the nucleus (a zero temperature--gradient boundary condition was applied 
at the lower boundary, placed several times below the seasonal skin depth). All models were run from aphelion (with an initial temperature of $120\,\mathrm{K}$) up to and including the October or November 2014 
master period under consideration, with $10\,\mathrm{s}$ time step.

\begin{table*}
\begin{center}
\begin{tabular}{||l|l|l|l|l||}
\hline
\hline
Quantity & Symbol & Value & Unit & Reference/Comment\\
\hline
Input parameters & & & &\\
\hline
Hertz factor & $h$  & & &\\
Volumetric ice fraction & $f_{\rm i}$ & & &\\
\hline
Constants & & & &\\
\hline
Bond albedo & $A$ & $0.0123\pm 0.007$ & & \citet{fornasieretal15}\\
Ice specific heat capacity & $c_{\rm i}$ & $1200$ & $\mathrm{J\,kg^{-1}\,K^{-1}}$ & \citet{klinger81}\\
 & & & & Average of $100\,\mathrm{K}$, $200\,\mathrm{K}$ values\\
Dust specific heat capacity & $c_{\rm i}$ & $420$ & $\mathrm{J\,kg^{-1}\,K^{-1}}$ & Forsterite, \citet{robieetal82}\\
 & & & & Average of $100\,\mathrm{K}$, $200\,\mathrm{K}$ values\\
Ice latent heat & $L_{\rm H_2O}$ & $2.66\cdot 10^6$ & $\mathrm{J\,kg^{-1}}$ & \citet{tancredietal94}\\
 & & & & $200\,\mathrm{K}$ value\\
Emissivity & $\varepsilon$ &  0.9 & & Standard\\
Ice conductivity & $\kappa_{\rm i}$ & $4.25$ & $\mathrm{W\,m^{-1}\,K^{-1}}$ & \citet{klinger80}\\
 & & & & Average of $100\,\mathrm{K}$, $200\,\mathrm{K}$ values\\
Dust conductivity & $\kappa_{\rm d}$ & $5.00$ & $\mathrm{W\,m^{-1}\,K^{-1}}$ & Forsterite, \citet{horai71}\\
Ice density & $\rho_{\rm i}$ & 917 & $\mathrm{kg\,m^{-3}}$ & \citet{weast74}\\
Dust density & $\rho_{\rm d}$ & 3250 & $\mathrm{kg\,m^{-3}}$ & Forsterite, \citet{horai71}\\
Porosity & $\psi$ & 0.7 & & \citet{patzoldetal19}\\
\hline
Dependent parameters & & & &\\
\hline
Volumetric heat capacity & $c=(1-\psi)(\rho_{\rm i}f_{\rm i}c_{\rm i}+\rho_{\rm d}f_{\rm d}c_{\rm d})$ & & $\mathrm{J\,m^{-3}\,K^{-1}}$ &\\
Heat conductivity & $\kappa=h(f_{\rm i}\kappa_{\rm i}+(1-f_{\rm i})\kappa_{\rm d})$ & & $\mathrm{W\,m^{-1}\,K^{-1}}$ &\\
Thermal inertia & $\Gamma=\sqrt{c\kappa}$ & & $\mathrm{J\,m^{-2}\,K^{-1}\,s^{-1/2}=MKS}$ &\\
Thermal skin depth & $\mathcal{L}=\sqrt{P\kappa/2\upi c}$ & & $\mathrm{m}$ &\\
\hline 
\hline
\end{tabular}
\caption{These parameters were applied in the Basic Thermophysical Model (\textsc{btm}) work. The only free parameters are $f_{\rm i}$ and $\Gamma$ (regulated through $h$). 
 For evaluating the thermal skin depth we applied a fixed approximate nucleus rotation period of $P=12.4\,\mathrm{h}$ \protect\citep{mottolaetal14}. Here, `ice' means $\mathrm{H_2O}$.}
\label{tab_BTM}
\end{center}
\end{table*}

\subsection{The advanced thermophysical model \textsc{nimbus}} \label{sec_models_NIMBUS}

The second thermophysical model applied in this paper is relatively complex. We use the Numerical Icy Minor Body evolUtion Simulator, or 
\textsc{nimbus}, that is fully described by \citet{davidsson21}. In its currently applied form, it considers a porous mixture of dust, 
crystalline water ice, and (in certain models) $\mathrm{CO_2}$ ice. \textsc{nimbus} has the capability to consider amorphous and cubic 
water ice, as well as $\mathrm{CO}$ ice (that partially may be trapped in any of the $\mathrm{H_2O}$ ices and/or the $\mathrm{CO_2}$ ice), 
but such material is not considered here (heating by short-- and long--lived radionuclides is switched off as well). The ices are considered finite resources, 
which means that they may form sublimation fronts that withdraw underground. Heat is transported both through solid--state conduction and through 
radiative transfer, using temperature--dependent heat conductivities that have been measured in the laboratory for all species under consideration 
(heat capacities are temperature--dependent as well). \textsc{nimbus} considers sub--surface sublimation, gas diffusion within the porous medium 
along temperature and vapour pressure gradients, gas venting to space at the upper surface, and recondensation at depth if sufficiently cool regions 
are encountered by the vapour. These processes consume (sublimation), release (condensation), and transports (advection) energy as well. 
In essence, \textsc{nimbus} solves a coupled system of differential equations describing energy and (vapour/ice) mass conservation. 

The full--scale \textsc{nimbus} code considers both radial and latitudinal internal transport of energy and vapour. However, the current work 
applies a specialised version (\textsc{nimbusd} with \textsc{d} for `dust') that sacrifices latitudinal energy and mass transport in order to enable 
erosion of the upper surface (injecting dust and potentially ice into the coma, while thinning the dust mantle). Such simplifications are acceptable because 
of the short duration of the current simulations (latitudinal energy and mass transport are important on geological time--scales). We apply the erosion rate (as function 
of heliocentric distance) of \citet{davidssonetal22}, based on \emph{Rosetta} observations of the 67P dust mass loss. Because there is no risk of confusion 
as to which version is being applied in the current paper, we here refer to \textsc{nimbusd}  simply as \textsc{nimbus}.

The governing equations and many auxiliary functions are described in detail by \citet{davidsson21} and are not repeated here. The initial composition is 
determined by assigning a certain mass ratio between refractories and water ice $\mu$, and a certain molar abundance of $\mathrm{CO_2}$ relative to water 
(see sections.~\ref{sec_results_Oct_NIMBUS} and \ref{sec_results_Nov_NIMBUS}). The initial porosity is determined by requiring a bulk density of $535\,\mathrm{kg\,m^{-3}}$ 
\citep{preuskeretal15}. Over time the porosity and bulk density change because of local sublimation of ice and recondensation of vapour. The heat conductivity and heat capacity 
depend on the continuously changing porosity and temperature, as well as on a porosity--dependent Hertz factor, obtained by using the method of \citet{shoshanyetal02}. 
This gives rise to a certain (range of) instantaneous thermal inertia during nucleus rotation. Deviations from this nominal Hertz factor correction (introduced when it is desirable 
to drastically change the thermal inertia) are described in sections.~\ref{sec_results_Oct_NIMBUS} and \ref{sec_results_Nov_NIMBUS} when necessary. Another important free parameter 
in \textsc{nimbus} in the context of the current simulations is the diffusivity \citep[regulated through tube lengths $L$, tube radii $r_{\rm p}$, and tortuosity $\xi$, see equation~(46) in][]{davidsson21}. 
\textsc{nimbus} is fed with the same illumination sequence as the \textsc{btm} (see section~\ref{sec_models_illum}). \textsc{nimbus} here uses radial grid cells that grow with geometric progression 
from $1\,\mathrm{mm}$ at the surface to $200\,\mathrm{m}$ at the core. It uses a dynamic time step that ensures that certain criteria regarding changes of energy and pressure 
are respected at all times.

\subsection{The radiative transfer model \textsc{themis}} \label{sec_models_THEMIS}

\subsubsection{Fundamentals and nominal procedures} \label{sec_models_THEMIS_basics}

The temperature as function of depth, obtained with \textsc{btm} or \textsc{nimbus}, is fed to a radiative transfer equation solver 
called \textsc{themis} (short for THermal EMISsion) that calculates the radiance emitted toward MIRO, that can be converted to a synthetic 
antenna temperature. \textsc{themis} is a Monte~Carlo--based parallel code implemented in C++/MPI by one of us (Davidsson), 
described here for the first time. \textsc{themis} generates a large number of `test photons' in 
proportion to the Planck function, evaluated at the local temperature and wavelength $\lambda$. Each photon is emitted into 
a random direction, and is followed individually through a sequence of transfers (with lengths chosen at random to statistically conform with a given 
extinction coefficient $E_{\lambda}$) and interactions with the solid medium. The interaction can be absorption (the photon is lost), scattering (with a probability determined 
by the scattering coefficient $S_{\lambda}=w_{\lambda}E_{\lambda}$, where $w_{\lambda}$ is the single--scattering albedo), or escape across the upper boundary. In case scattering takes place, 
a new direction of motion and flight distance to the next interaction are selected, and the process repeats until the photon is absorbed, escapes, 
or penetrates so deep into the medium that it most likely will not make it across the upper boundary. \textsc{themis} keeps track of the number of 
escaping photons, and the angle between the surface normal and their direction of travel at the time of escape. This information is used in order to 
calculate the radiance $R_{\lambda}(e)$ of emitted radiation from the comet surface, as function of emergence angle $e$. Given a certain temperature profile (that changes with 
time during nucleus rotation), the only free parameters of the model are the extinction coefficient $E_{\lambda}$ and the single--scattering albedo $w_{\lambda}$. 

In order to verify the correctness of \textsc{themis} radiances we cross--checked it against known solutions to the radiative transfer problem. Under the condition that the radiation 
source function can be written on the form
\begin{equation} \label{eq:R20}
U(\tau)=U_0+U_1\exp(-\tau/Y),
\end{equation}
(where $U_0$, $U_1$, and $Y$ are constants and $\tau$ is optical depth), and in the limit of isotropic scattering, \citet{hapke93} demonstrated that the equation of radiative transfer can be solved 
analytically, with the radiance given by
\begin{equation} \label{eq:R21}
I_{\lambda}(e)=\gamma H(\mu_{\rm e})\frac{U_0}{\upi}+\frac{Y}{Y+\mu_{\rm e}}\gamma^2H(Y)H(\mu_{\rm e})\frac{U_1}{\upi}
\end{equation}
where $\mu_{\rm e}=\cos(e)$,
\begin{equation} \label{eq:R22}
H(x)=\frac{1+2x}{1+2\gamma x},
\end{equation}
and
\begin{equation} \label{eq:R23}
\gamma=\sqrt{1-w_{\lambda}}.
\end{equation}

\begin{figure}
\centering
\scalebox{0.4}{\includegraphics{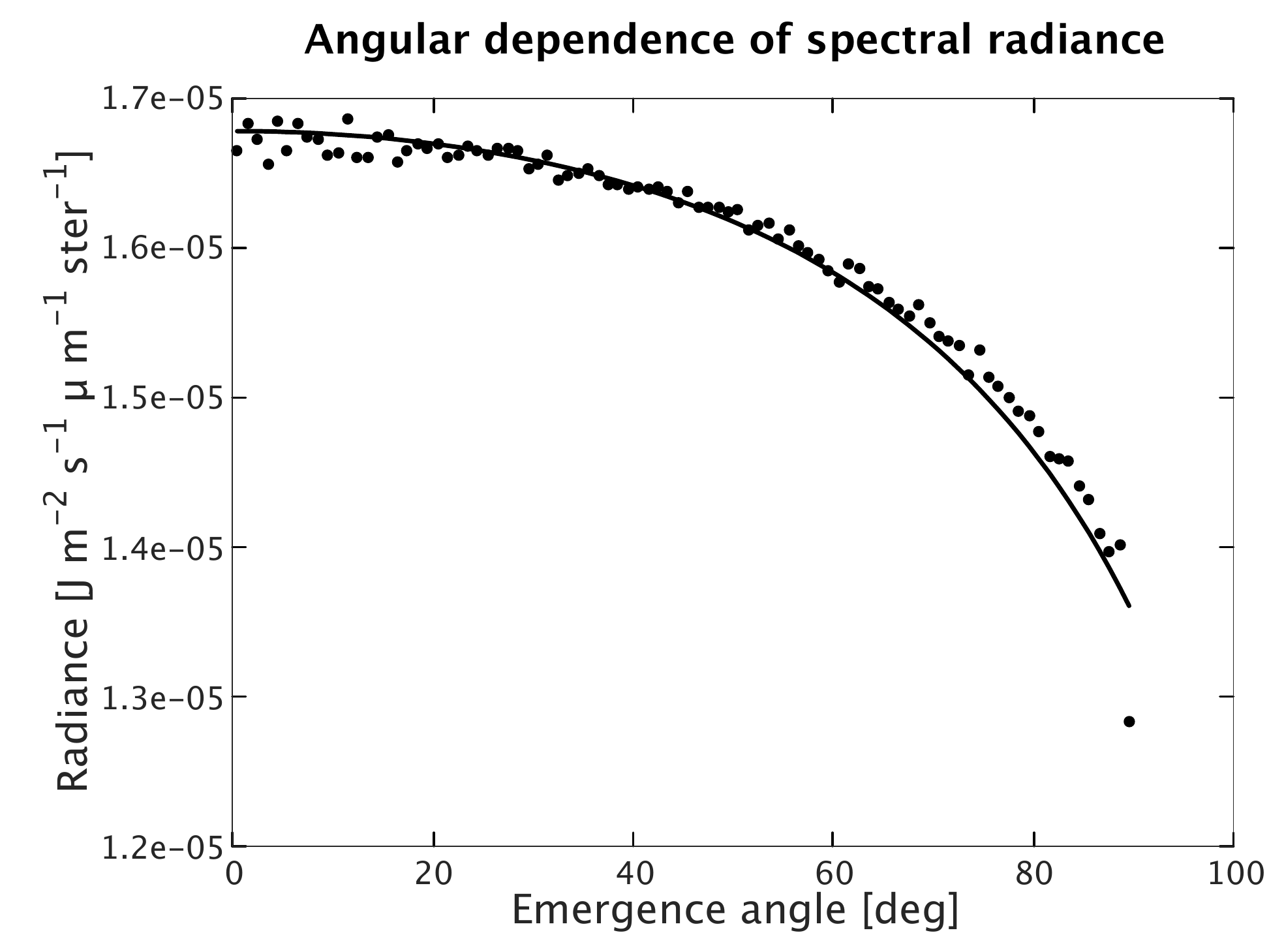}}
     \caption{Verification of the numerical radiative transfer equation solver \textsc{themis} (dotted curve) against the analytical 
Hapke solution, equation~(\ref{eq:R21}), as a solid curve.}
     \label{fig_hapi05b}
\end{figure}

In one specific verification attempt of \textsc{themis}, a temperature profile with $T=200\,\mathrm{K}$ at the surface, falling linearly to $T=150\,\mathrm{K}$ at 
a depth of $z_{\rm max}=0.15\,\mathrm{m}$ was considered. An extinction coefficient of $E_{\rm MM}=338\,\mathrm{m^{-1}}$ was applied for the 
relation between optical and physical depths of $\tau=E_{\rm MM}z$, and the Planck function was used to generate a radiation source function 
$P_{\rm MM}=P_{\rm MM}(\tau)$. The parameter $U_0=5.9875\cdot 10^{-5}\,\mathrm{J\,m^{-2}\,s^{-1}\,\mu m^{-1}}$ was obtained 
from $U_0=P_{\rm MM}(\tau=1.3)$, and the $\tau\leq 1.3$ part of $\ln (P_{\rm MM}(\tau)-U_0)$ was fitted with a linear curve in a least--squares sense, 
having a slope $Y^{-1}$ and crossing with the $\tau=0$ axis of $\ln U_1$. That allowed us to evaluate $U_1=3.3755\cdot 10^{-7}\,\mathrm{J\,m^{-2}\,s^{-1}\,\mu m^{-1}}$ 
and $Y=0.61532$. It was verified that equation~(\ref{eq:R20}) provided a reasonable fit to $P_{\rm MM}(\tau)$ down to an optical depth of unity. Assuming $w_{\rm MM}=0.5$, 
equation~(\ref{eq:R21}) was evaluated for these values of $\{U_0,\,U_1,\,Y\}$, as shown by the solid curve in Fig.~\ref{fig_hapi05b}. \textsc{themis} was run with 
$\lambda=1.594\cdot 10^{-3}\,\mathrm{m}$, $E_{\rm MM}=338\,\mathrm{m^{-1}}$, and $w_{\rm MM}=0.5$, using 150 depth bins, 90 emergence angle 
bins in the upper hemisphere and generating $4\cdot 10^8$ Monte Carlo test photons, of which $8.6\cdot 10^6$ escaped through the upper surface 
and contributed to the \textsc{themis} solution seen in Fig.~\ref{fig_hapi05b} as dots. The similarity between the \textsc{themis} solution and equation~(\ref{eq:R21}) 
is sufficiently high for us to have confidence in our numerical radiative transfer calculations. We emphasise that the reason for using \textsc{themis} instead of 
equation~(\ref{eq:R21}) for the simulations in this paper is that \textsc{themis} can handle any temperature versus depth function, whereas the analytical solution 
is limited to situations where equation~(\ref{eq:R20}) applies.

The \textsc{themis} simulations in this paper all used $9.6\cdot 10^7$ test photons, 90 emergence angle bins (each $1^{\circ}$ wide), 
and a slab thickness $d_{\rm z}$ that would result in radiation being attenuated to a fraction $\exp(-d_{\rm z}E_{\lambda})=10^{-6}$ when escaping at 
the surface, compared to the production at that depth. The calculated radiances $R_{\lambda}(e)$ are converted to a synthetic antenna temperature 
as follows \citep[see][but note that we here express the radiance per wavelength interval instead of per frequency $\nu$ interval, remembering that $R_{\lambda}(e)=c_0R_{\nu}(e)/\lambda^2$]{gulkisetal10}:
\begin{equation} \label{eq:01}
T_{\rm sy}=\frac{R_{\lambda}(e)\lambda^4}{2c_0k_{\rm B}}
\end{equation}
where $c_0$ is the speed of light in vacuum and $k_{\rm B}$ is the Boltzmann constant. The differing antenna temperatures at SMM and MM are obtained by applying 
the corresponding channel wavelength $\lambda$ (furthermore, the applied extinction coefficients $E_{\rm SMM}$ and $E_{\rm MM}$, as well as single--scattering 
albedos $w_{\rm SMM}$ and $w_{\rm MM}$ depend on $\lambda$). 

As seen from Fig.~\ref{fig_hapi05b}, the calculated radiance often depends on the emergence angle $e$ when temperature changes with depth. When calculating 
antenna temperature curves for the master periods, we applied the $e$--values valid at the time of MIRO observations, as seen for several bins in Fig.~\ref{fig_hapi05}. 
In order to obtain a continuous synthetic antenna temperature curve we applied interpolated $e$--values between bins. 

The solution provided by \textsc{themis} works for materials with substantial surface roughness on the millimetre--decimetre size scale, but not for perfectly flat medium/vacuum interfaces, 
for the following reason. When radiation within a solid medium reaches a flat boundary with vacuum, a fraction of the radiation is reflected and the remainder is transmitted 
according to the Fresnel law. Transmission dominates at small emergence angles, but for $e \stackrel{>}{_{\sim}} 50^{\circ}$ the reflected fraction grows rapidly and 
is additionally a function of polarisation: the reflection is stronger for radiation having its electric field oscillating perpendicularly to the plane of incidence than when the field oscillates parallel to it 
\citep[see, e.~g.,][]{lagerros96b}. This means that the emissivity (the ratio between the emerging radiance and the Planck function evaluated for the surface temperature) drops rapidly at 
$e \stackrel{>}{_{\sim}} 50^{\circ}$ (even for isothermal media). Furthermore, the MIRO SMM and MM receivers observe the nucleus in orthogonal polarisations, which potentially could introduce 
systematic differences in the observed intensity at large emergence angles. However, surface roughness randomises the orientation of locally planar boundaries. \citet{lagerros96b} demonstrates 
that surface roughness rapidly removes the part of the emergence--angle dependence of microwave transmission caused by refraction. For this reason, \textsc{themis} does not apply Fresnel expressions 
when evaluating the emerging radiance. The same is true for the \citet{hapke93} solution above: the reflection and transmission governed by the Fresnel expressions are applied on 
constituent grain level (when calculating the single--scattering albedo as function of the refractive index of the grain material) but not at the interface between the medium and the exterior vacuum. 
In fact, the \textsc{themis} solution is very close to that of \citet{hapke93} according to Fig.~\ref{fig_hapi05b}. The question is then whether the surface material at Hapi~D behaves more like a 
flat or a rough medium. Figure~\ref{fig_hapi05} (left) has two bins near $d_{\rm n}=287$ with $e=52^{\circ}$ and $e=77^{\circ}$. The fact that the measured MM antenna temperatures 
are virtually identical, and the SMM differing by merely $\sim 2\,\mathrm{K}$  speaks against a strong $e$--dependence for radiation transmission. Additionally, the antenna temperatures for observations with 
$e=64^{\circ}$ and $e=70^{\circ}$ near $d_{\rm n}=286.7$ are similar as well, at both SMM and MM. To fully account for roughness, we apply a beaming effect that increases $R_{\lambda}$ by 
a factor 1.04 \citep{mueller02}. We therefore consider \textsc{themis} adequate for modelling emission from the rough surface of Hapi~D.

\subsubsection{Testing the need for sub--FWHM temperature resolution} \label{sec_models_THEMIS_basics_test}

If the temperature varies drastically within the MIRO beams, it is necessary to calculate the radiance as function of position within the fields of view. 
However, a sizeable region at and around Hapi~D is nearly flat and similarly illuminated, so the emission within the MIRO beams is expected to be 
homogeneous on macroscopic level. 

To quantify the level of temperature dispersion within the beam footprints, all facets within  the MM FWHM$\times1.5$ footprint were identified (this $36\arcmin$ region collects 92 per 
cent of the observed power) for a given  October or November bin, and their illumination conditions were calculated for the relevant master period, including self heating and shadowing 
from the terrain facets. Temperatures 
versus time and depth were calculated individually for each facet using an ice--free \textsc{btm} model with $\Gamma=30\,\mathrm{MKS}$, for simplicity considering steady--state solutions. 
Radiances emerging from each facet were calculated, assuming $E_{\rm MM}=25\,\mathrm{m^{-1}}$. Those were used to calculate a weighted mean antenna temperature $\tilde{T}_{\rm MM}$, 
equivalent to what would have been observed with the MM channel. The weighting factors included both the projected facet areas as seen from along the line of 
sight, as well as a Gaussian beam power profile with a FWHM of $23.8\arcmin$ that reduces the relative importance of the radiance contribution with increasing distance from 
the footprint centre. This was compared to the corresponding antenna temperature emanating from the representative facet F\#1 itself, $\tilde{T}_{\rm F\#1}$, assumed to apply within the entire footprint. 

Table~\ref{tabnew} reports the emergence angle $e$, the  antenna temperature $\tilde{T}_{\rm MM}$, as well as the difference $\delta \tilde{T}_{\rm MM}=\tilde{T}_{\rm F\#1}-\tilde{T}_{\rm MM}$ 
for a selection of bins from both the October and November data sets (counted consecutively from left to right, as displayed in Fig.~\ref{fig_hapi05}). In October, bins \#1--\#3 constitute the beginning of the curve and the 
first peak, bins \#4 and \#5 are at the first dip, bins \#6, \#20, and \#26 represent the continuous stare,  \#27--\#28 constitute the second peak, and \#30 represents the end of the curve. 
The differences between full--beam and F\#1 synthetic antenna temperatures are admittedly somewhat larger than the $\pm 2.5\,\mathrm{K}$ uncertainties of the data. 
Yet, they are small compared to the level of variation in the observed data that we try to fit (and as it will turn out, compared to the $10$--$30\,\mathrm{K}$ level discrepancies between the data 
and the best models achievable prior to the introduction of the solid--state greenhouse effect and $\mathrm{CO_2}$ ice). Also note that the differences will be substantially smaller at SMM, because 
of the thinner beam.

Because the November footprint sizes tend to be smaller than the October ones (see section~\ref{sec_obs_miro_footprint}), we expected $\delta \tilde{T}_{\rm MM}$ to be 
smaller as well. To verify this expectation, we considered a selection of bins limited to the  sunlit part of the continuous stare (bins \#1--\#17). Indeed, the November 
$|\delta T_{\rm MM}|$ values are smaller, and are all below $2.5\,\mathrm{K}$. Note, again, that $|\delta T_{\rm SMM}|$ would be smaller still. This reinforces the visual impression that Hapi~D (and the extended 
region fitting within the MM footprint) is quasi--flat, thus similarly illuminated, and consequently rather isothermal. 

Because we need to consider illumination conditions with high ($\sim 20\,\mathrm{min}$) temporal resolution for an orbital arc stretching over several years, performing this type of evaluation 
(and the associated thermophysical modelling) for each facet in the MIRO footprints is not computationally feasible -- thus direct illumination, self heating, and 
thermophysical modelling is only made for the representative facet F\#1 (for which \textsc{themis} provides the radiance $R_{\lambda}(e)$). Based on the 
investigation presented in Table~\ref{tabnew}, we consider the observed terrains sufficiently flat and isothermal to allow for such a simplified treatment.

\begin{table}
\begin{center}
\small
\begin{tabular}{||r|l|l|r|r|l|l|r||}
\hline
Oct & & & & Nov & & & \\
Bin & $e$ & $\tilde{T}_{\rm MM}$ & $\delta \tilde{T}_{\rm MM}$ & Bin & $e$ & $\tilde{T}_{\rm MM}$ & $\delta \tilde{T}_{\rm MM}$\\
\hline
\#1 & 34.5 & 171.2 & $+0.5$ & \#1 & 61.7 & 177.1 & $+1.7$\\
\#2 & 69.8 & 181.8 & $-0.7$ & \#3 & 61.1 & 178.3 & $+1.3$\\ 
\#3 & 64.5 & 182.5 & $-1.6$ & \#5 & 59.9 & 179.3 & $+0.9$\\
\#4 & 71.6 & 163.4 & $+5.6$ & \#7 & 58.8 & 180.0 & $+0.6$\\
\#5 & 73.3 & 167.0 & $+3.7$ &  \#9 & 57.7 & 180.6 & $+0.2$\\
\#6 & 67.8 &184.4 & $-4.2$ & \#11 & 58.5 & 181.2 & $-0.1$\\
\#20 & 73.6 & 189.2 & $-3.1$ &  \#13 & 57.6 & 181.6 & $-0.3$\\
\#26 & 74.8 & 191.2 & $-3.5$ & \#15& 55.1 & 181.2 & $-0.5$\\ 
\#27 & 48.2 & 184.6 & $-2.0$ &  \#17 & 54.3 & 181.0 & $-0.7$\\
\#28 & 45.3 & 179.1 & $+1.6$ &  & &  & \\
\#30 & 77.4 & 173.4& $-1.1$ &  &  &  & \\
\hline
\end{tabular}
\caption{This table quantifies the difference between calculating the antenna temperature $\tilde{T}_{\rm MM}$ based on all facets within the MM FWHM$\times 1.5$ footprint, and 
just considering a single representative facet F\#1 with antenna temperature $\tilde{T}_{\rm F\#1}$. Footprint facets have individually calculated illumination conditions, 
\textsc{btm} thermophysical solutions (assuming ice--free media with $\Gamma=30\,\mathrm{MKS}$) and radiances (assuming $E_{\rm MM}=25\,\mathrm{m^{-1}}$), and contribute 
to  $\tilde{T}_{\rm MM}$ in proportion to their projected surface area and beam power at the relevant distance from the beam centre. The small $\delta\tilde{T}_{\rm MM}=\tilde{T}_{\rm F\#1}-\tilde{T}_{\rm MM}$ values 
indicate nearly isothermal footprints due to quasi--flat terrain.}
\label{tabnew}
\end{center}
\end{table}

\begin{figure*}
\centering
\begin{tabular}{cc}
\scalebox{0.4}{\includegraphics{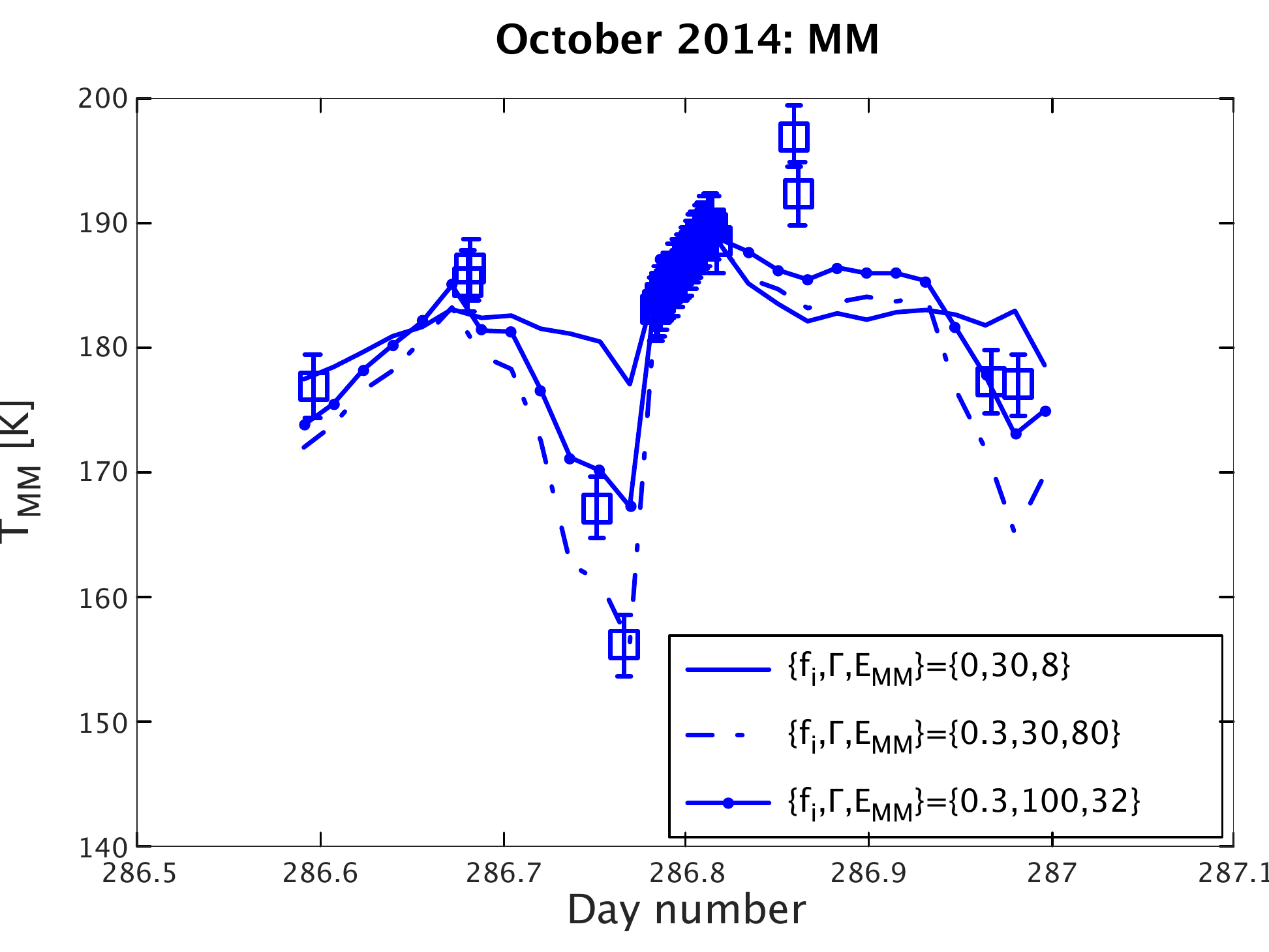}} & \scalebox{0.4}{\includegraphics{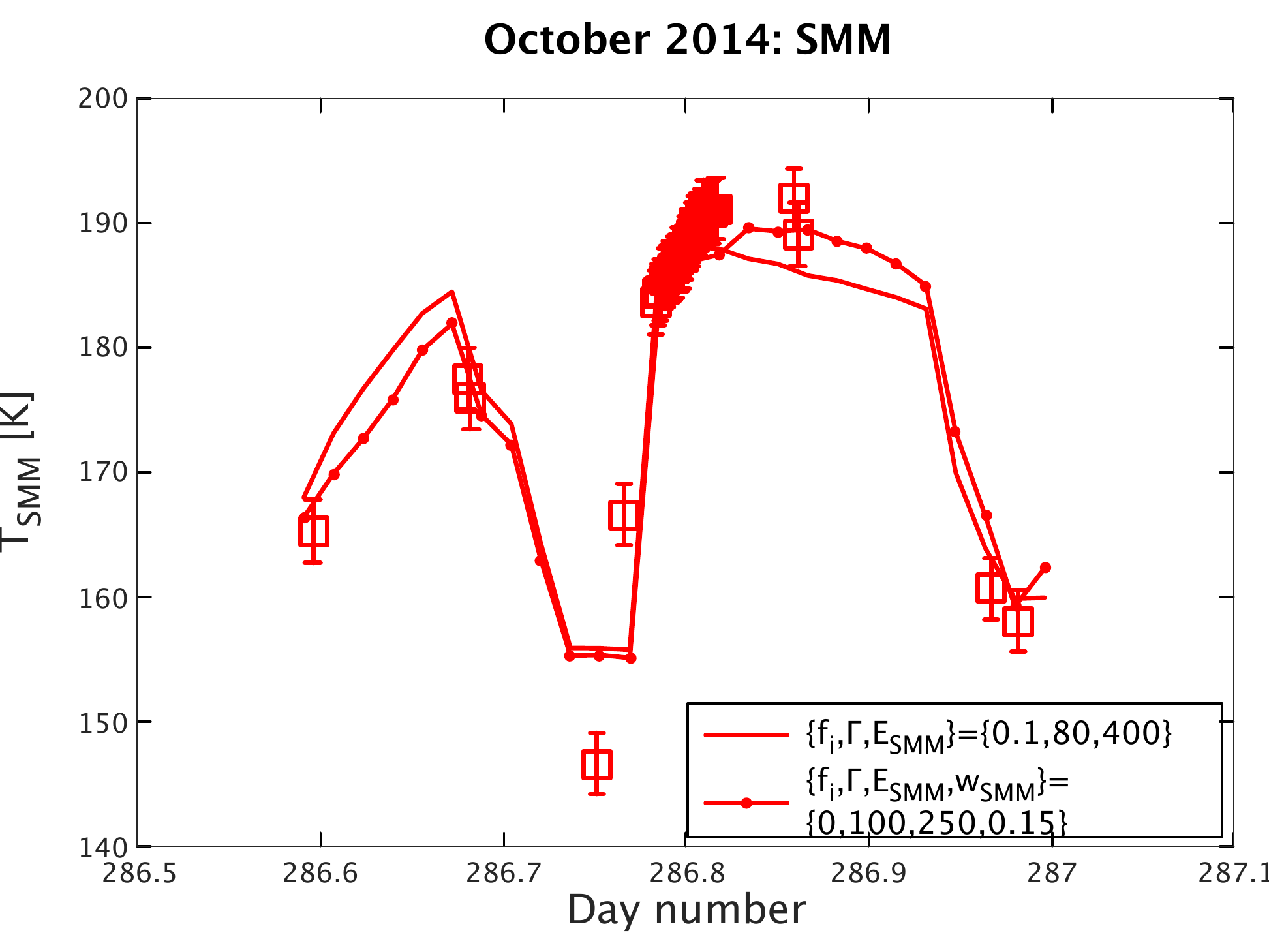}}\\
\end{tabular}
     \caption{\emph{Left}: \textsc{btm} solutions versus MIRO October 2014 MM data. Ice--free media are far too warm, unless an unrealistically low extinction 
coefficient is applied, which then makes the amplitude too small. Adding significant cooling due to water ice sublimation, and applying a low thermal inertia matches 
parts of the curve, but the highest antenna temperatures are not reproduced. No \textsc{btm} model managed to reproduce the MIRO MM data, regardless of  
ice abundance, thermal inertia, extinction coefficient, or single--scattering albedo. \emph{Right:} \textsc{btm} solutions versus MIRO October 2014 SMM data (assuming $B_{\rm e}=0.96$). The 
best available fits do not match the lowest antenna temperature. Furthermore, the low abundance and high thermal inertia is inconsistent with the MM solutions.}
     \label{fig_hapi06}
\end{figure*}

\subsubsection{Goodness--of--fit} \label{sec_models_THEMIS_basics_Q}

In order to quantify the goodness--of--fit of a synthetic antenna temperature curve $T_{\rm sy}(t)$ with respect to the observed MIRO antenna temperature $T_{\rm A}(t)$ with 
uncertainties $\Delta T_{\rm A}$ at $M$ specific instances (the bins), we calculate the incomplete gamma function \citep[see][]{pressetal02}. It is given by
\begin{equation} \label{eq:R24}
Q=Q(\beta,\omega)=\frac{\int_{\omega}^{\infty}e^{-z}z^{\beta-1}\,dz}{\int_0^{\infty}e^{-z}z^{\beta-1}\,dz}
\end{equation}
where $\beta=(M-\eta)/2$, $\eta$ is the number of free parameters in the model, and $\omega=\chi^2/2$, where
\begin{equation} \label{eq:R25}
\chi^2=\sum_M\left(\frac{T_{\rm sy}(t)-T_{\rm A}(t)}{\Delta T_{\rm A}}\right)^2
\end{equation}
is the chi--squared residual between measurements and model.

\emph{Assuming} that $T_{\rm sy}$ represents reality, a single attempt by MIRO to measure the diurnal antenna temperature curve at the $M$ bins would result in some 
residual $\chi^2$, because $T_{\rm A}$ at each bin would be off by some amount consistent with the standard deviation $\Delta T_{\rm A}$. Hypothetically, if MIRO could 
repeat the measurements of the same temperature curve a large number of times (each having a $\chi^2$--value corresponding to that particular attempt) one would 
obtain a distribution of $\chi^2$ values. $Q(\beta,\omega=\chi^2/2)$ is the probability that the particular $\chi^2$--value (for which it is evaluated)  
is exceeded by chance in a single measurement of $T_{\rm A}$. That is to say, the closer $Q$ is to unity, the higher the probability that any discrepancy between $T_{\rm A}$ and $T_{\rm sy}$ is due 
to `bad luck' (i.e., for a large number of $T_{\rm A}(t)$ measurements the averaged curve would approach $T_{\rm sy}$). Of course, $T_{\rm A}(t)$ is the measured 
reality, and $Q\ll 1$ will be interpreted as $T_{\rm sy}$ not being a good representation of that reality. It is customary that a theoretical curve is considered to provide 
a good fit to measured data if $Q\geq 0.01$  \citep{pressetal02}, and this is the criterion that will be considered in this paper.

As previously mentioned, we apply $\Delta T_{\rm A}=2.5\,\mathrm{K}$. The \textsc{btm} has $\eta=3$ at MM ($f_{\rm i}$, $\Gamma$, $E_{\rm MM}$ are free parameters) 
and $\eta=4$ at SMM (addition of the SMM single--scattering albedo $w_{\rm SMM}$). We nominally do not consider scattering at MM (i.~e., $w_{\rm MM}=0$) because 
it is not seen on the Moon \citep{garykeihm78} and we assume that the granular material of the comet is similar to lunar regolith in terms of optical properties. We applied 
the same $\eta$--values for \textsc{nimbus} to allow for a direct comparison between the models.

We noticed one flaw of this method. Some models fitted the dense swarms of bins (see Fig.~\ref{fig_hapi05}) at $d_{\rm n}\approx 286.80$ (October 2014) or 
$d_{\rm n}\approx 314.55$ (November 2014) extremely well, which yielded a high $Q$--value despite the fact that $T_{\rm sy}$ completely missed one or several 
bins elsewhere. In order not to bias the goodness--of--fit to that particular dense part of $T_{\rm A}(t)$, the swarms were replaced by three representative bins 
during $Q$--evaluation. By doing so, false--positives were avoided, and $Q\geq 0.01$ values were only obtained when the entire overall $T_{\rm A}(t)$ was 
well--represented by a $T_{\rm sy}$ model curve. We note that $Q$ is a strongly non--linear function. In cases where visual inspection would suggest 
that there was `almost a fit', the $Q$--value would still be very low ($\sim 10^{-8}$, and often orders of magnitude lower), unless $T_{\rm sy}$ was fully 
statistically consistent with the bins and their error bars. In the following, when stating that a model does not fit the data, it implies $Q\ll 0.01$ 
(for brevity we rarely report the actual value).

\section{Results} \label{sec_results}

\subsection{October 2014} \label{sec_results_Oct}

\subsubsection{October: \textsc{btm} results} \label{sec_results_Oct_BTM}

The \textsc{btm} has two free parameters: the volumetric ice fraction $f_{\rm i}$ and the thermal inertia $\Gamma$ (the specific heat capacity is assumed 
fixed at the values in Table~\ref{tab_BTM}, and only the heat conductivity is varied). A small and sparse grid was 
originally considered, focusing on low thermal inertia and small ice abundance, consistent with previous work \citep{schloerbetal15}. When finding 
solutions proved difficult, that grid was densified and extended into other regions of parameter space that appeared more promising. The final grid 
considered the $0\leq f_{\rm i}\leq 0.3$ interval with $\Delta f_{\rm i}=0.05$ resolution and $\Gamma=\{30,\,50,\,80,\,100,\,130,\,150,\,180,\,200,\,230\}\,\mathrm{MKS}$, 
plus a region with $0.35\leq f_{\rm i}\leq 0.55$ with the same $\Delta f_{\rm i}$ but $\Gamma=\{130,\,150,\,180,\,200,\,230\}\,\mathrm{MKS}$. Thus, a total 
of 88 thermophysical models were run. We first attempted to fit the MM data. For each \textsc{btm} solution, the \textsc{themis} code was run for different $E_{\rm MM}$--values while 
keeping $w_{\rm MM}=0$, in order to find the antenna temperature solution with the smallest possible residual with respect to the empirical data. 
A total of 812 such \textsc{themis} models were considered.

The simulations show that ice--free models are too warm for realistic $E_{\rm MM}$--values \citep[$E_{\rm MM}=26^{+11}_{-6}\,\mathrm{m^{-1}}$ on average, according to][]{schloerbetal15}. 
Generally, an increasing extinction coefficient increases the dayside antenna temperature. This is because a higher extinction coefficient means a larger number of absorbers per volume unit, which also 
translates to a larger number of emitters. At $\Gamma=30\,\mathrm{MKS}$ the extinction coefficient 
has to be reduced to $E_{\rm MM}=8\,\mathrm{m^{-1}}$ (if so, material down to $12\,\mathrm{cm}$ depth would contribute significantly to the signal) for 
the modelled antenna temperature to drop to the level of the data. But then the amplitude is not anywhere near that of the observed curve, as seen in the left panel of Fig.~\ref{fig_hapi06}. 
Increasing the thermal inertia for such models only reduces the amplitude further.

 \begin{figure*}
\centering
\begin{tabular}{cc}
\scalebox{0.4}{\includegraphics{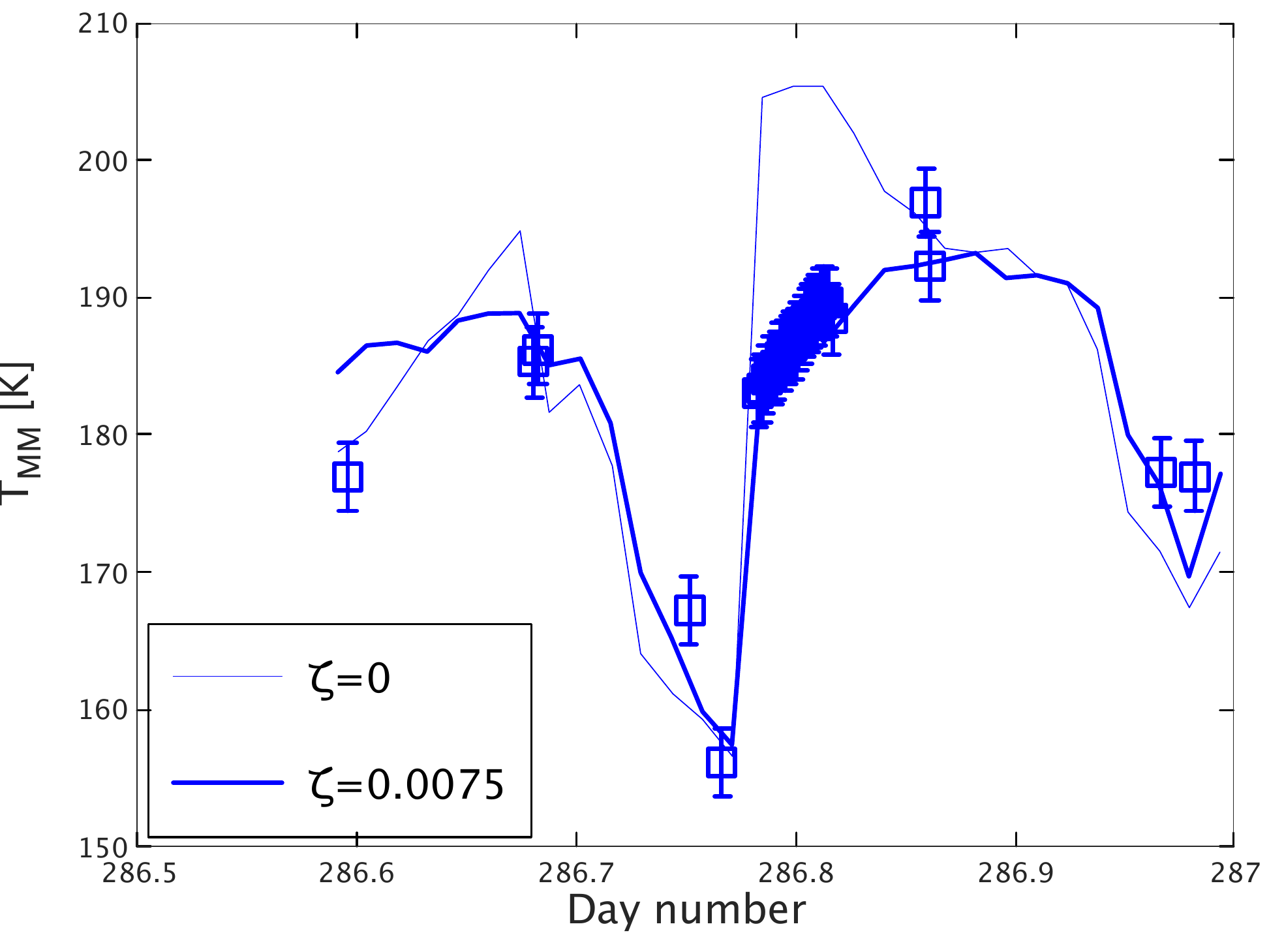}} & \scalebox{0.4}{\includegraphics{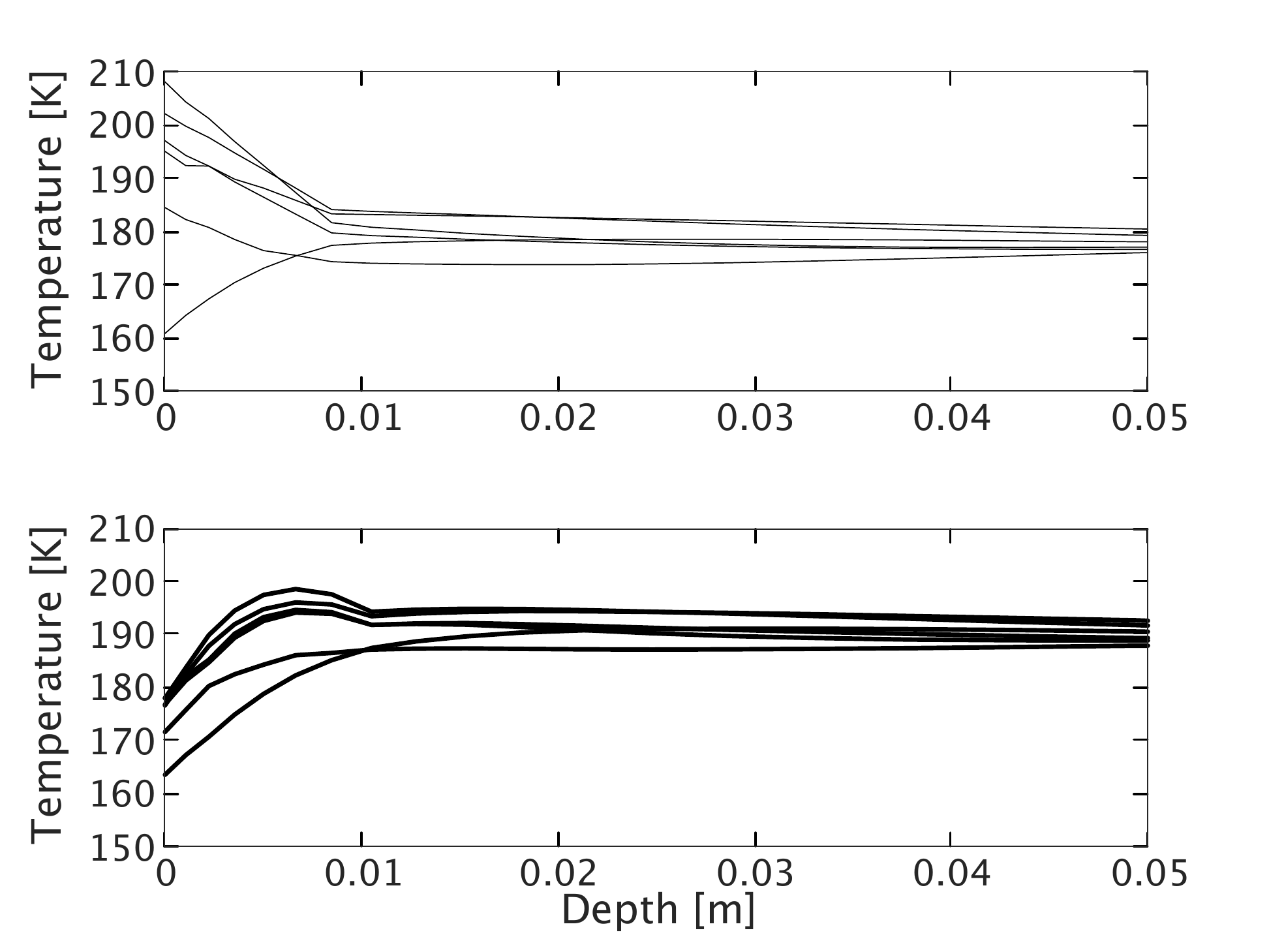}}\\
\end{tabular}
     \caption{\emph{Left:} \textsc{nimbus} solutions versus MIRO October 2014 MM antenna temperatures. The model with $\zeta=0$ had an $8\,\mathrm{mm}$ dust mantle and 
$\{L,\,r_{\rm p}\}=\{10,\,1\}\,\mathrm{mm}$. The model that included the solid--state greenhouse effect ($\zeta=7.5\cdot 10^{-3}\,\mathrm{m}$) had a $9.9\,\mathrm{mm}$ dust mantle 
and $\{L,\,r_{\rm p}\}=\{7.5,\,0.75\}\,\mathrm{mm}$. Both models were applying $E_{\rm MM}=100\,\mathrm{m^{-1}}$.The large temperature reduction near day number 
$d_{\rm n}=286.8$ when $\zeta$ increases is caused by a high sensitivity of the radiative transfer solution to the sign of the near--surface temperature gradient at large emergence angles. 
The significant improvement of the fit as $\zeta$ increases is taken as evidence of the solid--state greenhouse effect in smooth terrain on Comet 67P. \emph{Right:} Physical temperature 
versus depth at a selection of time instances during day and night for $\zeta=0$ (upper) and $\zeta=7.5\cdot 10^{-3}\,\mathrm{m}$ (lower). Note that the solid--state greenhouse effect leads 
to positive near--surface temperature gradients at all rotational phases, and sub--surface temperature maxima.}
     \label{fig_hapi07}
\end{figure*}

When introducing cooling due to sublimation of near--surface water ice, the models can be brought to the general level of the data using 
more reasonable extinction coefficients. Figure~\ref{fig_hapi06}, left panel, shows a case with $\{f_{\rm i},\,\Gamma,\,E_{\rm MM}\}=\{0.3,\,30\,\mathrm{MKS},\,80\,\mathrm{m^{-1}}\}$, 
that roughly has the right amplitude at the first peak and dip. However, there are severe problems at the second peak. The model climbs correctly towards 
the second peak and reproduces the dense collection of data bins, but instead of reaching the high temperatures at $d_{\rm n}\approx 286.85$ the model 
temperature drops. Model curves that are significantly warmer around $d_{\rm n}\approx 286.8$ than at $d_{\rm n}\approx 286.85$ (while the observed curve 
clearly indicates a continuously increasing antenna temperature beyond  $d_{\rm n}\approx 286.8$ that peaks at $d_{\rm n}\approx 286.85$) are here referred to as 
having a `peak shape problem'. Additionally, the second dip near $d_{\rm n}=287$ becomes far too cold. Increasing the thermal inertia, as illustrated in the left panel of Fig.~\ref{fig_hapi06} by the 
$\{f_{\rm i},\,\Gamma,\,E_{\rm MM}\}=\{0.3,\,100\,\mathrm{MKS},\,32\,\mathrm{m^{-1}}\}$ model, reduces the amplitude (thus ruining the fit at the first dip), without 
improving the situation at the second peak. There seemed to be a weak improvement by simultaneously increasing both $f_{\rm i}$ and $\Gamma$, but the 
best achievable model with $\{f_{\rm i},\,\Gamma,\,E_{\rm MM}\}=\{0.4,\,200\,\mathrm{MKS},\,330\,\mathrm{m^{-1}}\}$ performed as poorly as the 
$\{f_{\rm i},\,\Gamma,\,E_{\rm MM}\}=\{0.3,\,30\,\mathrm{MKS},\,80\,\mathrm{m^{-1}}\}$ model at the first dip and second peak, although the second dip 
was closer to the data.
 
Having failed to find convincing MM solutions with \textsc{btm} for October 2014 when using $w_{\rm MM}=0$, scattering was introduced in the \textsc{themis} 
modelling. We did this reluctantly because scattering is not expected at the MM wavelength, as mentioned previously. The effect of 
increasing $w_{\rm MM}$ is to lower the antenna temperature, so we applied $w_{\rm MM}>0$ for models that were too hot and increased the single--scattering 
albedo in steps of $\Delta w_{\rm MM}=0.05$ until reaching the level of the data, resorting to fine--tuning when it was deemed meaningful. We considered 777 $w_{\rm MM}>0$ 
models spread across the thermophysical model grid, bringing the total of \textsc{themis} MM runs to 1,589. The smallest residuals were obtained for 
$\{f_{\rm i},\,\Gamma,\,E_{\rm MM},\,w_{\rm MM}\}=\{0,\,130\,\mathrm{MKS},\,50\,\mathrm{m^{-1}},\,0.3\}$. Although most of the data bins could be 
fitted, the modelled temperature remained $10\,\mathrm{K}$ too warm at the first dip (despite the fact that the MM model assumed complete darkness at this point, even when some 
illumination of the actual footprint took place, see section.~\ref{sec_obs_miro_footprint}). In conclusion, no convincing reproduction of October 2014 MM 
data could be found using thermophysical solutions based on \textsc{btm}.

Next, the SMM case was considered. Another 18 thermophysical models were added to form a rectangular grid with $0\leq f_{\rm i}\leq 0.4$  ($\Delta f_{\rm i}=0.05$ resolution) and\\ 
$\Gamma=\{30,\,50,\,80,\,100,\,130,\,150,\,180,\,200,\,230\}\,\mathrm{MKS}$. Higher $f_{\rm i}$--values were not considered. For each $\{f_{\rm i},\,\Gamma\}$ combination, 
\textsc{themis} was run at the SMM wavelength to find the $E_{\rm SMM}$--value that minimised the residual between the synthetic antenna temperature curve and the empirical bins, 
assuming $w_{\rm SMM}=0$. In total, 208 such simulations were made. The closest match was found for $\{f_{\rm i},\,\Gamma,\,E_{\rm SMM}\}=\{0.1,\,80\,\mathrm{MKS},\,400\,\mathrm{m^{-1}}\}$. 
As seen in the right panel of Fig.~\ref{fig_hapi06}, that model is not sufficiently cold at the first dip, and failed to reproduce the second peak. Scattering is more likely to take place 
at SMM than at MM wavelengths, therefore another 258 \textsc{themis} simulations with $w_{\rm SMM}>0$ were considered, bringing the total number of simulations to 466. 
The best of these simulations, having $\{f_{\rm i},\,\Gamma,\,E_{\rm SMM},\,w_{\rm SMM}\}=\{0,\,100\,\mathrm{MKS},\,250\,\mathrm{m^{-1}},\,0.15\}$ improved the fit at the second 
peak modestly (though not performing well at the continuous stare), and still failed to match the first dip. After having completed this extensive investigation of the parameter space we had to conclude that 
the  Basic Thermophysical Model is not capable of reproducing the MIRO measurements and that its time--dependent temperature solutions,  at and below the surface, 
do not match those of the comet. We note that the discrepancies were far too large to realistically be attributed to any of the simplifications described in section~\ref{sec_models}.

\subsubsection{October: \textsc{nimbus} results}  \label{sec_results_Oct_NIMBUS}

As a starting point, a \textsc{nimbus} model with refractory to water ice mass ratio $\mu=2$, tube length and radius $\{L,\,r_{\rm p}\}=\{100,\,10\}\,\mathrm{\mu m}$, and tortuosity 
$\xi=1$ was considered \citep[these values were consistent with the pre--perihelion $\mathrm{H_2O}$ and $\mathrm{CO_2}$ production rates observed by \emph{Rosetta}/ROSINA, according 
to the \textsc{nimbusd} modelling by][]{davidssonetal22}. 
In order to prevent the thermal inertia from decreasing below $\sim 30\,\mathrm{MKS}$ at high porosities, a Hertz factor $h=0.0016$ ceiling was introduced (this constraint was later modified to 
consider lower thermal inertia in a controlled manner). The model was propagated from 
aphelion on 2012 May 23 to 2014 October 6 (at $3.23\,\mathrm{au}$). To illustrate the weak water--driven activity in this $\sim 2.3\,\mathrm{yr}$ segment of the orbit, the 
water withdrew merely $8.5\,\mathrm{mm}$ under the surface and the dust mantle was eroded by only $4\,\mathrm{mm}$. The temperature distribution at the final date was used 
as initial condition for a series of simulations with different model parameters. Thus, a time period of $\sim 8$ days 
for thermal re--adjustments due to the new parameters were applied, before reaching the October 14 master period.  A first series of 15 simulations systematically explored 
the effects of considering different dust mantle thicknesses ($2$--$8\,\mathrm{mm}$), thermal inertia ranges, and $\{L,\,r_{\rm p}\}$ values. The thinner the dust mantle, and the larger the $\{L,\,r_{\rm p}\}$ values, 
the more effective is the cooling due to water sublimation. The lower porosity below the dust mantle (because of the presence of ice), leads to enhanced heat conductivity 
and heat capacity in that region compared to the dust mantle. Furthermore, for $r_{\rm p}\stackrel{>}{_{\sim}} 1\,\mathrm{mm}$, radiative heat transport starts to add 
visibly to the instantaneous thermal inertia. For the largest diffusivity considered, $\{L,\,r_{\rm p}\}=\{28,\,6.6\}\,\mathrm{mm}$, the total thermal inertia is $\sim 25\,\mathrm{MKS}$ at 
night and $\sim 50\,\mathrm{MKS}$ at day.
 
Moving the sublimation energy sink from the top cell (as in the \textsc{btm}) to a sub--surface region surrounding the 
water ice sublimation front (as in \textsc{nimbus}) unfortunately did not introduce any significant qualitative changes to the synthetic antenna 
temperature curves. Applying temperature--dependent heat conductivities and heat capacities had no significant effect either. The peak shape problem 
remained severe, illustrated by the thin blue curve in the left panel of Fig.~\ref{fig_hapi07}. Note that \textsc{btm} produced curves of that type as well, when 
using higher $E_{\rm MM}$ values than in Fig.~\ref{fig_hapi06} in order to force a match with the warmest MIRO antenna temperatures at $d_{\rm n}\approx 286.85$. 
In short, a properly modelled dust mantle, overlaying a sublimating icy interior, did not provide a better match than the admittedly nonphysical \textsc{btm} scenario 
where all cooling takes place at the very surface.

We decided to try to understand the reason why the modelled antenna temperature, when applying a sufficiently large $E_{\rm MM}$ value to match the 
hottest measurements at $d_{\rm n}\approx 286.85$, yielded an even higher temperature at $d_{\rm n}\approx 286.8$ that exceeded the 
measurements by $\sim 20\,\mathrm{K}$ or more. First it was recognised that the MIRO observations near $d_{\rm n}\approx 286.8$ were 
acquired at very large emergence angles, sometimes exceeding $e=75^{\circ}$, as seen in Fig.~\ref{fig_hapi05}. Second, because of the relatively 
strong solar illumination at this point of the diurnal curve, the temperature gradient is very steep, having a high surface temperature and cooling 
rapidly with depth (upper right panel of Fig.~\ref{fig_hapi07}). The analytical solution to the equation of radiative transfer (equation~\ref{eq:R21}) was scrutinised 
to understand how the radiance would depend on emergence angle in such conditions. 
In the absence of scattering ($w_{\rm MM}=0$ and $\gamma=1$), the $H(x)$ function (equation~\ref{eq:R22}) becomes constantly equal to unity. 
It means that the first term in equation~(\ref{eq:R21}) is a constant, and that $H(\mu_{\rm e})$ in the second term no longer introduces any $e$--dependence. 
It is therefore only the term $Y/(Y+\mu_{\rm e})$ that is changing with emergence angle, taking a value of $Y/(Y+1)<1$ at nadir observations that 
grows towards unity as $e\rightarrow 90^{\circ}$ and $\mu_{\rm e}\rightarrow 0$. Because $U_1>0$ for a negative temperature gradient ($T$ falling with depth), 
the observed radiance is expected to increase strongly with $e$, growing from $I_{\lambda}(0)=(U_0+|U_1|Y/(Y+1))/\upi$ towards $I_{\lambda}(90^{\circ})=(U_0+|U_1|)/\upi$. 
Therefore, the unusually high modelled antenna temperature near $d_{\rm n}\approx 286.8$ appeared to be a consequence of a strongly negative temperature gradient 
in combination with very large $e$--values.

\begin{figure*}
\centering
\begin{tabular}{cc}
\scalebox{0.35}{\includegraphics{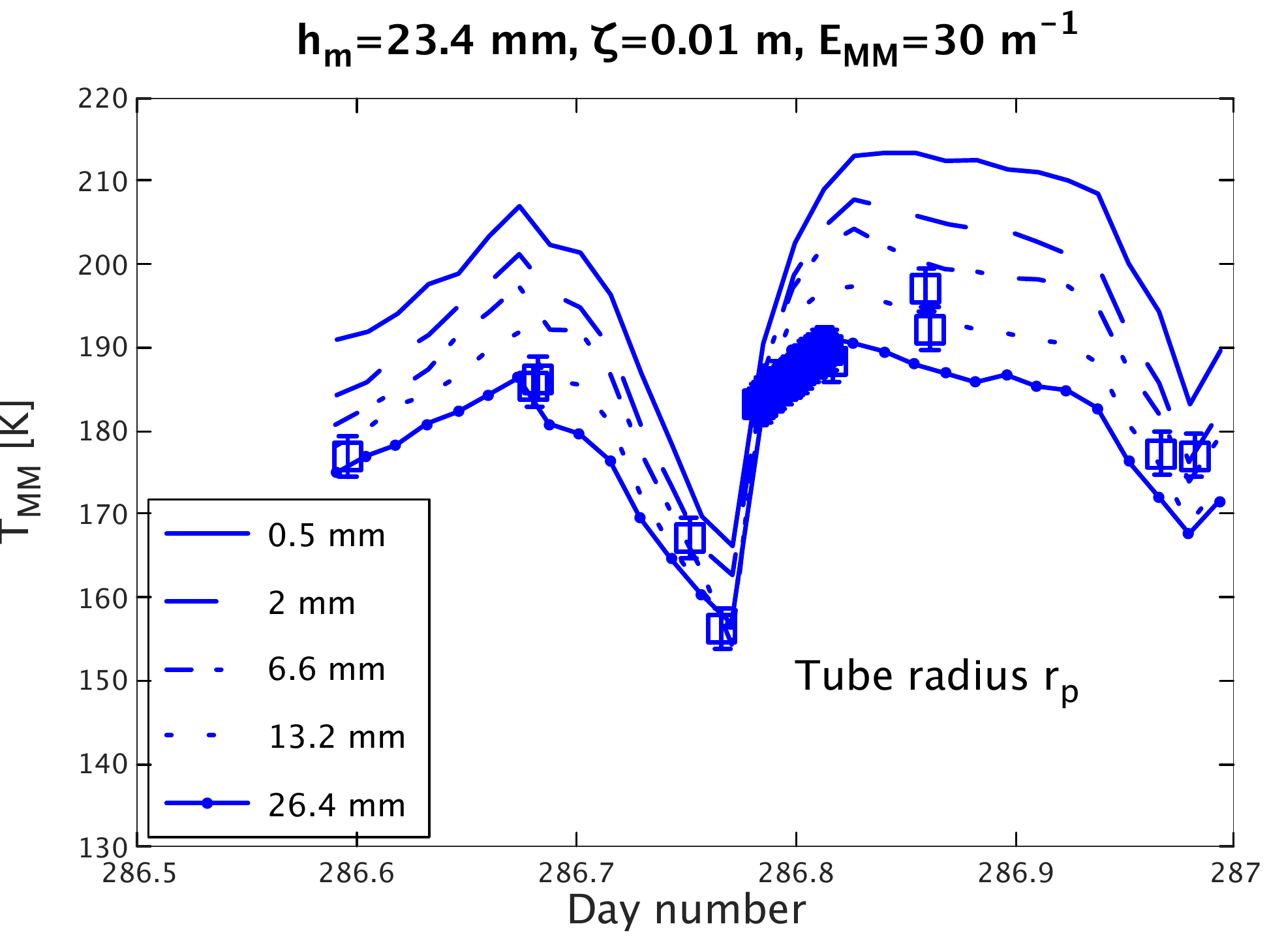}} & \scalebox{0.35}{\includegraphics{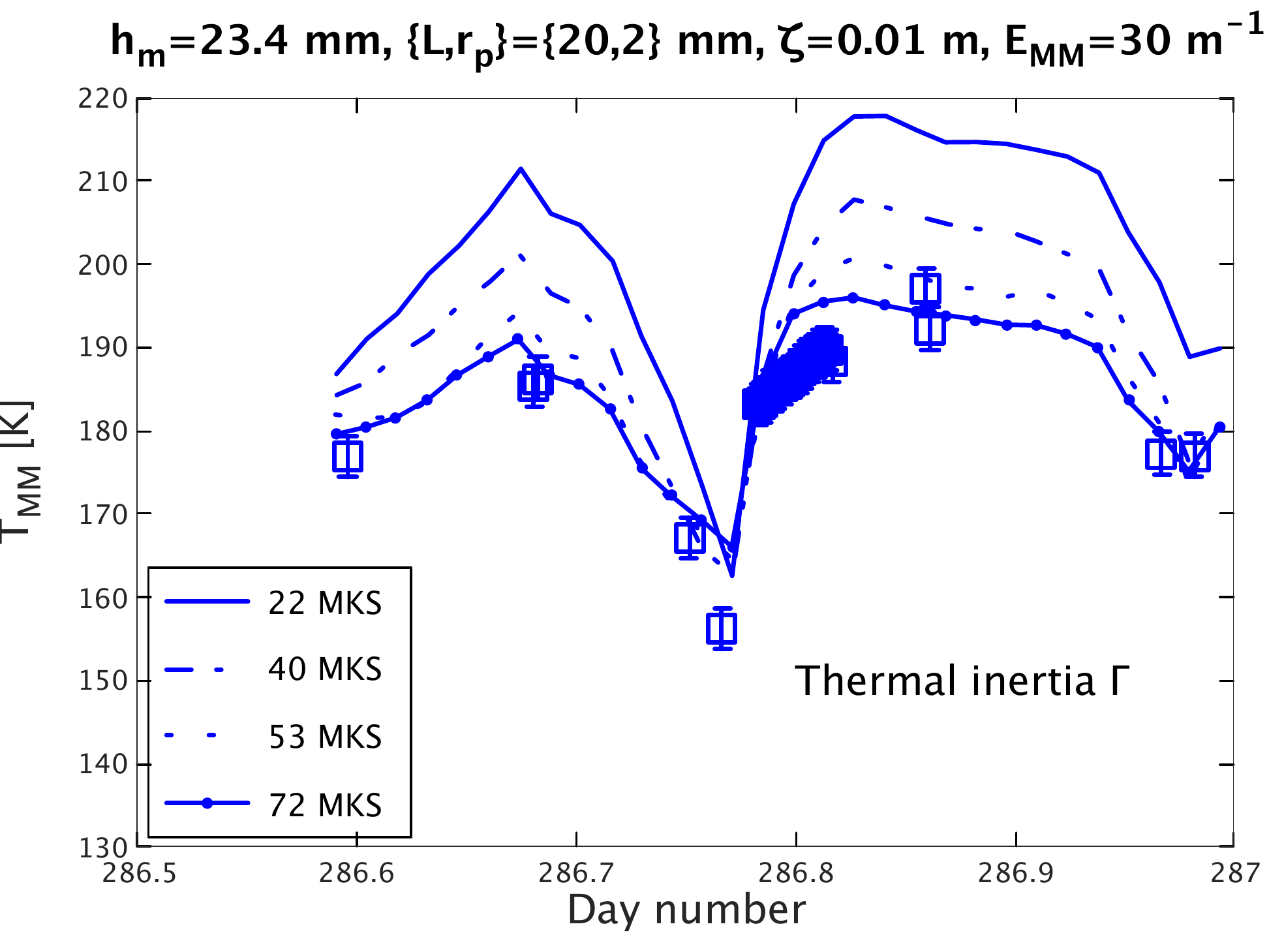}}\\
\scalebox{0.35}{\includegraphics{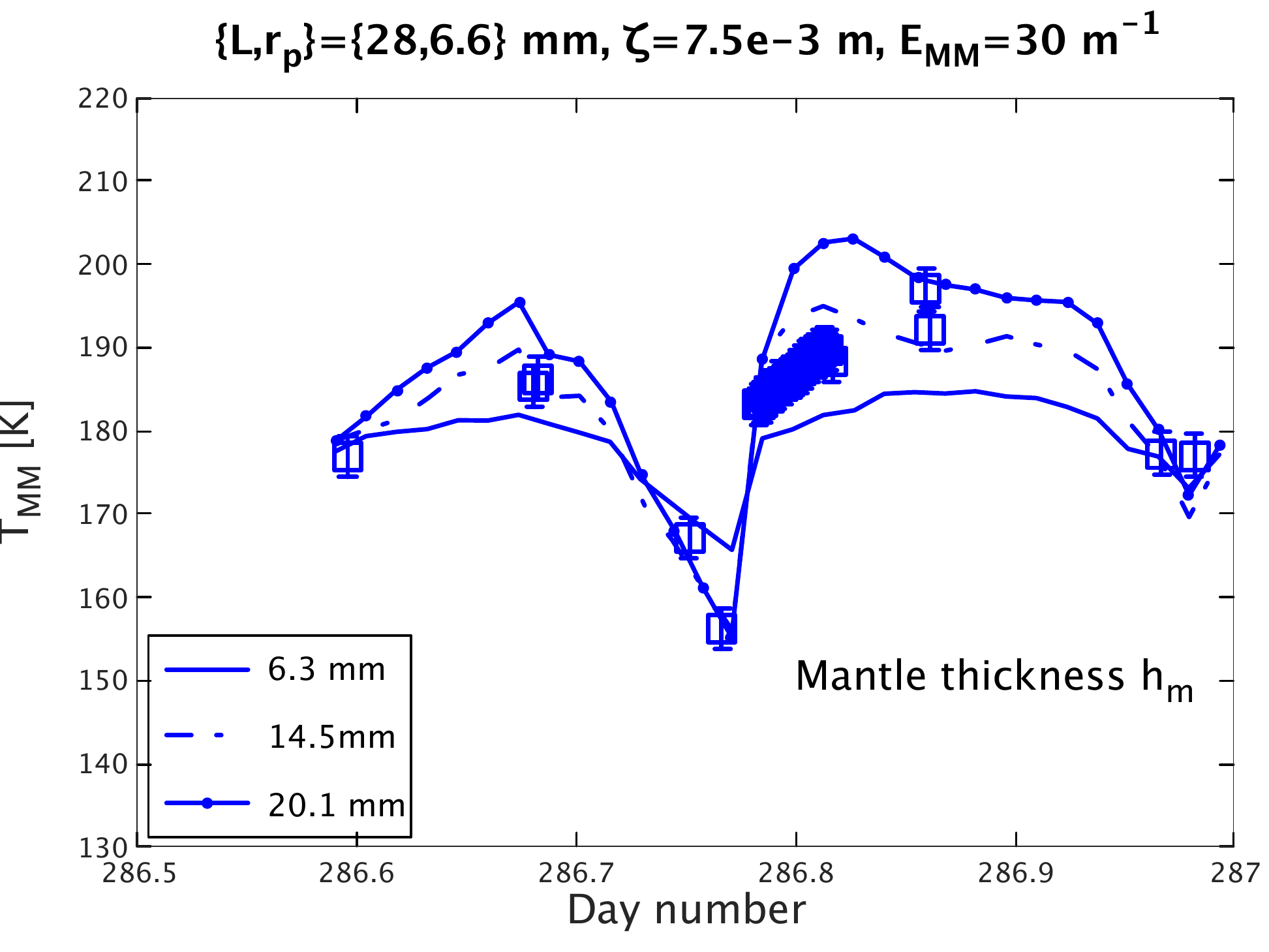}} & \scalebox{0.35}{\includegraphics{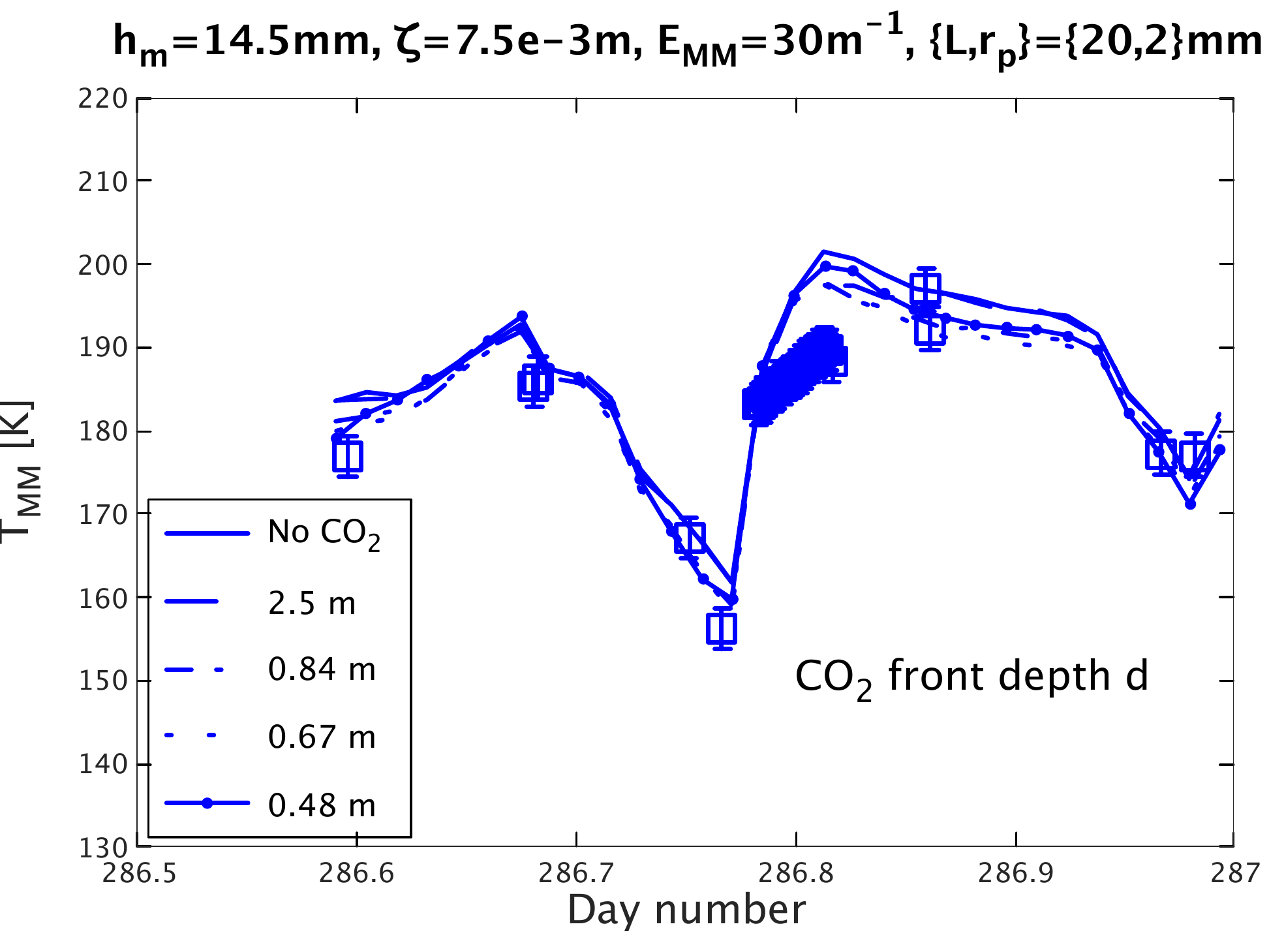}}\\
\end{tabular}
     \caption{The panels show MIRO October 2014 MM data, and how antenna temperature curves based on \textsc{nimbus} simulations change when one parameter is varied 
while all others are fixed in the thermophysical model. \emph{Upper left:} Effects of changing diffusivity (by varying the pore radius $r_{\rm p}$).  Note that the entire 
curves are being vertically displaced when $r_{\rm p}$ changes. \emph{Upper right:} Effects of changing thermal inertia, which are very small at the lowest antenna temperatures. 
\emph{Lower left:} Effects of changing the thickness of the dust mantle. Note that those effects are very small at the very beginning/end of the curves. \emph{Lower right:} Effects 
of changing the depth of the $\mathrm{CO_2}$ sublimation front (including one case without $\mathrm{CO_2}$. The biggest effects are seen at the beginning of the curve, and 
at the lowest/highest temperatures.}
     \label{fig_hapi08}
\end{figure*}

However, if the temperature gradient is positive ($T$ increasing with depth), then $U_1<0$ and equation~(\ref{eq:R21}) suggests that the observed 
radiance, as well as the antenna temperature, would fall strongly with increasing $e$, peaking at  $I_{\lambda}(0)=(U_0-|U_1|Y/(Y+1))/\upi$ at nadir 
and decreasing towards $I_{\lambda}(90^{\circ})=(U_0-|U_1|)/\upi$. That observation led to the suspicion that the MIRO measurements perhaps 
could be matched by a model that would have a positive temperature gradient on the day--side. Negative day--time temperature gradients 
are unavoidable in models that assume that the surface material is fully opaque to visual radiation, so that all solar energy is absorbed at the very 
surface. However, models that allow for a finite visual opacity (combined with opaqueness in the thermal infrared) absorb energy within a near--surface 
volume, as opposed to just the surface itself, while the radiative cooling from such depths is insignificant. Thus, day--side temperature profiles develop that peak at a certain depth, having 
temperatures that fall both towards the surface, and towards greater depths. An example is seen in the lower right panel of Fig.~\ref{fig_hapi07}. This phenomenon, referred to as `the solid--state greenhouse effect' 
has been thoroughly explored theoretically \citep[e.g.][]{brownandmatson87, clow87, matsonandbrown89, komleetal90, davidssonandskorov02b}, experimentally \citep[e.g.][]{kaufmannetal06, kaufmannetal07, 
kaufmannandhagermann15}, and has been observed in space \citep[e.g.][]{urquhartandjakosky96}.

Accordingly, \textsc{nimbus} was modified to accept a flux profile $\exp(-z/\zeta)$ of incident radiation, where $\zeta$ is the $e$--folding scale of light penetration. 
A total of 30 \textsc{nimbus} models were run, considering $\zeta=5\cdot 10^{-3}\,\mathrm{m}$, $\zeta=7.5\cdot 10^{-3}\,\mathrm{m}$, and $\zeta=10^{-2}\,\mathrm{m}$, with different dust mantle 
thicknesses on the range $6$--$27\,\mathrm{mm}$ and tube dimensions on the range $\{100,\,10\}\,\mathrm{\mu m}\leq\{L,\,r_{\rm p}\}\leq \{28,\,6.6\}\,\mathrm{mm}$. 
Most models had $\Gamma\approx 30\,\mathrm{MKS}$, except one low--$r_{\rm p}$ model for which $\Gamma\approx 50\,\mathrm{MKS}$ was forced, and 
several high--$r_{\rm p}$ models that reach $\Gamma\approx 50\,\mathrm{MKS}$ naturally through a significant radiative contribution to heat transfer. 
The peak shape problem was not resolved for $\zeta=5\cdot 10^{-3}\,\mathrm{m}$, but drastic improvements were seen for larger $e$--folding scales as long 
as the mantle was not too thick and the $\{L,\,r_{\rm p}\}$--values were not too large. An example of peak shape improvement (thick blue curve) caused by the 
solid--state greenhouse effect is seen in the left panel of Fig.~\ref{fig_hapi07}, along with remaining issues that could not be resolved. Near $d_{\rm n}=286.8$ the antenna temperature 
drops from $\sim 205\,\mathrm{K}$ when the medium is considered opaque ($\zeta=0$) to $\sim 185\,\mathrm{K}$ when light is absorbed with a penetration length scale of 
$\zeta=7.5\cdot 10^{-3}\,\mathrm{m}$. At $d_{\rm n}=286.85$, where the 
sensitivity to the properties of the near--surface temperature gradient are low because of the modest $e$--values, the synthetic antenna temperature is modified 
rather little. Therefore, the model that includes the solid--state greenhouse effect reproduces the entire second peak. A temperature reduction is seen at the first peak as well, 
for the same reason (see $e$--values in  Fig.~\ref{fig_hapi05}). Temperature increases are seen at the beginning and end of the curve, as well as just before the first dip. 
These are likely secondary--effects, caused be the higher efficiency by which solar energy is driven into the surface material.

Because of the drastic improvement seen for $\zeta\geq 7.5\cdot 10^{-3}\,\mathrm{m}$ we consider the existence of a solid--state greenhouse effect consistent 
with MIRO measurements. We think that the prominent solid--state greenhouse effect is caused 
by the nature of airfall material, consisting of loosely assembled mm--cm--sized chunks that form a medium with substantial macro porosity \citep{pajolaetal16,pajolaetal17b,davidssonetal21}. Solar radiation is 
being absorbed gradually with depth, creating a sub--surface temperature peak and a positive temperature gradient in the uppermost layer that causes a clearly 
observable signature in the MIRO data: an unusually low antenna temperature at strongly illuminated and warming regions observed at large emergence angles.

\begin{figure*}
\centering
\begin{tabular}{cc}
\scalebox{0.4}{\includegraphics{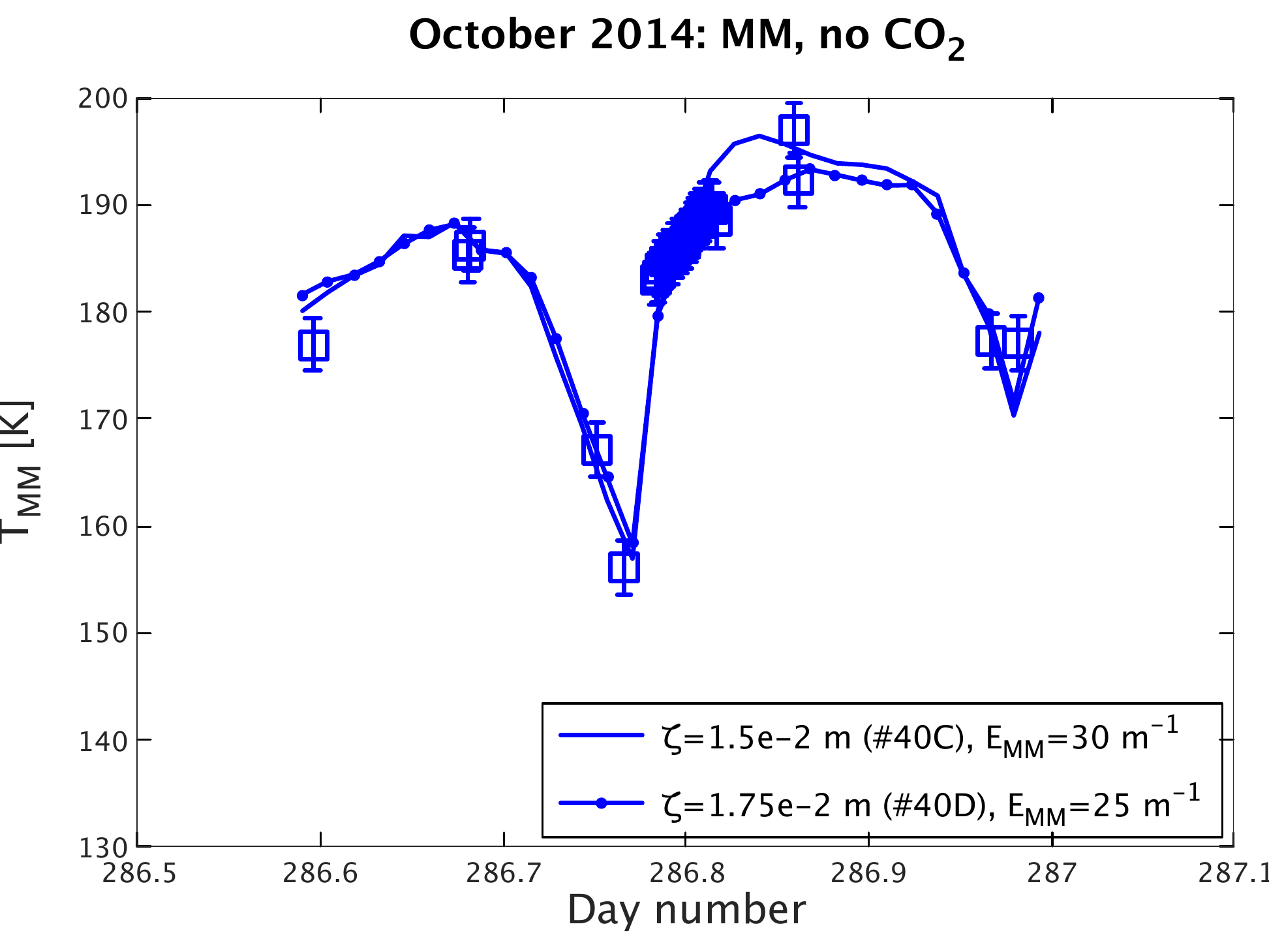}} & \scalebox{0.4}{\includegraphics{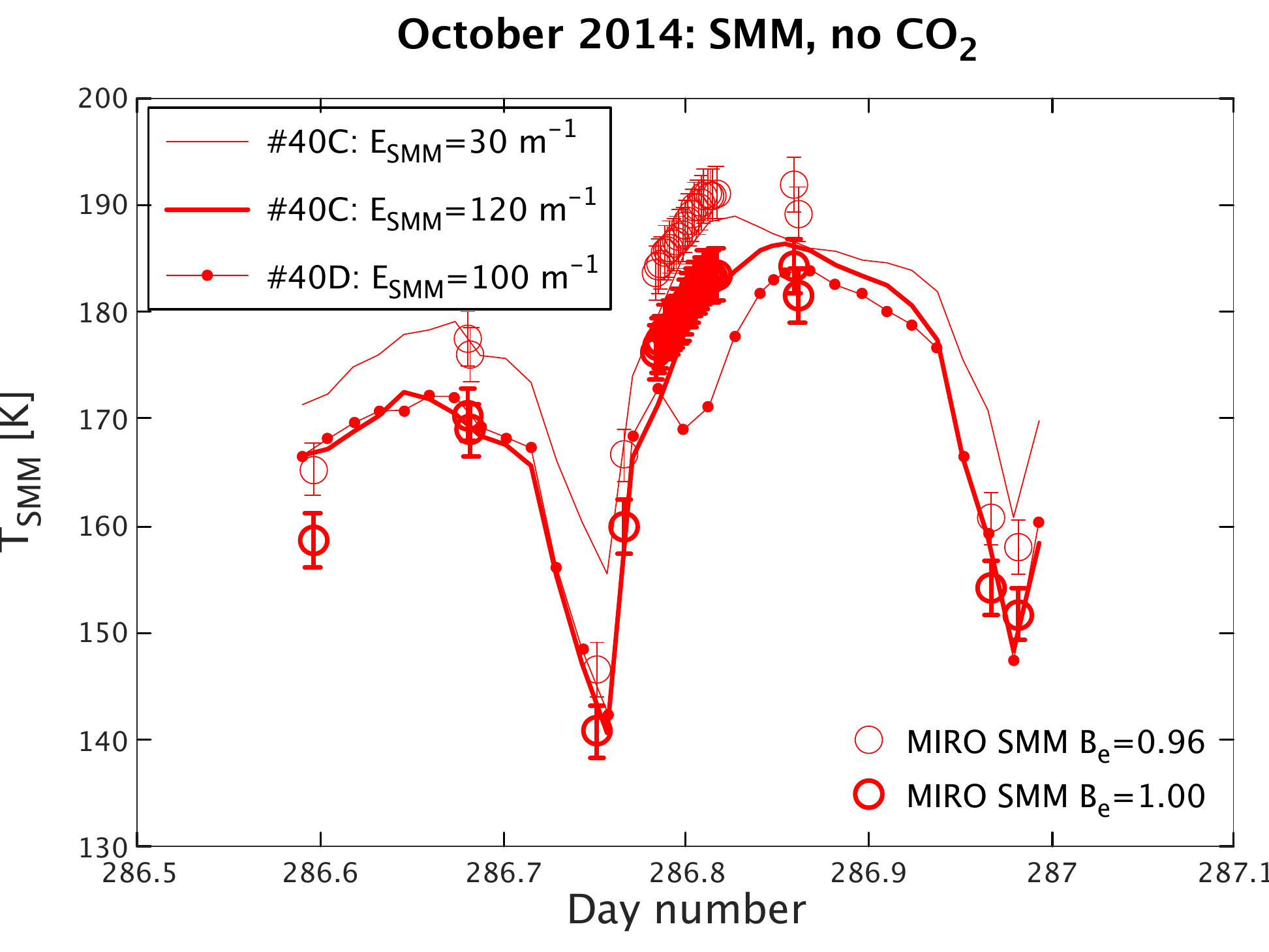}}\\
\end{tabular}
     \caption{The best available \textsc{nimbus}--based reproductions of the MIRO October 2014 MM (left) and SMM (right) $T_{\rm A}$ for models \emph{without} sublimating $\mathrm{CO_2}$ ice. 
These models have $\{L,\,r_{\rm p}\}=\{28,\,6.6\}\,\mathrm{mm}$ and $h_{\rm m}=20.1\,\mathrm{mm}$, but different light penetration length scales ($\zeta=1.5\cdot 10^{-2}\,\mathrm{m}$  for \#40C, but 
$\zeta=1.75\cdot 10^{-2}\,\mathrm{m}$ for \#40D). Model \#40C requires $E_{\rm MM}=30\,\mathrm{m^{-1}}$, but \#40D needs  $E_{\rm MM}=25\,\mathrm{m^{-1}}$. The models are too warm 
at the beginning of the curve. At SMM, model \#40C needs an unrealistically low 
$E_{\rm SMM}=30\,\mathrm{m^{-1}}$ if $B_{\rm e}=0.96$ (and completely misses the first dip)  but would increase towards $E_{\rm SMM}\rightarrow 120\,\mathrm{m^{-1}}$ and match most data 
(except being much too warm at $d_{\rm n}=286.6$) if $B_{\rm e}\rightarrow 1$. 
The higher $\zeta$ of model \#40D yields sufficiently low near--surface temperatures for night--time frost formation, and the daytime sublimation of this frost produces an unacceptable $T_{\rm sy,SMM}$ reduction 
at the bin cluster (red bullet curve). That cooling  is barely detectable in $T_{\rm sy,MM}$.}
     \label{fig_hapi09}
\end{figure*}

Encouraged by these initial models, another 51 \textsc{nimbus} simulations were performed, to further expand the considered set of 
mantle thicknesses, tube dimensions, and $\zeta$--values, for different thermal inertia. Many simulations aimed at a thermal inertia near $\sim 30\,\mathrm{MKS}$. 
About half of those simulations considered $\{L,\,r_{\rm p}\}=\{20,\,2\}\,\mathrm{mm}$, resulting in a relatively small radiative contribution to heat transport. 
A lower limit on the Hertz factor was placed so that the thermal inertia would not be reduced much below $30\,\mathrm{MKS}$ even for very high porosities. 
That resulted in the thermal inertia primarily being governed by solid--state conduction, with the total value typically varying diurnally within the range $30\leq\Gamma\leq 35\,\mathrm{MKS}$. 
The other half considered $\{L,\,r_{\rm p}\}=\{28,\,6.6\}\,\mathrm{mm}$, resulting in a relatively large radiative contribution to heat transport. In this case the 
nominal \citet{shoshanyetal02} porosity--correction of heat conductivity was used, leading to negligible solid--state conduction for the high porosity characterising the dust mantle. That resulted in the thermal 
inertia primarily being governed by radiative transfer, typically varying diurnally within the range $25\leq\Gamma\leq 40\,\mathrm{MKS}$. Two methods were used in the remaining models to consider 
higher thermal inertia. The first method considered $\{L,\,r_{\rm p}\}=\{20,\,2\}\,\mathrm{mm}$ but multiplied the Hertz factor (having the previously mentioned enforced lower limit) 
with factors 2, 4, or 6 in order to push the thermal inertia (dominated by solid--state conduction) into the ranges $50\leq\Gamma\leq 55\,\mathrm{MKS}$, $65\leq\Gamma\leq 75\,\mathrm{MKS}$, 
and $80\leq\Gamma\leq 90\,\mathrm{MKS}$, respectively. There were also cases where the Hertz factor lower limit was reduced to consider $15\leq\Gamma\leq 25\,\mathrm{MKS}$. 
The second method applied the nominal \citet{shoshanyetal02} Hertz factor but boosted the tube dimensions to $\{L,\,r_{\rm p}\}=\{56,\,13.2\}\,\mathrm{mm}$ 
(yielding $35\leq\Gamma\leq 60\,\mathrm{MKS}$) or $\{L,\,r_{\rm p}\}=\{112,\,26.4\}\,\mathrm{mm}$ (yielding $55\leq\Gamma\leq 90\,\mathrm{MKS}$). Note that heat transport 
dominated by radiation has a stronger temperature dependence, thus a wider diurnal thermal inertia range, than does solid--state heat conduction. We considered both options to 
make sure our modelling is not biased towards one particular type of behaviour. We considered mantle thicknesses ranging from $6.3\,\mathrm{mm}$ to $6.8\,\mathrm{cm}$ and 
$e$--folding scales $7.5\cdot 10^{-3}\leq\zeta\leq 2.5\cdot 10^{-2}\,\mathrm{m}$.

The large combined set of 81 \textsc{nimbus} models allows for systematic studies of how the antenna temperature curve changes 
when one given parameter changes and all other conditions are held fixed. Such a study is valuable in order to better understand how to improve a given 
model curve towards a given empirical data set, by adjustments of its model parameters. The upper left panel of Fig.~\ref{fig_hapi08} 
exemplifies how the synthetic antenna temperature curve changes with tube dimensions $\{L,\,r_{\rm p}\}$. Larger pores, resulting 
in a larger net sublimation and cooling, lowers the antenna temperature curve, as expected. It is interesting to note that this 
reduction is substantial both at day and at night. To first order, changing $\{L,\,r_{\rm p}\}$ therefore leads to a vertical 
displacement of the entire curve. Changes due to the thermal inertia are illustrated in the upper right panel of Fig.~\ref{fig_hapi08}. 
Increasing the thermal inertia reduces the amplitude of the curve. Interestingly, the temperature at the coolest part of the 
curve barely changes at all. Therefore, $\Gamma$--adjustments have a large effect on the day--time temperatures while 
the dip temperature is very stable. Changes to the thickness of the dust mantle are illustrated in the lower left panel of 
Fig.~\ref{fig_hapi08}. The thicker the mantle, the larger is the amplitude of the antenna temperature curve. In this case, 
the temperatures at the beginning and at the end of the curve hardly changes, while there are  substantial shifts elsewhere. 
Also note that the peak shape problem is prominent in most of these curves. As previously stated, this happens when the 
mantle is either too thick or the $\{L,\,r_{\rm p}\}$--value is too large for the considered $\zeta$--value. In fact, there is 
a substantial improvement of the shape of the second peak in the lower left panel of Fig.~\ref{fig_hapi08}, as the mantle 
thickness is reduced.

\begin{figure*}
\centering
\begin{tabular}{cc}
\scalebox{0.4}{\includegraphics{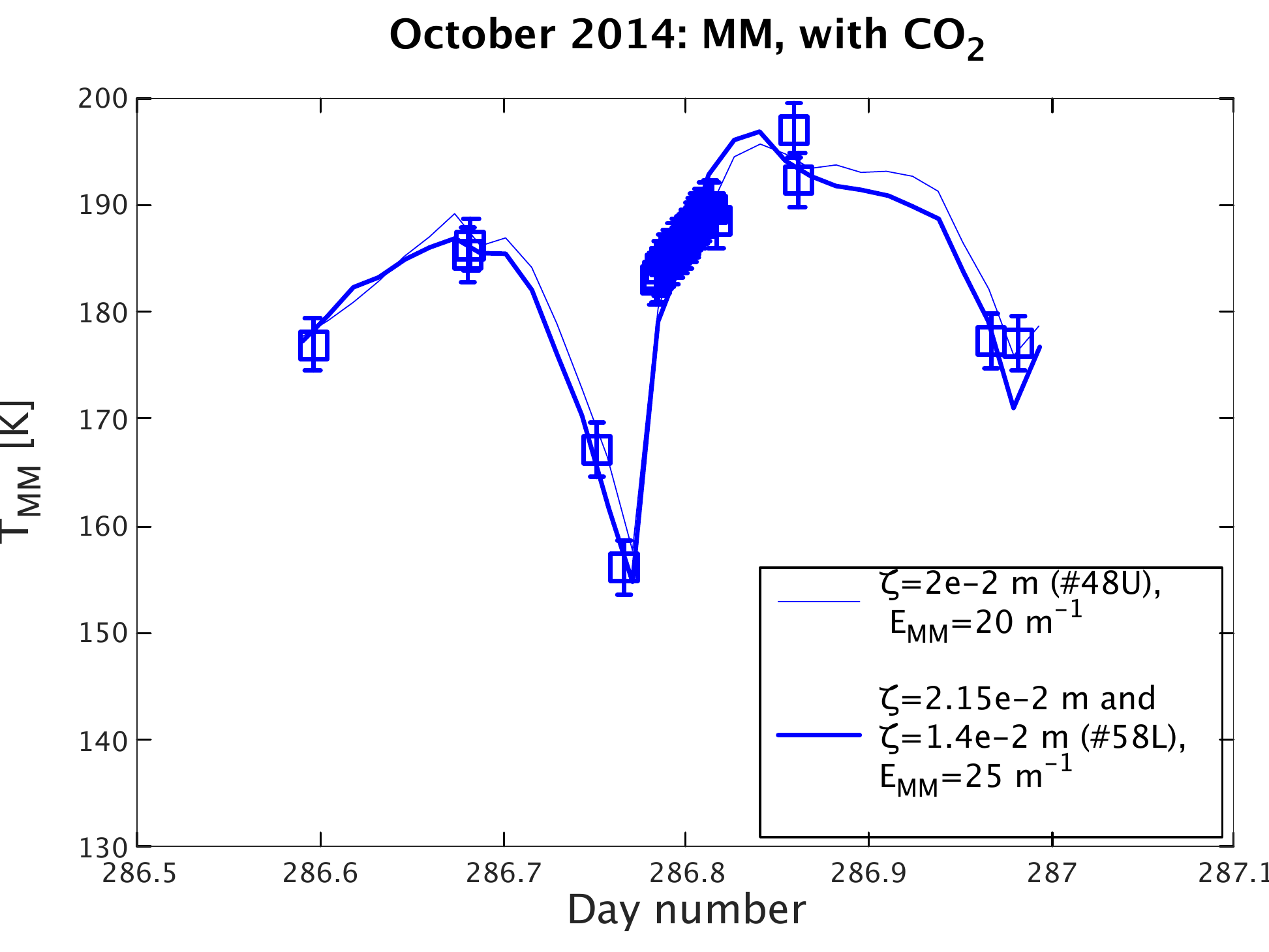}} & \scalebox{0.4}{\includegraphics{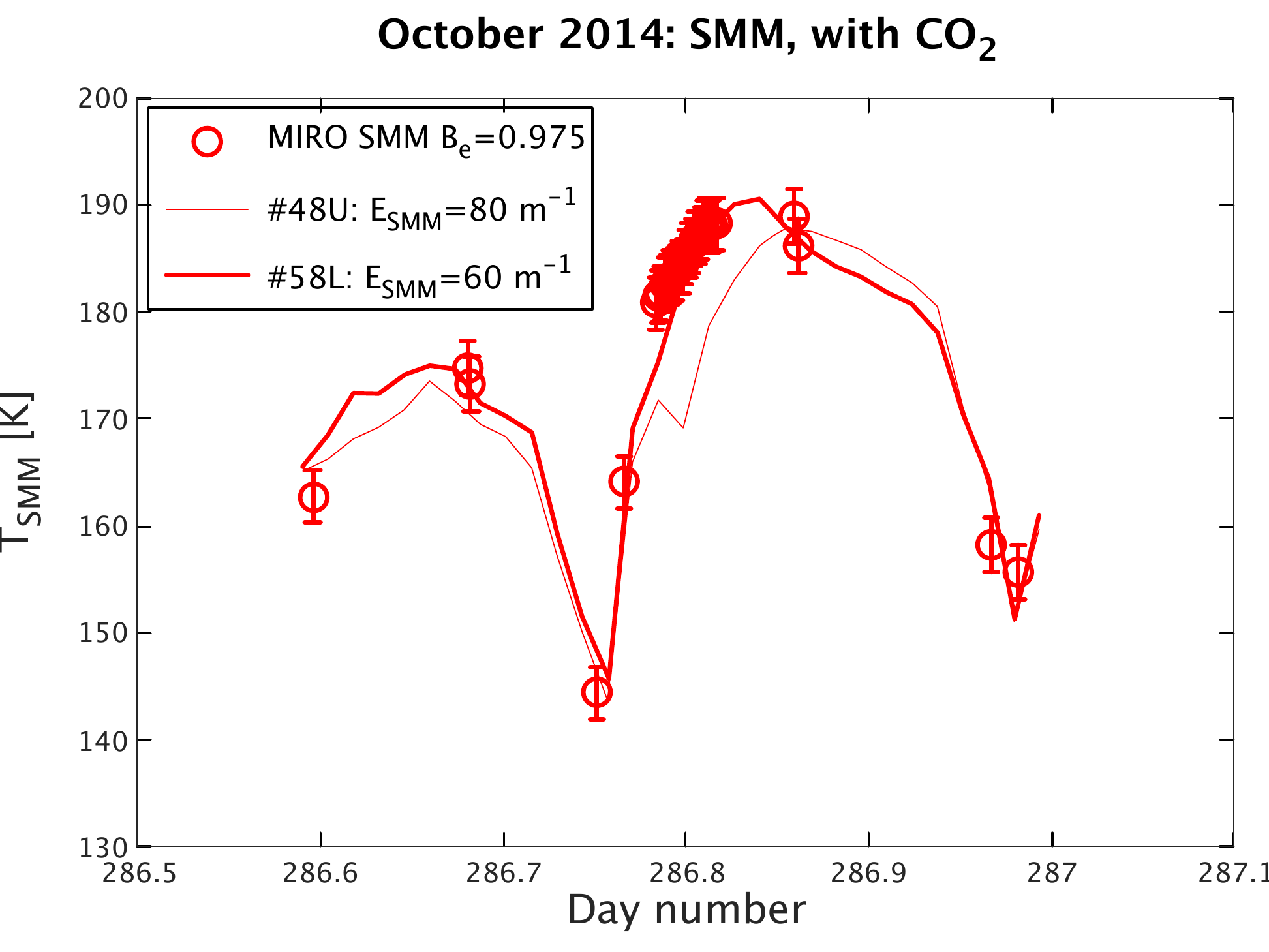}}\\
\end{tabular}
     \caption{The best available \textsc{nimbus}--based reproductions of the MIRO October 2014 MM (left) and SMM (right) for models \emph{including} sublimating $\mathrm{CO_2}$ ice. 
Model \#48U with tube dimensions $\{L,\,r_{\rm p}\}=\{28,\,6.6\}\,\mathrm{mm}$, light penetration length scale $\zeta=2.0\cdot 10^{-2}\,\mathrm{m}$, dust mantle thickness 
$h_{\rm m}=27.1\,\mathrm{mm}$, and $\mathrm{CO_2}$ at a depth of $0.48\,\mathrm{m}$ performed best at MM (for $E_{\rm MM}=20\,\mathrm{m^{-1}}$, light blue), but 
was less satisfactory at SMM (for $E_{\rm SMM}=80\,\mathrm{m^{-1}}$, light red). A better SMM fit (model \#58L, thick red) was obtained for a variable light penetration length scale 
(nominally $\zeta=2.15\cdot 10^{-2}\,\mathrm{m}$, but reducing to $1.4\cdot 10^{-2}\,\mathrm{m}$ in 
the upper centimetre at $d_{\rm n}=286.767$--$286.85$) in combination with a somewhat thicker mantle ($h_{\rm m}=23.4\,\mathrm{mm}$), while tube dimensions and $\mathrm{CO_2}$ depth remained 
the same (using $E_{\rm SMM}=60\,\mathrm{m^{-1}}$).  The corresponding MM model (\#58L with  $E_{\rm MM}=25\,\mathrm{m^{-1}}$, thick blue) also performed well.  Note the significant improvement 
at the beginning of the curves with respect to Fig.~\ref{fig_hapi09}, that considered $\mathrm{CO_2}$--free models. The best--fit model parameters are given in Table~\ref{tab1}.}
     \label{fig_hapi10}
\end{figure*}

As can be seen from Fig.~\ref{fig_hapi08}, many modelling attempts were unsuccessful, leading to curve shapes and absolute values that are  
incompatible with the measured data. Only four models were formally consistent with the data ($Q>0.01$ for $\pm 2.5\,\mathrm{K}$ error bars). 
The two best MM solution are seen in the left panel of Fig.~\ref{fig_hapi09}. The first model (\#40C) had $\{L,\,r_{\rm p}\}=\{28,\,6.6\}\,\mathrm{mm}$ and 
the original \citet{shoshanyetal02} Hertz factor, resulting in a radiation--dominated thermal inertia varying between $23$--$39\,\mathrm{MKS}$ 
during rotation. That model had a light penetration length scale of $\zeta=1.5\cdot 10^{-2}\,\mathrm{m}$, a $h_{\rm m}=20.1\,\mathrm{mm}$ dust mantle thickness, and 
the best fit was reached for $E_{\rm MM}=30\,\mathrm{m^{-1}}$, having $Q=0.059$. The other model (\#40D) was identical except that $\zeta=1.75\cdot 10^{-2}\,\mathrm{m}$, having $Q=0.039$ 
at $E_{\rm MM}=25\,\mathrm{m^{-1}}$. Those models suggest the MIRO MM channel would sample radiation in the top $E_{\rm MM}^{-1}=3.3$--$4.0\,\mathrm{cm}$ of 
the surface. The largest residual is at the first bin ($d_{\rm n}\approx 286.6$),  where the models were somewhat too warm. It was 
also notoriously difficult to reach a sufficiently low temperature at the dip ($d_{\rm n}\approx 286.75$) when the fit was reasonable elsewhere.

In order to qualify as possible solutions, the same physical model should reproduce the MIRO SMM data as well. That comparison is complicated by 
the error margins on the beam efficiency, $B_{\rm e}$. If $B_{\rm e}=0.96$, the MIRO SMM measurements would suggest relatively high antenna temperatures, 
shown in the right panel of Fig.~\ref{fig_hapi09} by the thin red circles. Pushing the \#40C SMM  solution (thin red curve) to reproduce the 
continuous--stare bin cluster of the $B_{\rm e}=0.96$ data requires $E_{\rm SMM}=30\,\mathrm{m^{-1}}$. However, because the transparency of comet analogue 
material is expected to increase with the wavelength \citep[$E_{\rm MM}<E_{\rm SMM}$;][]{garykeihm78} that is not a satisfactory 
solution \citep[for 67P, the average is $E_{\rm SMM}=100^{+100}_{-33}\,\mathrm{m^{-1}}$ according to][]{schloerbetal15}. 
Furthermore, that model has a significant peak shape problem, and the dip temperature is far too high. Assuming $B_{\rm e}=1$ for the 
SMM channel yields the somewhat cooler thick red circles, with model \#40C providing the smallest residuals for $E_{\rm SMM}=120\,\mathrm{m^{-1}}$, 
where $Q=0.0039$ (thick red curve).  As was the case for the MM channel the biggest residual is at the first bin ($d_{\rm n}\approx 286.6$). Model \#40D (red bullet curve), having a 
larger $\zeta$--value, is very similar to \#40C, except at the second peak where the temperature dips temporarily when reaching $\sim 170\,\mathrm{K}$ (at the bin cluster of the continuous stare) 
and remaining substantially below the measurements until the peak tip. The reason for this phenomenon is frost formation within the dust mantle during the dip, and cooling because of water ice sublimation once 
the Sun illuminates Hapi~D. All $\zeta>0$ models have rather low surface temperatures at day, and even lower ones at night (see Fig.~\ref{fig_hapi07} to the lower right), 
sometimes forcing vapour from the warmer interior to condense near the surface. Model \#40C accumulates a maximum of $11.0\,\mathrm{kg\,m^{-3}}$ worth of 
water ice during the first dip (peaking $\sim 2\,\mathrm{mm}$ below the surface), that is removed rather quickly. However, \#40D accumulates $18.6\,\mathrm{kg\,m^{-3}}$ (corresponding to a volumetric 
abundance of $\sim 10$ per cent relative refractories, for an assumed porosity of $\psi=0.8$), resulting 
in more substantial cooling. It is interesting to note that MIRO likely would have been capable of detecting frost removal, had it been an important process at 
the sampled parts of the temperature curve. VIRTIS did detect $3\,\mathrm{\mu m}$ absorption due to near--surface water ice in other regions of Hapi in September 2014, suggesting 5--14 per cent 
ice by volume \citep{desanctisetal15}. Such ice was only seen near shadows, and disappeared after $\sim 15\,\mathrm{min}$ of illumination, indicating it 
was frost that condensed during darkness. We note that only the \textsc{nimbus} models with $\zeta>0$ display frost formation at night and during shadowing, typically reaching 5--10 per cent 
abundances (i.~e., closely matching the levels inferred from VIRTIS observations). The lack of frost signatures at Hapi~D according to MIRO, and its presence elsewhere at Hapi according to VIRTIS, 
could mean that the frost--formation efficiency varies strongly by location. Because $\zeta>0$ seem to be one of the requirements for frost formation, this suggests that VIRTIS provides 
additional support for the existence of a solid--state greenhouse effect, extending beyond Hapi~D.

We consider \#40C and \#40D marginal fits, at best. Realistically, $B_{\rm e}\leq 0.98$, which means $E_{\rm SMM}$ could be substantially lower than $120\,\mathrm{m^{-1}}$ 
(although perhaps not as low as $30\,\mathrm{m^{-1}}$). The excess temperature near $d_{\rm n}\approx 286.6$ (of about $10\,\mathrm{K}$) does not appear to be correctable with the means available. 
Figures~\ref{fig_hapi07} and \ref{fig_hapi08} show that changes to the dust mantle thickness and the solid--state greenhouse effect parameter $\zeta$ have very weak 
influences on that part of the curve. The most effective way of modifying the $d_{\rm n}\approx 286.6$ temperature is to change the tube dimensions 
and/or the thermal inertia, but that leads to large modifications elsewhere and does not solve the problem. The models with the over--all best performance 
also had trouble to fully reach the lowest temperature at the dip. Increasing the net sublimation and cooling capacity of the water sublimation front requires a 
high diffusivity, i.e., large $\{L,\,r_{\rm p}\}$ values. But large pores also enhances the thermal inertia because of the efficient radiative heat transfer, 
which tends to reduce the amplitude of the antenna temperature curve. The best compromise between sufficient cooling and not having excessive thermal inertia 
appeared to be $\{L,\,r_{\rm p}\}=\{28,\,6.6\}\mathrm{mm}$. We do not think it is possible to further reduce the residuals between model curves and the data, 
within the limitations of the currently considered model.

Thinking of ways to increase the cooling without having to increase the diffusivity, we decided to test the effect of having an additional sublimation front at larger depth, caused by $\mathrm{CO_2}$. 
Initial tests, seen in the lower right panel of Fig.~\ref{fig_hapi08}, showed no measurable differences between a model without $\mathrm{CO_2}$, and another 
model that had $\mathrm{CO_2}$ at a depth of $2.5\,\mathrm{m}$. However, models with $\mathrm{CO_2}$ at depths of $0.48$--$0.84\,\mathrm{m}$ caused 
a reduction of the antenna temperature at the dip of a few Kelvin, which potentially could improve the model fits. More importantly, the addition of $\mathrm{CO_2}$ 
had another unexpected and desirable effect: it caused a distinct drop of temperature at the beginning of the curve around $d_{\rm n}\approx 286.6$. This 
was seen as a potential way of removing the excess temperature problem seen in $\mathrm{CO_2}$--free models, without causing too large changes elsewhere. 
Encouraged by these initial results we ran 97 models including $\mathrm{CO_2}$ (increasing the total of  \textsc{nimbus} models to 192  for October 2014). 
These models considered different combinations of tube dimensions, $\zeta$--values, dust mantle thicknesses, and $\mathrm{CO_2}$ front depths, primarily targeting thermal inertia 
in the $20$--$40\,\mathrm{MKS}$ range dominated either by solid--state conduction or radiative heat transfer.

Among these $\mathrm{CO_2}$ \textsc{nimbus} models, nine had higher $Q$--values in the MM than the best non--$\mathrm{CO_2}$ \textsc{nimbus} models. 
The best of these (\#48U), shown in the left panel of Fig.~\ref{fig_hapi10} as a thin blue curve, had $\{L,\,r_{\rm p}\}=\{28,\,6.6\}\,\mathrm{mm}$ and the original 
\citet{shoshanyetal02} Hertz factor, resulting in a radiation--dominated thermal inertia varying between $23$--$38\,\mathrm{MKS}$ during rotation. 
That model had $\zeta=2.0\cdot 10^{-2}\,\mathrm{m}$, a $h_{\rm m}=27.1\,\mathrm{mm}$ dust mantle, $\mathrm{CO_2}$ at a depth of $0.48\,\mathrm{m}$, and 
$Q=0.140$ for $E_{\rm MM}=20\,\mathrm{m^{-1}}$. As can be seen, the synthetic curve reproduces the measured data well. Compared to the best $\mathrm{CO_2}$--free 
solution in the left panel of Fig.~\ref{fig_hapi09}, the fit near the first bin has been substantially improved, and the lowest temperature at the dip is reproduced 
more convincingly. The corresponding SMM curve seen as a thin red curve in the right panel of Fig.~\ref{fig_hapi10} fits well, except at the data cluster near $d_{\rm n}=286.8$. 
Here, the temperature is too low, because of the removal of some frost that accumulated near the surface of the dust mantle during the shadowing period. The SMM signal is 
very sensitive to the temperature profile in the uppermost layer, and the difference between the model and the data is obvious. The fact that the same model conforms very 
well with the data in the MM channel illustrates how the local cooling quickly becomes undetectable when a thicker slab contributes to the measured signal. 

\begin{figure*}
\centering
\begin{tabular}{cc}
\scalebox{0.4}{\includegraphics{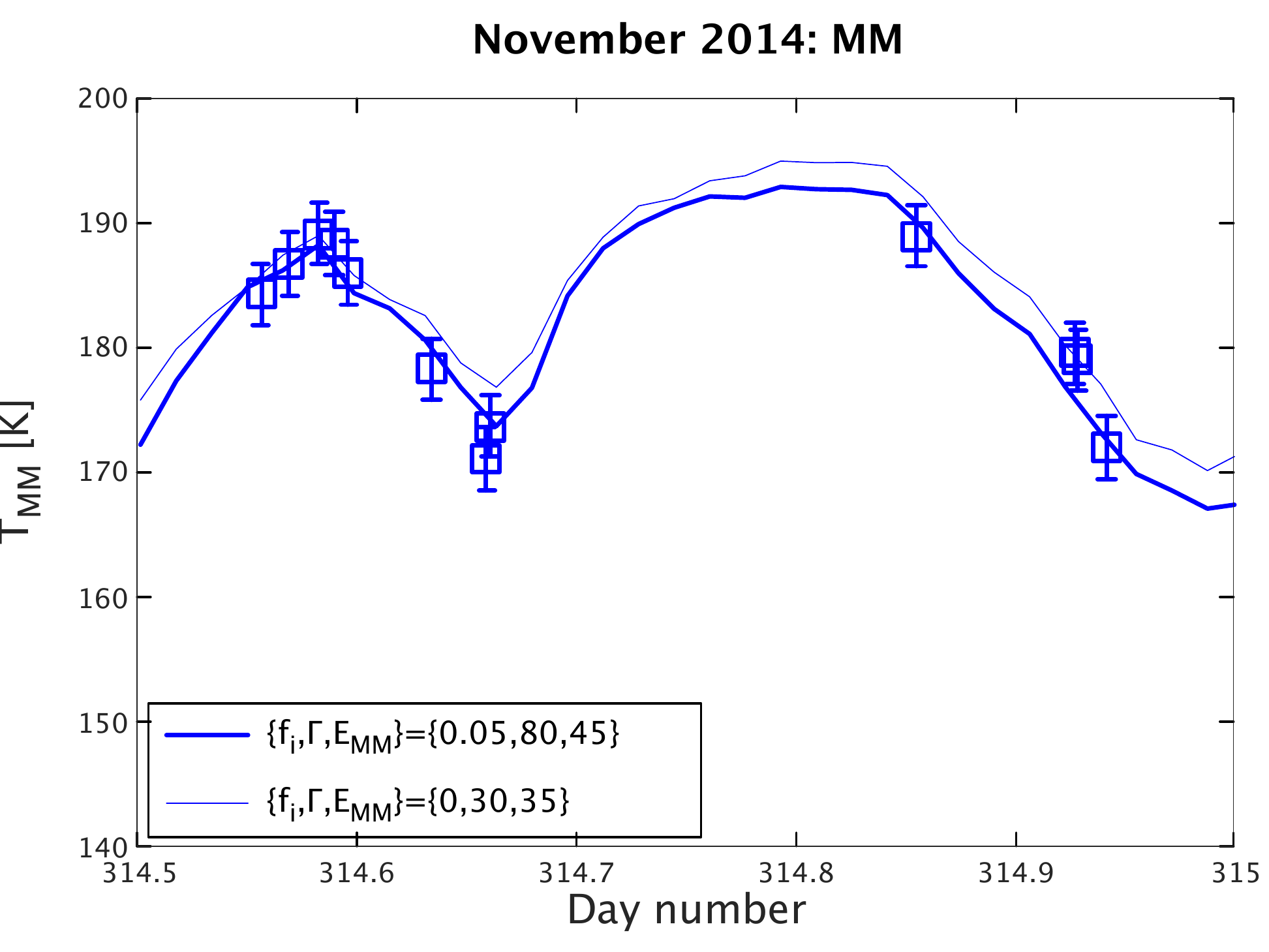}} & \scalebox{0.4}{\includegraphics{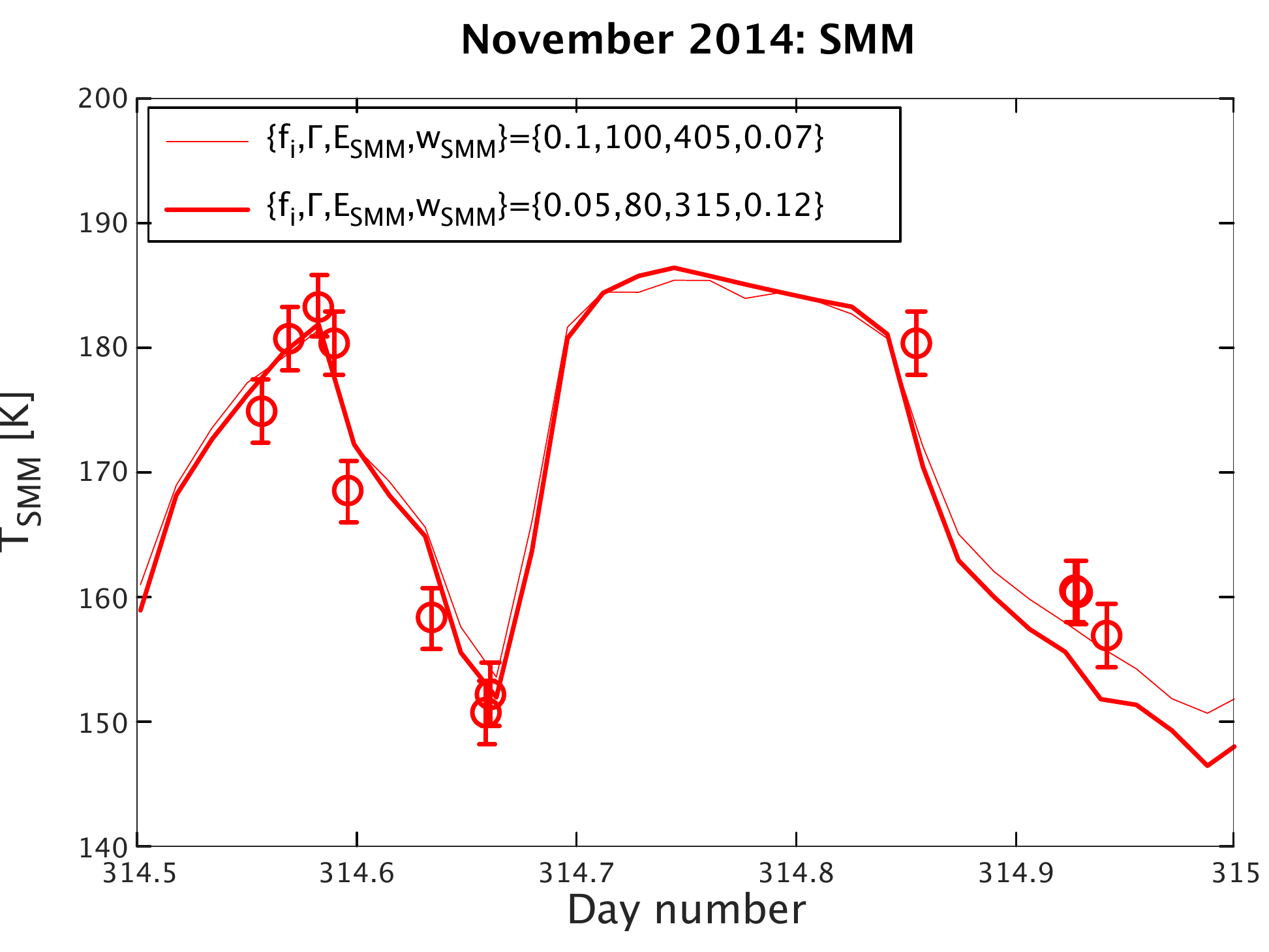}}\\
\end{tabular}
     \caption{The best available \textsc{btm} solutions versus MIRO November 2014 MM (left) and SMM (with $B_{\rm e}=0.96$, right) data (note that just a handful of the bins at the first 'continuous stare' peak 
were plotted to ease a comparison between models and data). Though not perfect, the \textsc{btm} performed substantially 
better in November compared to October 2014. This suggests that some change has taken place in the near--surface material of Hapi~D, so that the basic thermophysical 
model is a better representation of reality. This change seems to be associated with a reduction of cooling by sublimation and an increase of the thermal inertia.}
     \label{fig_hapi11}
\end{figure*}

At this point, we speculated if the very presence of frost could have an effect on the capability of visual radiation to penetrate into the surface material. We therefore experimented with 
a radiation $e$--folding scale being reduced to $1.4\cdot 10^{-2}\,\mathrm{m}$ within the upper centimetre during the $d_{\rm n}=286.767$--$286.85$ part of the curve. 
This would concentrate heating to the upper part of the dust mantle and remove the frost more quickly. In those experiments, we found the best solution for $\{L,\,r_{\rm p}\}=\{28,\,6.6\}\,\mathrm{mm}$, 
$22\leq\Gamma\leq 41\,\mathrm{MKS}$, $\zeta=2.15\cdot 10^{-2}\,\mathrm{m}$ (except for the brief reduction to $1.4\cdot 10^{-2}\,\mathrm{m}$), $h_{\rm m}=23.4\,\mathrm{mm}$, 
and $\mathrm{CO_2}$ at a depth of $0.48\,\mathrm{m}$. In this case the SMM model (\#58L) had $Q=0.019$ for $E_{\rm SMM}=60\,\mathrm{m^{-1}}$ when assuming $B_{\rm e}=0.975$, 
seen in the right panel of Fig.~\ref{fig_hapi10} as a thick red curve. The corresponding MM model, seen in the left panel of Fig.~\ref{fig_hapi10} as a thick blue curve had 
$Q=0.086$ for $E_{\rm MM}=25\,\mathrm{m^{-1}}$. This is marginally worse than model \#48U, but still an acceptable fit. We consider these solutions the best achievable 
with the available models, and believe that the corresponding parameters collected in Table~\ref{tab1} are representative of the physical properties of Hapi~D in October 2014.

\begin{table*}
\begin{center} {\bf Best--fit physical properties of Hapi~D} \end{center}
\begin{center}
\small
\begin{tabular}{||l|r|r||}
\hline
Physical quantity & October 2014 & November 2014\\
\hline
Thermal inertia $\Gamma$ of dust mantle & $22$--$41\,\mathrm{MKS}$ &  $110\,\mathrm{MKS}$\\
 & & ($65\,\mathrm{MKS}$ in top $8\,\mathrm{mm}$)\\
Dust mantle thickness $h_{\rm m}$ & $2.3\,\mathrm{cm}$ & $21\,\mathrm{cm}$ \\
Tube dimensions $\{L,\,r_{\rm p}\}$ & $\{28,\,6.6\}\,\mathrm{mm}$ &  $\{10,\,1\}\,\mathrm{\mu m}$\\
Light penetration $e$--folding scale $\zeta$ & $2.15\cdot 10^{-2}\,\mathrm{m}$ &  0\\
 & ($1.4\cdot 10^{-2}\,\mathrm{m}$ when frost) & \\
MM extinction coefficient $E_{\rm MM}$ & $25\,\mathrm{m^{-1}}$ & $80\,\mathrm{m^{-1}}$\\
SMM extinction coefficient $E_{\rm SMM}$ & $60\,\mathrm{m^{-1}}$ & $600\,\mathrm{m^{-1}}$\\
SMM single--scattering albedo $w_{\rm SMM}$ & 0 & 0.1\\
$\mathrm{CO_2}$ sublimation front depth & $0.48\,\mathrm{m}$ &  $0.44\,\mathrm{m}$\\
\hline
\end{tabular}
\caption{The physical properties of the surface material of Hapi~D in October and November 2014, based on best--fit \textsc{nimbus} 
models to MIRO MM and SMM data. Note that a refractories/water--ice mass ratio $\mu=2$ was used for October and $\mu=1$ was used for November. 
Both values are compatible with outgassing from airfall material, with a preference for the former \protect\citep{davidssonetal22}. The molar 
$\mathrm{CO_2}$ abundance was 30 per cent relative to water, as found by \protect\citet{davidssonetal22}.}
\label{tab1}
\end{center}
\end{table*}

It is significant that the same physical solution that works at SMM also reproduces the MM data. The SMM footprint primarily samples Hapi~D. 
The fact that the larger MM footprint includes some additional terrain, and yet is consistent with the same physical conditions, indicates 
the presence of a solid--state greenhouse effect and shallow $\mathrm{CO_2}$ ice deposits in a region that is larger than the pits themselves.

\subsection{November 2014} \label{sec_results_Nov}

\subsubsection{November: \textsc{btm} results} \label{sec_results_Nov_BTM}

The November dataset was the first to be analysed, and the methodology was somewhat different from the final one, applied for the October dataset. 
The initial efforts focused on a subset of the November data: the 24 bins constituting the (time--shifted) $d_{\rm n}=314.55$--$314.60$ continuous stare at Hapi~D. 
The first 16 bins track a MM antenna temperature increase from $184\,\mathrm{K}$ to $190\,\mathrm{K}$, during which Hapi~D is 
illuminated. At that point, Hapi~D moves into shadow and the remaining 8 bins of the continuous stare track a temperature reduction to $185\,\mathrm{K}$. The rate of 
cooling after the sudden switch--off of strong illumination was considered a particularly important indicator of the thermophysical properties of 
the surface material. A total of 16 \textsc{btm} models were run for a grid of volumetric ice abundances $f_{\rm i}=\{0,\,0.05,\,0.1,\,0.2\}$ 
and thermal inertia--values $\Gamma=\{30,\,50,\,80,\,100\}\,\mathrm{MKS}$. Then, a total of 108 \textsc{themis} models were run (assuming $w_{\rm MM}=0$), in order 
to find the $E_{\rm MM}$--values that yielded the smallest residuals between synthetic and measured MM antenna temperatures, for each \textsc{btm} model. 
The best match was $\{f_{\rm i},\,\Gamma,\,E_{\rm MM}\}=\{0.05,\,100\,\mathrm{MKS},\,30\,\mathrm{m^{-1}}\}$ with $Q=0.98$ (this early work assumed error bars of 
$\pm 1\,\mathrm{K}$ at MM and $\pm 1.5\,\mathrm{K}$ at SMM, i.~e., the $Q$--values would be somewhat larger if applying our standard $\pm 2.5\,\mathrm{K}$ value).

For the SMM investigation of this limited dataset we assumed $B_{\rm e}=0.94$ and made 52 searches with \textsc{themis} 
for the best $E_{\rm SMM}$--value, assuming $w_{\rm SMM}=0$, and additionally 202 models for which $w_{\rm SMM}>0$. The best $w_{\rm SMM}=0$ solution 
had $\{f_{\rm i},\,\Gamma,\,E_{\rm SMM}\}=\{0.05,\,50\,\mathrm{MKS},\,200\,\mathrm{m^{-1}}\}$ with $Q=0.090$. 
The $\{f_{\rm i},\,\Gamma\}=\{0.05,\,100\,\mathrm{MKS}\}$ model (that performed best at MM) had $Q=0.045$ for $E_{\rm SMM}=400\,\mathrm{m^{-1}}$. 
Allowing for $w_{\rm SMM}>0$ the best solution was $\{f_{\rm i},\,\Gamma,\,E_{\rm SMM},\,w_{\rm SMM}\}=\{0,\,50\,\mathrm{MKS},\,200\,\mathrm{m^{-1}},\,0.15\}$ with $Q=0.824$. 
This limited investigation suggested a lower level of sublimation cooling and potentially a higher thermal inertia in November compared to October.

The investigation was then extended to the full November 2014 dataset (i.~e., also including the 8 bins outside the continuous stare). \textsc{themis} simulations were run for all combinations $f_{\rm i}=\{0,\,0.05,\,0.1\}$ 
and $\Gamma=\{30,\,50,\,80,\,100\}\,\mathrm{MKS}$, primarily considering the $E_{\rm MM}$ values that had worked best for the limited dataset, though 
some additional $E_{\rm MM}$ values were considered (25 models in total). The best model was  $\{f_{\rm i},\,\Gamma,\,E_{\rm MM}\}=\{0.05,\,80\,\mathrm{MKS},\,45\,\mathrm{m^{-1}}\}$ with 
$Q=0.624$, seen in the left panel of Fig.~\ref{fig_hapi11} as a thick blue curve. Two other models provided temperature curves similar to this one;  
$\{f_{\rm i},\,\Gamma,\,E_{\rm MM}\}=\{0.05,\,100\,\mathrm{MKS},\,30\,\mathrm{m^{-1}}\}$ with $Q=0.543$ (i.e., the best fit for the $d_{\rm n}=314.55$--$314.60$ peak), 
$\{f_{\rm i},\,\Gamma,\,E_{\rm MM}\}=\{0.1,\,100\,\mathrm{MKS},\,45\,\mathrm{m^{-1}}\}$ with $Q=0.226$. A fourth model, shown in the left panel of Fig.~\ref{fig_hapi11} 
as a thin blue curve, had $\{f_{\rm i},\,\Gamma,\,E_{\rm MM}\}=\{0,\,30\,\mathrm{MKS},\,35\,\mathrm{m^{-1}}\}$ with $Q=0.100$, for which discrepancies with respect to 
the data start to become obvious. All other models performed substantially worse.

Another 25 \textsc{themis} models were run at SMM, assuming $B_{\rm e}=0.96$. The best of these models, seen as a thin red curve in the right panel of Fig.~\ref{fig_hapi11} 
had $\{f_{\rm i},\,\Gamma,\,E_{\rm SMM},\,w_{\rm SMM}\}=\{0.1,\,100\,\mathrm{MKS},\,405\,\mathrm{m^{-1}},\,0.07\}$ with $Q=3\cdot 10^{-11}$, i.e., it did not formally fit the data 
(though this $\{f_{\rm i},\,\Gamma\}$--combination worked well at MM). The best case from the MM study, $\{f_{\rm i},\,\Gamma\}=\{0.05,\,80\,\mathrm{MKS}\}$, had the smallest 
residuals at SMM  ($Q=2\cdot 10^{-16}$) for $\{E_{\rm SMM},\,w_{\rm SMM}\}=\{315\,\mathrm{m^{-1}},\,0.12\}$, seen as a thick read curve in the right panel of Fig.~\ref{fig_hapi11}. 
In both cases, the main problem is a too slow cooling after the first peak, and a too rapid cooling after the second peak.

In summary, the \textsc{btm} produced very convincing fits for the November 2014 MIRO MM data, suggesting 5--10 per cent water ice by volume and a thermal inertia 
of $80\leq\Gamma\leq 100\,\mathrm{MKS}$. Although such models provided SMM antenna temperature curves with an amplitude and shape that were about right, 
they were not statistically compatible with the measurements. Yet, it is interesting that \textsc{btm} performed far better in November than in October. This suggests 
that the Hapi surface material may have evolved significantly, in a manner that makes the \textsc{btm} a better representation of reality. Although the \textsc{btm} 
did not represent the MM October 2014 data well, the model in Fig.~\ref{fig_hapi06} with reasonable amplitude and absolute values 
(having $\{f_{\rm i},\,\Gamma\}=\{0.3,\,30\,\mathrm{MKS}\}$) suggests that the cooling by sublimation was reduced, and the thermal inertia increased, when 
going from October to November 2014. We now turn to \textsc{nimbus} simulations to investigate whether that model confirms that conditions have changed, and if so, 
what those new conditions might be.

\subsubsection{November: \textsc{nimbus} results}  \label{sec_results_Nov_NIMBUS}

A total of 56 \textsc{nimbus} simulations were performed in order to better understand the November 2014 MIRO observations. 
The models assumed (initial values) refractories to water ice mass ratio $\mu=1$ \citep[also consistent with the inbound water production rate;][]{davidssonetal22}, 30 per cent $\mathrm{CO_2}$ with respect 
to water by number, and a porosity of 63 per cent (in order to obtain a bulk density of 
$535\,\mathrm{kg\,m^{-3}}$). One model was run from aphelion to November 8, 2014 (about 2.5 days before the start of the master period) in order to have 
a starting--point that reasonably accounted for previous evolution (in terms of temperature, stratification, \emph{etc}). All other models continued that model with new 
parameter combinations and used the 2.5 days (about 5 revolutions) as a relaxation period to adjust to the new conditions. 

\begin{figure*}
\centering
\begin{tabular}{cc}
\scalebox{0.4}{\includegraphics{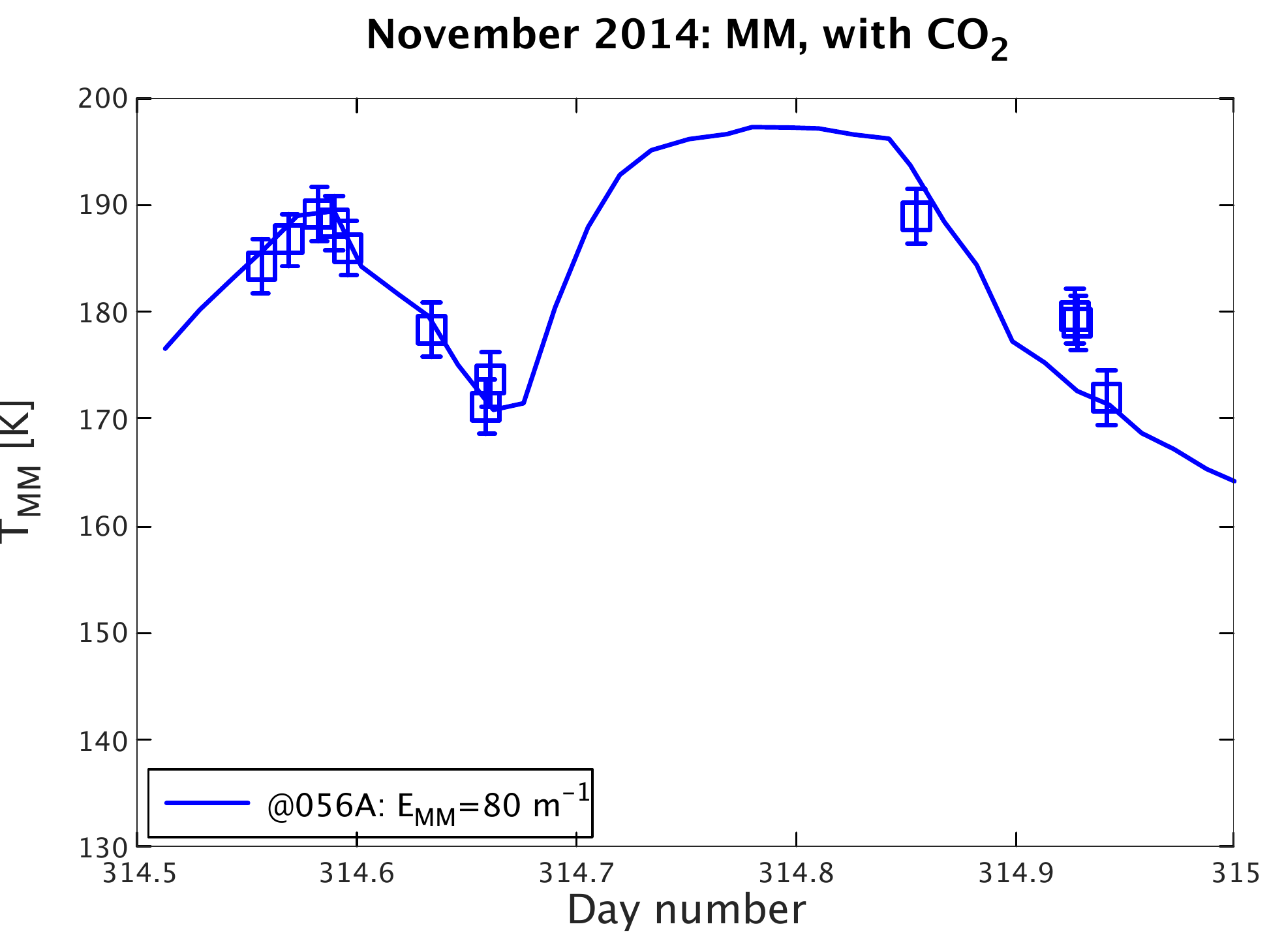}} & \scalebox{0.4}{\includegraphics{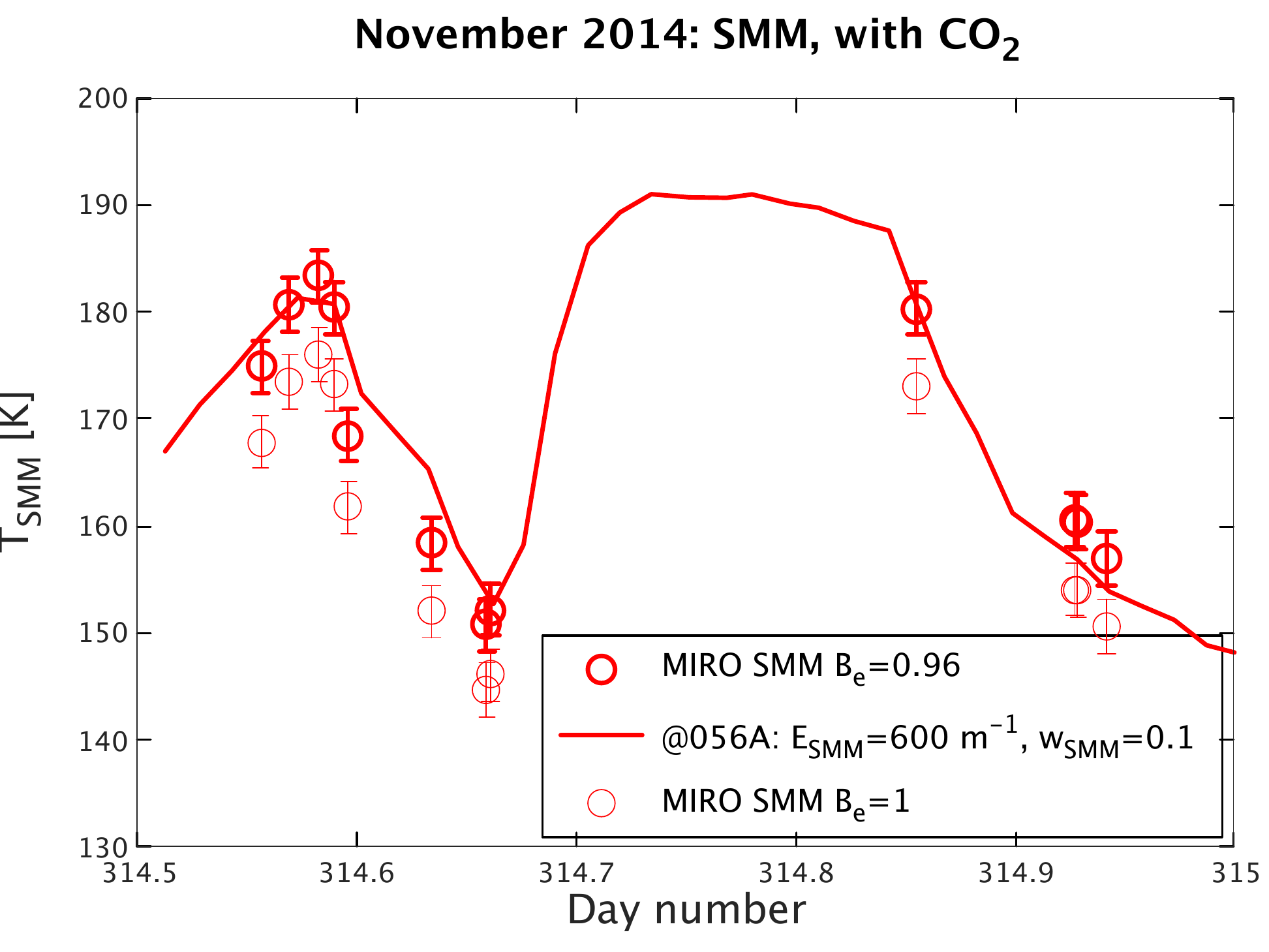}}\\
\end{tabular}
     \caption{The best available \textsc{nimbus}--based antenna temperature solutions (model @056A) compared with MIRO November 2014 MM (left) and SMM (right) data. 
This model has a thick ($h_{\rm m}=21\,\mathrm{cm}$) and opaque ($\zeta=0$) dust mantle with $\sim 65\,\mathrm{MKS}$ in the top $8\,\mathrm{mm}$ and 
$\sim 110\,\mathrm{MKS}$ below. Tubes are small, $\{L,\,r_{\rm p}\}=\{10,\,1\}\,\mathrm{\mu m}$, and $\mathrm{CO_2}$ is at a depth of $0.44\,\mathrm{m}$. 
The extinction coefficients are $E_{\rm MM}=80\,\mathrm{m^{-1}}$ and $E_{\rm SMM}=600\,\mathrm{m^{-1}}$, with a non--zero SMM single--scattering albedo of $w_{\rm SMM}=0.1$. 
These solutions confirm the differences at Hapi~D in November compared to the previous month, hinted at by the \textsc{btm} solutions. The cooling has been reduced because the 
dust mantle is thicker and the diffusivity is orders of magnitude lower, the thermal inertia and extinction coefficients have increased, the solid--state greenhouse effect 
is no longer detectable, and there may be some multiple--scattering at the SMM wavelength.}
     \label{fig_hapi12}
\end{figure*}

The first eleven models all considered conditions similar to the October 2014 solution ($2.5\,\mathrm{cm}$ dust mantle, $\zeta=2.15\cdot 10^{-2}\,\mathrm{m}$, 
$\mathrm{CO_2}$ at $0.44\,\mathrm{m}$ depth), but having smaller or higher Hertz factors compared to nominal conditions (to achieve different thermal inertia 
in the $40$--$100\,\mathrm{MKS}$ range) and $\{L,\,r_{\rm p}\}$ values ($\{10,\,1\}\,\mathrm{\mu m}$ or $\{28,\,6.6\}\,\mathrm{mm}$). This was done because 
of the suspected increase of thermal inertia and reduction of sublimation cooling (by lowering the diffusivity) with respect to October 2014. None of these 
models worked well, primarily because the MM antenna temperature amplitude was too large. The biggest difference compared to the rather successful \textsc{btm} simulations turned out 
to be the heat capacity of the dust mantle, that was 3.4 times smaller in the \textsc{nimbus} simulations. 

Therefore, 19 additional models were considered where the cell dust masses were increased 3.5--4.7 times (which also decreased the porosity), with 
the overall effect of producing thermal inertia in the $85$--$130\,\mathrm{MKS}$ range. These had $\{L,\,r_{\rm p}\}=\{10,\,1\}\,\mathrm{\mu m}$ to minimise 
the sublimation cooling, and considered different combinations of dust mantle opacities ($\zeta$ values) and dust mantle thicknesses 
(ranging $2.5$--$16.4\,\mathrm{cm}$). Despite the low diffusivity, most of these models were too cold. The exception was the case (model @021A) with the thickest 
dust mantle ($h_{\rm m}=16.4\,\mathrm{cm}$) that had $\Gamma=110$--$130\,\mathrm{MKS}$ and $\zeta=10^{-2}\,\mathrm{m}$, which yielded $Q=0.068$ 
for $E_{\rm MM}=100\,\mathrm{m^{-1}}$. However, when that model was tested at the SMM wavelength, the antenna temperature amplitude was far too small. 

At this point, 14 models with 3.5 times the original dust mass, and low thermal inertia ($30$--$60\,\mathrm{MKS}$) were run for different combinations of 
$\{L,\,r_{\rm p}\}$, $\zeta$, and $3.2\leq h_{\rm m}\leq 9.4\,\mathrm{cm}$, primarily to investigate if models with low thermal inertia are viable, despite the 
indications from \textsc{btm} that the thermal inertia is high. All these models had $Q<0.01$, and the best cases (with $Q$ on a $\sim 10^{-3}$ level) both had 
relatively thick dust mantles. It therefore seemed that the recipe for success (as suggested by the \textsc{btm} work) indeed is to consider a relatively high 
thermal inertia ($\sim 100\,\mathrm{MKS}$) and a substantially smaller level of cooling (heavily reducing the diffusivity is not sufficient, but the water 
sublimation front has to be comparably deep as well). 

Therefore, the final twelve models considered thick dust mantles ($14.4\leq h_{\rm m}\leq 27.3\,\mathrm{cm}$), dust mass enhancements that yielded 
thermal inertia in the $80$--$140\,\mathrm{MKS}$ range and $\{L,\,r_{\rm p}\}=\{10,\,1\}\,\mathrm{\mu m}$. Because several of these models were 
convincing at MM wavelength, the focus shifted more to finding ways to fit the SMM MIRO observations. The primary problem were the SMM amplitudes, 
although substantial improvements were seen compared to other SMM cases. The best overall solution (that somewhat degraded MM performance while 
substantially improving the SMM fit) was obtained by considering a somewhat lower dust mass enhancement in the top $8\,\mathrm{mm}$, yielding 
a somewhat lower thermal inertia ($\sim 65\,\mathrm{MKS}$) near the surface, compared to deeper parts of the mantle ($\sim 110\,\mathrm{MKS}$). 

This model (@056A) had a thick dust mantle ($h_{\rm m}=21\,\mathrm{cm}$), was opaque ($\zeta=0$), and had a best--fit extinction coefficient of 
$E_{\rm MM}=80\,\mathrm{m^{-1}}$ (resulting in $Q=0.048$). In the SMM this model had $Q=0.002$ for a beam efficiency of $B_{\rm e}=0.96$, when 
$E_{\rm SMM}=600\,\mathrm{m^{-1}}$ and the single--scattering albedo was $w_{\rm SMM}=0.1$. Though formally not statistically consistent with the data, this solution is still an 
improvement upon the best \textsc{btm} model (section~\ref{sec_results_Nov_BTM}). These solutions are shown in Fig.~\ref{fig_hapi12} and listed in Table~\ref{tab1}. If the MIRO SMM 
beam efficiency is closer to unity (as illustrated by the thin red circles in the right panel of Fig.~\ref{fig_hapi12}), this lowering of the SMM antenna temperature can be accommodated 
by increasing the single--scattering albedo a bit further.

The \textsc{nimbus} simulations seem to confirm the \textsc{btm} results -- the physical and chemical properties of Hapi~D in the top decimetres is distinctively 
different in November 2014 compared to the previous month. The dust mantle has become substantially thicker, and the properties of this mantle have changed. 
The thermal inertia has increased, the diffusivity has decreased substantially, there is no longer a measurable solid--state greenhouse effect, the 
extinction coefficients at both MM and SMM have increased significantly (i.e., the medium is less transparent), and there appears to be a need for multiple 
scattering at SMM wavelengths that was not present in October. All these changes are consistent with a significant compaction of the granular 
medium (see section~\ref{sec_results_context} for a further discussion).

As was the case for October, the fact that the same thermophysical model provides acceptable fits simultaneously at SMM and MM wavelengths 
speaks against a sharp difference in physical properties between Hapi~D and its immediate surroundings. Rather subtle differences in dust 
mantle strength and/or the exact depth of the $\mathrm{CO_2}$ could have been responsible for the more obvious morphological changes 
at Hapi~D in late December 2014, compared to the weaker alterations at Hapi~C revealed by the change of the spectral slope observed by OSIRIS.

\subsection{Contextual simulations} \label{sec_results_context}

The description of the physical and chemical properties of Hapi~D in October and November 2014 (summarised in Table~\ref{tab1}) are based on 
interpretations of MIRO observations with thermophysical and radiative transfer models. These results give rise to a number of questions: 1) under 
what conditions, if any, are the physical conditions in October 2014 a natural consequence of thermophysical evolution of this region in the previous 
years?; 2) what is causing the drastic change observed between October and November 2014, and is the short one--month timescale for these changes 
realistic?; 3) how is the region expected to change after November 2014, and can an explanation be found as to why pit formation was initiated at the 
end of December 2014?; 4) why did the pit changes stop at some point between late February and early March in 2015? In order to address these questions, we 
performed \textsc{nimbus} simulations from the May 2012 aphelion until mid--March 2015.

Table~\ref{tab1} states that $\mathrm{CO_2}$ is present at shallow depths, and we first need to investigate if and how the supervolatile can remain relatively 
close to the surface over extended periods of time. We first note, that the total amount of energy absorbed by Hapi~D 
between perihelion to aphelion (according to the model in section~\ref{sec_models_illum}) is $1.9\cdot 10^9\,\mathrm{J\,m^{-2}}$, the energy absorbed between the 2012 May 23 aphelion and late September 2014 is 
$3.1\cdot 10^9\,\mathrm{J\,m^{-2}}$, and a total of $4.3\cdot 10^9\,\mathrm{J\,m^{-2}}$ is absorbed from aphelion to perihelion. This means that 
Hapi~D receives $\sim 1.6$ times more energy between aphelion and the first considered MIRO observations, compared to that absorbed by the 
newly deposited airfall material near perihelion, on its way to aphelion. Therefore, we here only model the part of the orbit after May 2012, 
but will later discuss the effect of having been processed by an additional $\sim 60$ per cent of energy since the previous perihelion.

A first set of \textsc{nimbus} simulations considered model parameters consistent with our reproduction of October 2014 MIRO data ($\mu=2$, 30 per cent $\mathrm{CO_2}$ by 
number relative to water, $\{L,\,r_{\rm p}\}=\{28,\,6.6\}\,\mathrm{mm}$, $\zeta=2.15\cdot 10^{-2}\,\mathrm{m}$). Five models with $\mathrm{CO_2}$ fronts initially located 
at depths between $0.19$--$0.50\,\mathrm{m}$ at aphelion were considered. These models were stopped after one year, because the $\mathrm{CO_2}$ had already 
withdrawn to $0.79$--$1.11\,\mathrm{m}$, i.e., significantly below the targeted $0.48\,\mathrm{m}$ depth (Table~\ref{tab1}). It is clear that preservation of $\mathrm{CO_2}$ 
within half a metre of the surface requires special conditions, if at all possible.

\citet{davidssonetal22} performed global modelling of Comet 67P with \textsc{nimbusd} and simultaneously reproduced the global $\mathrm{H_2O}$ and $\mathrm{CO_2}$ 
production rates measured by \emph{Rosetta}/ROSINA, both pre-- and post--perihelion. They found that successful reproduction of the post--perihelion branch (when fresh airfall 
material deposited during the perihelion polar night is first exposed to sunlight) required a very large diffusivity on the northern hemisphere, consistent with 
$\{L,\,r_{\rm p}\}=\{0.1,\,0.01\}\,\mathrm{m}$. Such a high diffusivity would be expected in material dominated by 
loosely packed chunks with sizes ranging millimetres to decimetres. However, the pre--perihelion branch required a diffusivity corresponding to $\{L,\,r_{\rm p}\}=\{100,\,10\}\,\mathrm{\mu m}$. 
Such values are more consistent with particles ranging micrometres to millimetres. \citet{davidssonetal22} speculated 
that this three orders--of--magnitude drop in diffusivity taking place somewhere around aphelion was due to significant fragmentation in a top layer, caused by thermal fatigue and 
fracturing in brittle low--temperature material already weakened by water ice loss. Macroscopic chunks were pulverised and transformed to a low--porosity layer of fine grains. 
Such a layer could be millimetres to centimetres thick, yet provide an efficient gas diffusion barrier. It would overlay an interior still dominated by large chunks and substantial macro--porosity. 

We therefore considered the same \textsc{nimbus} models as before, 
but applied $\{L,\,r_{\rm p}\}=\{100,\,10\}\,\mathrm{\mu m}$ (and $\zeta=0$ because the conditions for efficient light penetration would temporarily be removed as well). Those 
simulations were run for more than twice as long (until the end of September 2014) compared to the first set, yet the $\mathrm{CO_2}$ ice only withdrew to $0.70$--$0.79\,\mathrm{m}$ 
(meanwhile, water ice withdrew from the surface to a depth of $8.4\,\mathrm{mm}$ and the total dust mantle erosion was merely $2\,\mathrm{mm}$). This was a substantial improvement, 
yet not satisfactory. A third set considered $\{L,\,r_{\rm p}\}=\{10,\,1\}\,\mathrm{\mu m}$ and additionally switched to $\mu=1$ (in order to increase the $\mathrm{CO_2}$ bulk density 
from $105\,\mathrm{kg\,m^{-3}}$ to $144\,\mathrm{kg\,m^{-3}}$ for a fixed 30 per cent abundance relative to water). That kept $\mathrm{CO_2}$ ice at $0.56$--$0.63\,\mathrm{m}$ in late 
September 2014 (increasing erosion to $3.4\,\mathrm{mm}$ because of the dust mantle bulk density reduction, and reducing the mantle thickness to $4\,\mathrm{mm}$). Note that 
the front propagation rate only is sensitive to diffusivity when sublimation is weak \citep[for an explanation of the lack of such sensitivity during strong sublimation, see section ~3.2 in][]{davidssonetal21}.

A final test lowered the Hertz factor by a factor 4, so that the typical dust mantle thermal inertia became $\sim 15\,\mathrm{MKS}$ instead of $\sim 30\,\mathrm{MKS}$. Such a 
reduction could take place if the quenching top layer has relatively few and small points of contact with the coarse substrate. Under such optimised conditions, $\mathrm{CO_2}$ ice withdrew 
from $0.19\,\mathrm{m}$ to $0.35\,\mathrm{m}$ between aphelion and the end of September 2014 (or by $16\,\mathrm{cm}$). 
If the front initially was at $0.31\,\mathrm{m}$, simulations showed that it withdrew by just $8\,\mathrm{cm}$ to $0.39\,\mathrm{m}$. In this case, a $3.6\,\mathrm{mm}$ layer was eroded, and the dust mantle 
thickness became $4\,\mathrm{mm}$. Considering that substantially less energy was available outbound, we therefore find the following sequence plausible: 1)  the $\mathrm{CO_2}$ ice was located to 
within $\sim 0.2\,\mathrm{m}$ of the surface when Hapi~D emerged from polar night after the 2009 perihelion; 2) it withdrew to $\sim 0.3\,\mathrm{m}$ by the time the comet reached aphelion in 2012; 3) it withdrew 
further to $\sim 0.4\,\mathrm{m}$ by the end of September 2014. 

\begin{figure*}
\centering
\begin{tabular}{cc}
\scalebox{0.4}{\includegraphics{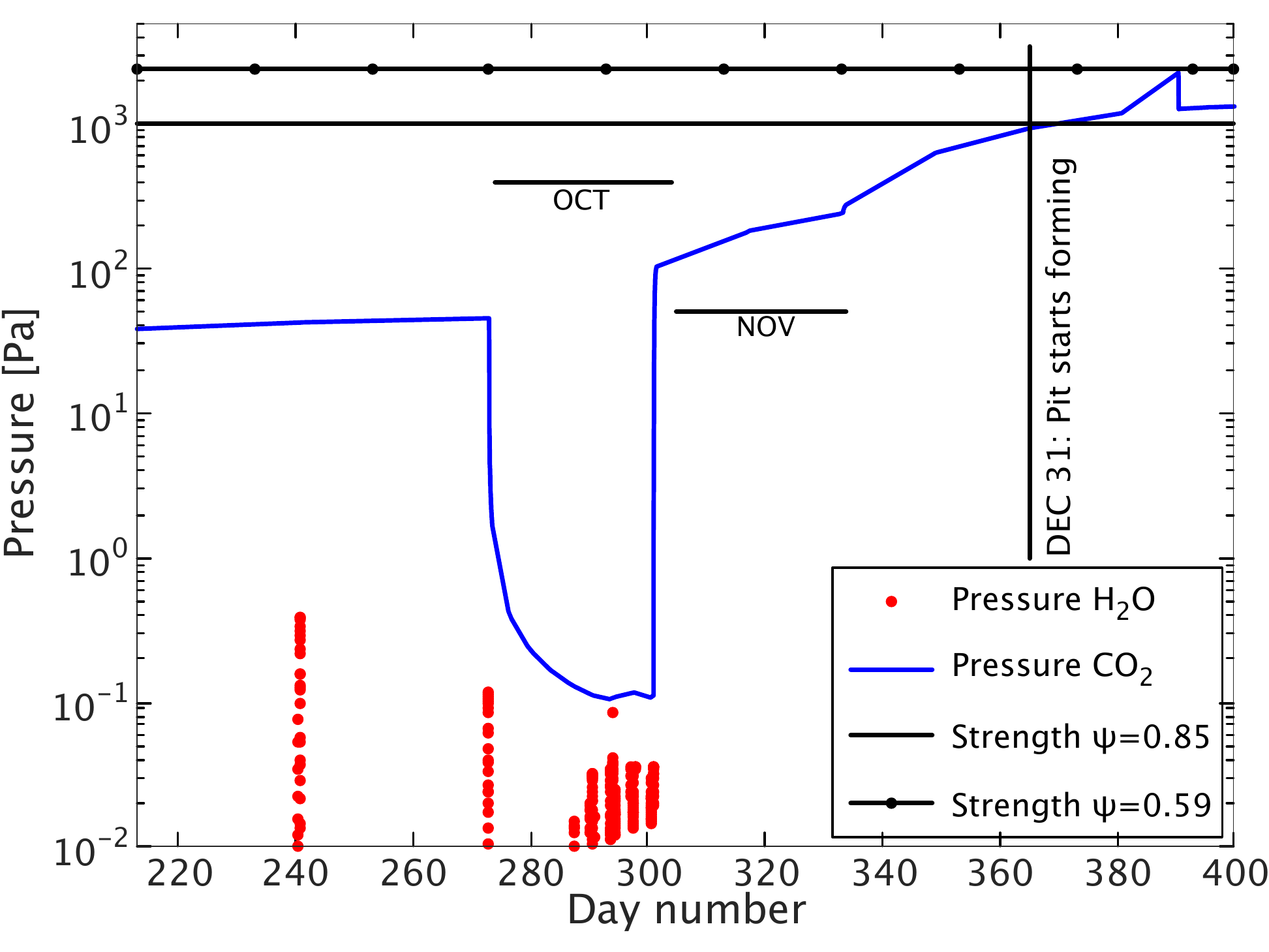}} & \scalebox{0.4}{\includegraphics{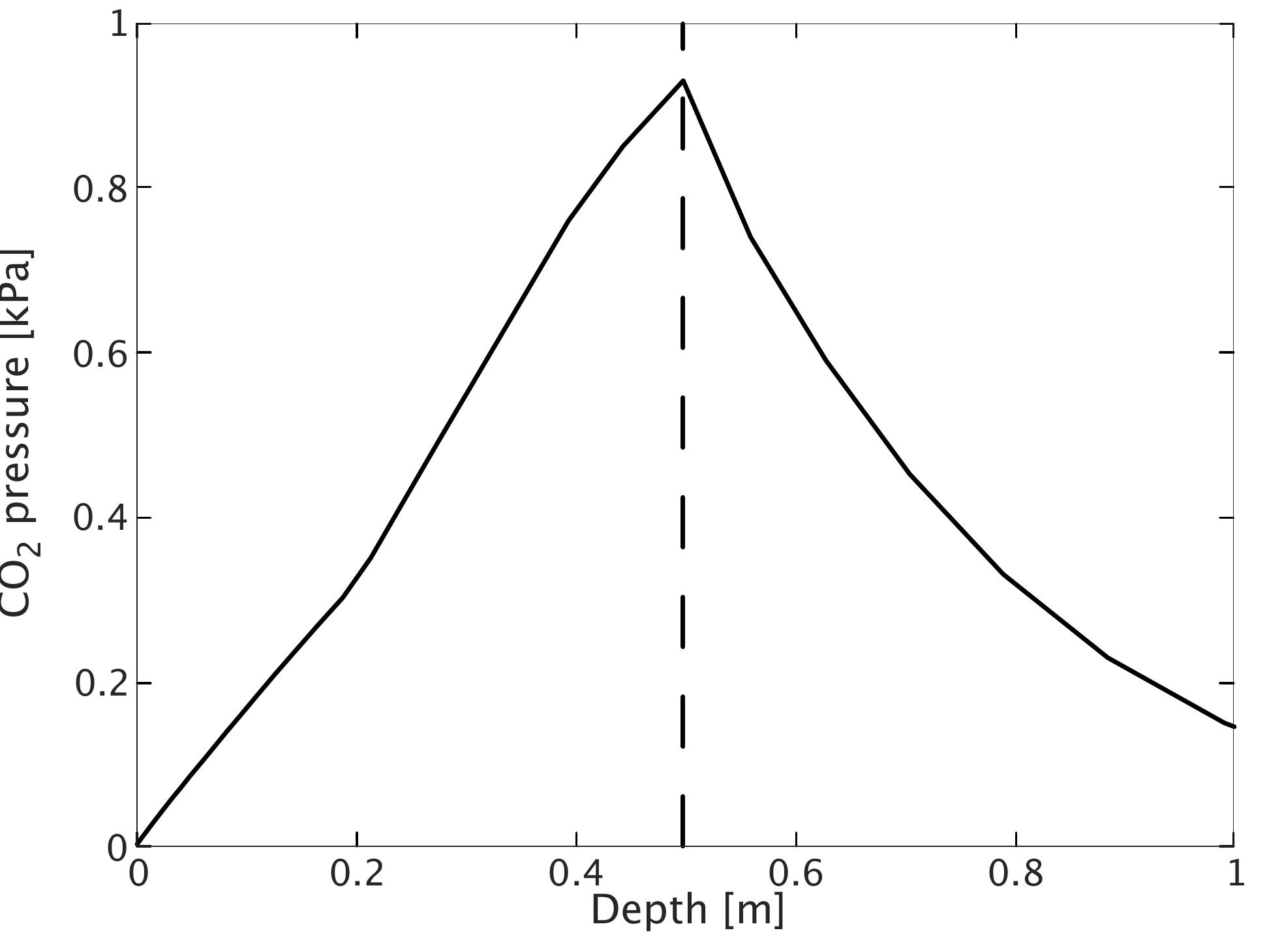}}\\
\end{tabular}
     \caption{\emph{Left:} The peak vapour pressures of $\mathrm{CO_2}$ and $\mathrm{H_2O}$, from early August 2014 (arrival of \emph{Rosetta} at Comet 67P) to late January 2015.  
The pressure of the shallowly located $\mathrm{H_2O}$ displays substantial diurnal variations (the value at every $10^{\circ}$ of rotational angle at selected revolutions are shown; the pressure 
is $<10^{-2}\,\mathrm{Pa}$ as of November 2014). The deeper $\mathrm{CO_2}$ yields a quasi--constant pressure on rotation--period time--scales. The dip in early October is due 
to the removal of an assumed low--diffusivity surface layer, and the return to a strong $\mathrm{CO_2}$ pressure in early November is due to an inferred fragmentation, collapse, and 
compaction of solids in the upper $\sim 0.5\,\mathrm{m}$. The pressure difference between the $\mathrm{CO_2}$ front and the surface eventually exceeds the tensile strength of 
porous assemblages of $\mathrm{\mu m}$--sized grains measured in the laboratory by \protect\citet{guttleretal09} and represented by the black lines: $\sim 1\,\mathrm{kPa}$ (strength of a $\psi=0.85$ medium) in late 
December when the first signs of pit formation were observed, reaching $\sim 2.4\,\mathrm{kPa}$ three weeks later (strength of a $\psi=0.59$ medium). \emph{Right:} The $\mathrm{CO_2}$ 
pressure versus depth $z$ in late December 2014. The pressure peaks near the $\mathrm{CO_2}$ sublimation front (located at the dashed vertical line), and falls off both inwards and outwards.}
     \label{fig_hapi13}
\end{figure*}

We therefore believe we have identified the conditions that would allow for the existence of shallow $\mathrm{CO_2}$ ice at Hapi~D in October 2014 ($\sim 0.48\,\mathrm{m}$ according to our 
interpretations of MIRO measurements). It requires a combination of low diffusivity, low heat conductivity, and a high $\mathrm{CO_2}$ concentration (yet, all parameters being within reasonable limits). 
Additionally, the $\mathrm{CO_2}$ ice would have to be shallow when emerging from near--perihelion polar night on the previous orbit, suggesting substantial removal of dust and perhaps water ice 
during an earlier pre--perihelion pit formation event. 

We postulate that the isolating and quenching layer was removed in early October 2014 by intensified erosion (most of the nominal $3.6\,\mathrm{mm}$ erosion takes place during September, and in reality, 
it may have been a factor of a few more substantial). Thereby, underlying coarse material was exposed. That removal would quickly increase the diffusivity by four orders of magnitude 
(here, from $\{L,\,r_{\rm p}\}=\{10,\,1\}\,\mathrm{\mu m}$ to $\{L,\,r_{\rm p}\}=\{28,\,6.6\}\,\mathrm{mm}$), and restore the nominal Hertz factor. That would speed up the withdrawal of both the 
$\mathrm{H_2O}$ and $\mathrm{CO_2}$ fronts. As the dust mantle thickened, the near--surface opacity would decrease and $\zeta$ increase. 

We ran a \textsc{nimbus} model throughout October 2014 that initially had $\mathrm{H_2O}$ and $\mathrm{CO_2}$ fronts at  $4\,\mathrm{mm}$ 
and  $0.35\,\mathrm{m}$, with $\zeta=0$ during the first week, $\zeta=4\cdot 10^{-3}\,\mathrm{m}$ during the second week, and $\zeta=2.15\cdot 10^{-2}\,\mathrm{m}$ during the remainder of the time 
(except if the dust mantle developed water frost at a $>0.1\,\mathrm{kg\,m^{-3}}$ level, which triggered temporary reinstatement of $\zeta=4\cdot 10^{-3}\,\mathrm{m}$). We found that the dust 
mantle thickness grew to $1.3\,\mathrm{cm}$, $1.9\,\mathrm{cm}$, $4.0\,\mathrm{cm}$, and $5.8\,\mathrm{cm}$ during weeks \#1 through \#4. We also found that the $\mathrm{CO_2}$ ice 
withdrew from $0.35\,\mathrm{m}$ to $0.44\,\mathrm{m}$. Considering that the October 2014 MIRO observations were performed during a 15--day period centred at the middle of the month 
(see section~\ref{sec_obs_miro}), our inferred dust mantle thickness of $2.3\,\mathrm{cm}$ (Table~\ref{tab1}) is consistent with the predicted dust mantle growth. In order to match a $\mathrm{CO_2}$ depth of 
$\sim 0.48\,\mathrm{m}$ (Table~\ref{tab1}), the initial depth in early October should have been somewhat deeper than the $0.35\,\mathrm{m}$ used here (perhaps at 
$\sim 0.40\,\mathrm{m}$, as mentioned previously). However, we think this test shows that the $\mathrm{CO_2}$ 
likely would have stayed within centimetres of the inferred depth throughout the 15--day period, despite the high diffusivity and resulting net sublimation rate. 

The contextual simulations indicate a $\sim 6\,\mathrm{cm}$ dust mantle at the end of October 2014, while the \textsc{nimbus} best--fit of November 2014 suggests a $\sim 21\,\mathrm{cm}$ 
mantle at that time. Possible explanations of this discrepancy include: 1) further growth during the remaining 10 days until the first November 2014 MIRO observation, as well as to the November 20 
mid--point of that measurement series; 2) more rapid growth if the dust--to--water--ice mass ratio is closer to $\mu=2$ than the assumed $\mu=1$ 
\citep[as suggested by][for airfall material]{davidssonetal22}; 3) the potential presence of an ice--free layer somewhere in the $6$--$21\,\mathrm{cm}$ region. For example, a 
$\stackrel{<}{_{\sim}} 14\,\mathrm{cm}$ ice--free mantle may have covered $\mathrm{H_2O}$-- and $\mathrm{CO_2}$--rich material at Hapi~D as it went into polar--night 
during the previous apparition, to be covered by a $\stackrel{>}{_{\sim}} 6\,\mathrm{cm}$ layer of fresh ice--rich airfall during the September 2009 perihelion. If so, that water ice could 
have been gradually removed during the time leading up to the end of October 2014. That could explain the jump in dust mantle thickness from $\sim 6\,\mathrm{cm}$ to $\sim 21\,\mathrm{cm}$, 
as October transited into November 2014, provided that our interpretation of MIRO measurements is correct. 

As previously pointed out, there is a distinct difference between the inferred October and November 2014 best--fit solutions to MIRO data (Table~\ref{tab1}). We suggest, that the 
water ice that was rapidly removed during October 2014, left behind a fragile dust mantle made up of eroded chunks that would have been weakened by the loss of the icy `glue' that held them 
intact. Continuously under stress and with forced relative movements caused by the $\mathrm{CO_2}$ and $\mathrm{H_2O}$ vapour welling up from underneath, it is likely that the chunks would 
 rub against one another and eventually crumble. We propose that the dust mantle chunks fragmented and decomposed into their smallest building--blocks, and that these tiny grains settled into a compact low--porosity layer in 
mid--November 2014. Such a collapse would explain the significant increases in thermal inertia and the MM/SMM extinction coefficients, as well as the four orders--of--magnitude drop in 
diffusivity,  and the removal of the solid--state greenhouse effect ($\zeta=0$), seen in Table~\ref{tab1}. Furthermore, local grain concentrations in this fine powder \citep[forming optically 
active sub--units;][]{hapke93}  may explain why the SMM channel seems to display signs of multiple--scattering ($w_{\rm SMM}>0$), as seen in lunar regolith \citep{garykeihm78}. 

We continued the contextual simulations, using November 2014 parameters and variants thereof. First, the dust bulk density was boosted a factor 4.7 (except in the top $\sim 8\,\mathrm{mm}$ that was 
boosted a factor 4) in order to account for the presumed compaction. Additionally, we set $\{L,\,r_{\rm p}\}=\{10,\,1\}\,\mathrm{\mu m}$, $\zeta=0$, and moved the water sublimation front to a depth of $21\,\mathrm{cm}$. 
That simulation was run throughout November and December 2014, this time focusing on the $\mathrm{CO_2}$ vapour pressure profile $p_{\rm CO_2}(z)$. We are interested in the $\mathrm{CO_2}$ 
vapour pressure due to its potential role in pit formation, because it exerts a substantial outward force on the near--surface material. We found, that by the end of 
December 2014 (when pit formation first became visible, see Fig.~\ref{fig_hapi01}, lower panels), the $\mathrm{CO_2}$ ice had withdrawn to $0.57\,\mathrm{m}$, and the vapour 
pressure at the $\mathrm{CO_2}$ sublimation front had reached $p_{\rm CO_2}(z=0.57\,\mathrm{m})\approx 0.3\,\mathrm{kPa}$.

In an additional simulation we decreased the diffusivity further by an order of magnitude during December 2014, by increasing the tortuosity from unity to $\xi=\sqrt{10}\approx 3.2$. 
The peak $\mathrm{H_2O}$ and $\mathrm{CO_2}$ pressures as functions of time for the total chain of contextual simulations are shown in Fig.~\ref{fig_hapi13} (left panel). The $\mathrm{CO_2}$ pressure 
first exceeds $0.01\,\mathrm{kPa}$ at $5.4\,\mathrm{au}$ inbound in March 2013, and remains below $0.045\,\mathrm{kPa}$ until late September 2014. Because of the postulated 
loss of the quenching top layer and the strong increase in diffusivity, the $\mathrm{CO_2}$ pressure plummets to $10^{-4}\,\mathrm{kPa}$ during October 2014. The $\mathrm{CO_2}$ 
is deep below the thermal skin depth, hence the blue curve shows no diurnal variations. However, the water ice is close to the surface, and the corresponding water vapour peak pressure 
(red) shows strong day/night variations. Except when $\{L,\,r_{\rm p}\}=\{28,\,6.6\}\,\mathrm{mm}$ in October 2014, the $\mathrm{CO_2}$ pressure is orders of magnitude stronger 
than that of $\mathrm{H_2O}$ (note that the water pressure drops below $10^{-2}\,\mathrm{Pa}$ after October 2014 because of the thickening dust mantle, and falls outside the plot). 
The compaction of the dust mantle in November 2014, reinstates a high $\mathrm{CO_2}$ pressure and it continues to grow with time thanks to the higher thermal conductivity, as 
solar radiation intensifies. With $\xi=\sqrt{10}\approx 3.2$ after November 2014, the $\mathrm{CO_2}$ ice withdraws to $0.50\,\mathrm{m}$, and the vapour pressure at the $\mathrm{CO_2}$ 
sublimation front reaches $p_{\rm CO_2}(z=0.50\,\mathrm{m})\approx 1\,\mathrm{kPa}$ at the end of December 2014. 

Figure~\ref{fig_hapi13} (left panel) also shows the tensile strengths of `dust cakes' consisting of $1\,\mathrm{\mu m}$ silica grains measured in the laboratory by \citet{guttleretal09}. 
These comet dust mantle analogues have a tensile strength of $\sim 1\,\mathrm{kPa}$ when the dust cake porosity is $\psi=0.85$, and a tensile strength of $2.4\,\mathrm{kPa}$ when $\psi=0.59$. 
We note that the peak $\mathrm{CO_2}$ pressure reaches $\sim 1\,\mathrm{kPa}$ at the end of December in the \textsc{nimbus} model with $\xi\approx 3.2$, and reaches $2.4\,\mathrm{kPa}$ 
three weeks later. We note the similarity between calculated $\mathrm{CO_2}$ vapour pressures and measured dust mantle analogue tensile strengths, and emphasise 
that the timing when this similarity took place coincides with the observed start of pit formation. We therefore propose that the pit formation observed by OSIRIS at Hapi~D at the end of 
December 2014, was caused by the sublimation of shallow ($\sim 0.5\,\mathrm{m}$) $\mathrm{CO_2}$ ice, that became sufficiently strong to start ejecting the mantle to space. 

The right panel of Fig.~\ref{fig_hapi13} shows $p_{\rm CO_2}(z)$ at the end of December 2014. The $\mathrm{CO_2}$ vapour pressure peaks at the sublimation front 
and falls off both towards the surface and towards the deep interior (vapour diffuses both upwards and downwards, following local gradients in pressure and temperature). 
Note that the tensile failure would occur at the depth where the mantle is weakest, having a strength that is smaller than the local $\mathrm{CO_2}$ pressure. This depth is not 
necessarily coinciding with the $\mathrm{CO_2}$ sublimation front, but could be more shallow. That suggests that pit deepening may have been gradual, i.e., shallower layers 
were ejected first and additional material was ejected later, as the pressure function adjusted to the new conditions. That is to say, the entire $\sim 0.5\,\mathrm{m}$ layer 
of dust was probably not ejected at once. Removal of decimetres of dust also means that energy can reach buried $\mathrm{CO_2}$ ice laterally near the rim, 
in addition to conduction from above. That may explain why the pits were spreading laterally over time. Finally, the depths of the pits ($\sim 0.5\,\mathrm{m}$) would be a 
natural consequence of ejecting loose dust and old airfall material (but not necessarily indigenous material, still held together by water ice).

The pits of Hapi~D seem to have stopped expanding at some point between 2015 February 28 and March 17 (Fig.~\ref{fig_hapi03}). This may be understood by using the 
illumination simulations (see section~\ref{sec_models_illum}) and calculating the total energy absorbed by Hapi~D per nucleus rotation (rot). This amount of energy peaks 
on 2015 February 15, at $2.98\cdot 10^6\,\mathrm{J\,m^{-2}\,rot^{-1}}$. Between 2015 February 28 and March 17, this energy fell rapidly from $2.87\cdot 10^6\,\mathrm{J\,m^{-2}\,rot^{-1}}$ 
to $2.63\cdot 10^6\,\mathrm{J\,m^{-2}\,rot^{-1}}$, reaching the same level that had prevailed in early June 2014. It is therefore clear, that pit growth proceeded at Hapi~D until 
the comet reached $r_{\rm h}=2.06$--$2.21\,\mathrm{au}$, but at that point (by virtue of the spin axis orientation and shadows caused by nucleus topography around the Hapi valley), 
the daily solar heating suddenly fell to a level as low as in June 2014 (when the comet had been at $r_{\rm h}=3.9\,\mathrm{au}$). Therefore, detectable morphology changes at Hapi~D stopped. 

We note that Hapi is a large region, and changes were recorded elsewhere at later times. Importantly, the `aeolian ripples' \citep{thomasetal15b} started disappearing in April 2015, and were 
replaced by an expanding pit, until the ripples reformed in December 2015 \citep[see Fig.~S10 in the supplementary material of][]{elmaarryetal17}. Whereas Hapi~D is located close to the 
north pole, the aeolian ripples are found near the equator. They were strongly illuminated around the time of the May 2015 equinox, which explains their later onset with respect to Hapi~D.

\section{Discussion} \label{sec_discuss}

Our understanding of the evolution of Hapi~D that has emerged through this work is summarised as follows. The region likely experienced excavation of several 
decimetres of material in February and March 2009, similarly to what OSIRIS observed on the following orbit. It entered polar night with a $\sim 0.1\,\mathrm{m}$ 
dust mantle, overlaying an icy interior, rich in both $\mathrm{H_2O}$ and $\mathrm{CO_2}$ ice. During the 2009 perihelion another $\sim 0.1\,\mathrm{m}$ 
was added in the form of mm--dm--sized airfall chunks consisting of refractories and water ice. On the way toward aphelion, 
$\mathrm{CO_2}$ withdrew from $\sim 0.2\,\mathrm{m}$ to $\sim 0.3\,\mathrm{m}$, while the upper few mm--to--cm lost its water ice and crumbled into a low--diffusivity top 
layer, having poor thermal contact with the substrate. This isolating and quenching top layer slowed the $\mathrm{CO_2}$ withdrawal, and it was removed, through intensified erosion, 
around September 2014. When MIRO started observing Hapi~D in October, the antenna temperature revealed a solid--state greenhouse effect (caused by the coarse near--surface material), 
and measurable signatures of shallow ($\sim 0.5\,\mathrm{m}$) $\mathrm{CO_2}$ ice. Rapid dust mantle thickening followed by collapse and compaction of the fragile dust layer in late October or early November 2014, 
caused significantly increased heat conductivity, optical opacity, and microwave extinction, as well as a drastic drop in diffusivity. Those modifications caused measurable changes to the thermal 
emission observed by MIRO. This scenario is also supported by the OSIRIS spectrophotometry (section~\ref{sec_obs_osiris_spectra}), that suggested that the top decimetres were not 
particularly ice--rich. The relatively high heat conductivity and low diffusivity, caused a gradually increasing $\mathrm{CO_2}$ vapour pressure and steepening 
pressure gradients. In late December 2014, the tensile strength of the mantle was exceeded and pits started to form, as observed by OSIRIS. Pit formation removed the 
upper $\sim 0.5\,\mathrm{m}$ of cometary material. These shallow depressions grew laterally (with a terminal velocity of $\sim 1.5\,\mathrm{m\,d^{-1}}$) until the daily solar energy input 
fell below an activity threshold in early March 2015. Towards the end of growth, gas drag was not sufficiently strong to eject the $2.1\pm 0.4\,\mathrm{m}$ boulder B3. When pit growth 
stopped at Hapi~D, the stratification may have been similar to that in early 2009. If so, pit formation may be a cyclic behaviour that repeats every orbit. We note that the escarpment 
stopped moving near a point where there previously was a ridge (compare Fig.~\ref{fig_hapi03}, lower right, with Fig.~\ref{fig_hapi01}, upper right). That ridge possibly marks the 
location where the 2009 escarpment came to a halt. 

We now discuss various aspects of this scenario in the light of other investigations in the literature. \citet{cambianicaetal20} attempted to determine the thickness of 
the material deposited in Hapi during one perihelion passage by measuring the length of shadows cast by boulders. Their average for ten boulders suggests 
the addition of a $1.4\,\mathrm{m}$ thick layer, but unfortunately the error bars for all individual boulders are as large or larger than the reported deposition thickness.  
We therefore consider our proposed deposition ($\sim 0.1\,\mathrm{m}$) consistent with the measurements of \citet{cambianicaetal20}. 

The compaction we propose to have taken place in late October or early November has not been observed in the form of a measurable subsidence. 
However, \citet{davidssonetal22c} demonstrate that the single--scattering albedo at Hapi~D was reduced between 2014 August 30 and December 10. 
They suggest that this darkening (presumably caused by a decrease of porosity and an increased coherent effect, as small brighter grains started acting as larger and darker 
optically effective particles) is a manifestation of the compaction that actually was observable by OSIRIS. 

\citet{davidssonetal22} found that $\mathrm{CO_2}$ ice on average is located $\sim 4\,\mathrm{m}$ below the surface on the northern hemisphere. It means that 
the $\mathrm{CO_2}$ ice at Hapi~D is unusually shallow. That may explain why pit formation is a localised phenomenon. The $\mathrm{CO_2}$ sublimation front depth probably  
varies strongly within the Hapi valley, so that some areas evolve more calmly, while others are subjected to more violent morphological changes. If roundish and laterally expanding 
depressions in smooth terrain are indicative of shallow $\mathrm{CO_2}$ deposits, it automatically means that the local airfall coverage is thin (because airfall chunks are not expected to 
carry $\mathrm{CO_2}$ -- that substance much be located in the native comet material, below the airfall layer). Mapping of regions with or without 
pit formation may therefore offer a method of `tomography' that probes the thickness of smooth material in such terrain. 

\citet{davidssonetal22} also found that the mass ratio of refractories to water ice was $\mu\approx 1$ on the strongly active southern hemisphere, and that the water 
abundance of airfall material is somewhat lower ($\mu\approx 2$). That increase of the refractory to water ice mass ratio is consistent with the level of water loss from cm--dm--sized chunks that are fully 
exposed to solar radiation in the coma during transfer times of $12\,\mathrm{h}$, according to \textsc{nimbus} calculations by \citet{davidssonetal21}. The current study 
confirms that a water abundance corresponding to $\mu=1$--2 is consistent with the dust mantle thickness and its variation with time, as inferred from MIRO observations. 
\citet{davidssonetal22} showed that the total comet production rates of $\mathrm{H_2O}$ and $\mathrm{CO_2}$ vapours measured by \emph{Rosetta}/ROSINA places strong 
constraints on the depths of the sublimation fronts of both species. They also found that reproduction of the high $\mathrm{CO_2}$ production rate right 
after perihelion, required a high $\mathrm{CO_2}$ abundance, perhaps as large as 30 per cent relative to water by number. This is consistent with what we find in the current work. 
Keeping $\mathrm{CO_2}$ ice as close to the surface as inferred from the October 2014 MIRO observations (even in the limit of very low diffusivity and conductivity) requires a 
$\mathrm{CO_2}$ concentration of $\stackrel{>}{_{\sim}} 100\,\mathrm{kg\,m^{-3}}$. For a bulk density of $535\,\mathrm{kg\,m^{-3}}$ (i.~e., the nucleus average), and a mass 
ratio of refractories to water ice of $\mu=1$--2, this corresponds to a molar $\mathrm{CO_2}$ abundance of at least 30 per cent with respect to water. That is somewhat 
high compared with the range of 10--23 per cent measured in massive protostars \citep{gerakinesetal99}, but is consistent with the $32\pm 2$ per cent range measured for 
low--mass protostars \citep{pontoppidanetal08}, assuming that the near--surface abundance is representative of the bulk. The latter should be more relevant analogues of the Solar System. 

\citet{birchetal19} proposed a scenario to explain the presence of roundish expanding features in smooth terrain. Sloped surfaces are illuminated directly from 
above by the Sun, but additionally, by infrared self--heating from surrounding flat terrain. Dust is more readily removed from an inclined surface compared to one that 
is perpendicular to local gravity. In their view, the combination of preferential heating and facilitated water ice exposure causes the slopes to evolve into moving escarpments through a 
higher erosion rate than for surrounding material. We do not reject this hypothesis, that may accurately describe the origin of certain moving escarpments on 67P. 
However, we do not think this mechanism primarily is causing the formation of the particular pits studied in this paper. Our thermophysical modelling shows that superficial water ice 
barely is active at the prevailing illumination conditions. The MIRO antenna temperatures seem to require the presence of an additional cooling agent besides water. 
$\mathrm{CO_2}$ activity offers a substantially more compelling explanation of the observed dramatic phenomenon, owing to its volatility. The engine that drives escarpment movement 
is buried decimetres under ground, and is more sensitive to reductions of that depth, than to near--surface topography. Once the mantle has been ejected at one point, 
$\mathrm{CO_2}$ activity near the escarpment base will intensify and rapidly remove additional material, until the escarpment enters a region where the 
$\mathrm{CO_2}$ ice is located too deeply to lift material, or the level of illumination becomes too low. 

Another scenario was proposed by \citet{bouquetyetal22}. They measured a range of morphometrical parameters describing the
dimensions, shapes, and orientations for 131 depressions on 67P, and found an analogy with terrestrial alases and martian
scalloped depressions. Such structures on Earth and Mars form when water ice is evacuated from a soil, and the remaining porous
solid loses its mechanical integrity and collapses. \citet{bouquetyetal22} therefore propose a similar formation scenario for the
depressions on 67P. These have depths ranging $0.4$--$16\,\mathrm{m}$, with a mean and standard deviation of $4.8\pm
4.5\,\mathrm{m}$. Additionally, \citet{thomas20} report depth measurements of a shallow depression in Anubis, and found that its 
elevation was reduced by $\sim 2\,\mathrm{m}$ between September 2014 and June 2016. If the \citet{bouquetyetal22} scenario is correct, 
it implies large local variations in the dust mantle thickness. The average thickness required to explain the observed water production rate,
 is $\sim 0.02\,\mathrm{m}$ \citep{davidssonetal22}, and the switch--off of dust jets (that are common across the entire nucleus) just beyond the terminator requires a water ice 
sublimation front at $\sim 0.006\,\mathrm{m}$ according to \citet{shietal16}. In pit--forming regions, the dust mantle would have to be 
$\sim 5\,\mathrm{m}$ thick on average. We note that a $\mathrm{CO_2}$ sublimation front average depth of $\sim 4\,\mathrm{m}$ is 
necessary to explain the carbon dioxide production rate curve of 67P according to \citet{davidssonetal22}. The similarity between the average pit 
depth and the average $\mathrm{CO_2}$ front depth suggests that ejection of dust and water ice due to $\mathrm{CO_2}$ activity is an 
alternative to the water loss and mantle collapse proposed by \citet{bouquetyetal22}. Further analysis is needed to demonstrate whether the $\mathrm{CO_2}$--driven
ejection of a $\geq 5\,\mathrm{m}$ thick layer indeed is possible, i.~e., if sufficient $\mathrm{CO_2}$ pressure can be reached at
such depths to overcome the tensile strength of the overlying ice--dust mixture. Whereas \citet{bouquetyetal22} see subsidence as 
the major pit--forming mechanism, we here see the collapse as a smaller prelude, that does not lead to detectable morphological changes 
\citep[put perhaps leads to darkening, as previously mentioned;][]{davidssonetal22c}. In the currently proposed scenario, morphological changes come after 
compaction and darkening, and are due to ejection of material, not subsidence. 

In order to further investigate the role of $\mathrm{CO_2}$ ice in pit formation and escarpment expansion on 67P, the MIRO database should be 
searched for observations at the time and place of other prominent examples of morphological changes in smooth terrain. It would be particularly interesting 
to perform an analysis of the Imhotep region prior to, and during, the May--July 2015 events documented by \citet{groussinetal15b}. We hope that 
the current paper serves as an inspiration and guide on how to perform such an investigation. It would be important to better understand whether 
near--surface $\mathrm{CO_2}$ ice is common within smooth terrain on 67P, or if the pits in Figs.~\ref{fig_hapi01}--\ref{fig_hapi03} are unique. 
Such an investigation could prove extremely valuable in the context of a cryogenic comet sample--return mission. Retrieval of $\mathrm{CO_2}$--rich material 
at a few decimetres depth is substantially easier and cheaper than being forced to drill several metres. Knowing where to sample is another difficult practical problem 
that needs to be solved. If escarpments in smooth terrain prove to be indicative of shallow $\mathrm{CO_2}$ deposits, such visible surface expressions of past activity could be 
exploited during reconnaissance prior to sampling. The unique capability of microwave instruments to measure sub--surface temperatures from orbit could further 
facilitate the search for accessible supervolatiles. 

As a final comment on our work, we note that Fig.~\ref{fig_hapi13} (right) shows an interesting phenomenon: the $\mathrm{CO_2}$ vapour pressure \emph{below} the $\mathrm{CO_2}$ sublimation 
front falls to very small values over a distance that is comparable to the depth of the front below the surface. In addition to the pressure difference between the front and the surface that 
is responsible for ejecting mantle material, there is also a strong pressure difference between the front and deeper regions, that would strive to displace material downwards. It means that the $\mathrm{CO_2}$ 
vapour not only can eject shallow material into the coma, but it might also be capable of compressing material at depth. 

The structural changes of a porous medium with its pores filled with a pressurised gas or liquid are studied in the branch of continuum mechanics known as 
poroelasticity, first formulated in detail by \citet{biot41}. The equation of motion for the solid describes its coupling to the gas;
\begin{equation} \label{eq:biot}
\rho\frac{\partial^2 u}{\partial t^2}=G\frac{\partial \sigma}{\partial z}-\frac{2(1+\nu_{\rm p})G}{3(2-\nu_{\rm p})H}\frac{\partial p}{\partial z}+\frac{G}{1-2\nu_{\rm p}}\frac{\partial\epsilon_{\rm v}}{\partial z}
\end{equation}
where $\rho$ is density, $u$ is the displacement, $t$ is time, $z$ is depth, $G$ is the shear modulus, $\sigma=K\partial u/\partial z$ is the stress ($K$ is the bulk modulus), $\nu_{\rm p}$ is the 
Poisson ratio, $p$ is the gas pressure, $H^{-1}$ is the poroelastic expansion coefficient, and $\epsilon_{\rm v}$ is the volume change of the solids. As long as the tensile material 
strength (first right--hand term) balances the pressure force (second right--hand term), the solid is static and the other terms are zero. Once the material yields, particle acceleration 
begins and the medium is deformed according to equation~(\ref{eq:biot}) until balance is restored anew. If equation~(\ref{eq:biot}) is integrated over a slab of thickness $dz$, the force due to 
gas pressure is proportional to the pressure difference between the slab walls, which explains the importance of the pressure profile in Fig.~\ref{fig_hapi13} (right): the prerequisites for 
ejection above the front and compression below the front, are co--existing. In this context, we note that thermophysical comet nucleus models typically treat dust mantle ejection by requiring 
that the combined gas drag force and centrifugal force overcome nucleus gravity on a grain--by--grain basis \citep[e.~g.,][]{shulman72,rickmanetal90,espinasseetal93,oroseietal95}. 
However, in the seminal paper by \citet{fanaleandsalvail84}, the dust mantle ejection criterion was formulated for  an entire mantle slab of thickness $dz$ (as discussed above), and they 
indeed applied the pressure difference over the slab in their criterion. In the context of sub--surface compaction, we are interested in understanding which magnitude the pressure difference might 
reach, compared to the compressive strength of the solids.

A numerical experiment was performed with \textsc{nimbus}, where the model in Fig.~\ref{fig_hapi13} was propagated to mid February 2015. At that point, the diurnal illumination profile 
was scaled up to peak at the flux expected at the perihelion sub--solar point of Comet 67P. Such conditions are not relevant for Hapi~D but could have been for other parts of the comet. 
After two weeks of cycling at those flux levels, the $\mathrm{CO_2}$ vapour pressure peaked at $4.9\,\mathrm{kPa}$, which may be considered the highest $\mathrm{CO_2}$ vapour pressure achievable for 67P at a 
depth of $\sim 0.5\,\mathrm{m}$ \citep[but note that $\mathrm{CO_2}$ at $\sim 0.2\,\mathrm{m}$ at the south pole reached a pressure of $19\,\mathrm{kPa}$ at perihelion;][]{davidssonetal22}. According to the work of \citet{guttleretal09} on compressive strength, $1\,\mathrm{\mu m}$--grain silica powder would compress to a porosity $\psi=0.73$ if the pressure 
reaches $4.9\,\mathrm{kPa}$. Mixtures of dust and ices are stronger, and by applying the method described by \citet{davidsson21} for an ice volumetric fraction of 40 per cent, 
such a mixture compacts to $\psi=0.76$ at $4.9\,\mathrm{kPa}$. These values are similar to the bulk porosity of 0.75-0.85 inferred for the nucleus itself \citep{kofmanetal15}. 
Taken at face value, this would not suggest significant additional compression of the nucleus material because of the $\mathrm{CO_2}$ vapour pressure. 

However, the measured porosity--pressure relation is based on short--term strength and ignores the creep deformation that would take place when the medium is subjected to a continuous stress for weeks and months. 
Laboratory investigations of creep in icy soils subjected to long--term loads show that the strain (i.~e., the degree of deformation, here equivalent with compaction) increases with 
time \citep[e.~g.,][]{fish83,gardneretal84,hampton86}. Measurements of the strain for terrestrial snow \citep[which constitutes an upper limit in terms of strength, because ice/dust mixtures are 
weaker than pure ice;][]{loreketal16} shows that 
\begin{equation}
\epsilon=\frac{\sigma}{33\,\mathrm{GPa}}\left(1-\frac{T}{273\,\mathrm{K}}\right)^{-0.65}\left(\frac{\rho}{0.9\,\mathrm{Mg/m^3}}\right)^{-9}t^{0.6}
\end{equation}
with $t$ measured in hours \citep{meussenetal99}. Considering compaction from an assumed bulk porosity of $\psi=0.8$ to $\psi=0.6$ (equivalent to a strain $\epsilon=0.75$), 
and assuming $\sigma=4.9\cdot 10^{-6}\,\mathrm{GPa}$, $T=130\,\mathrm{K}$, and $\rho=0.3\,\mathrm{Mg\,m^{-3}}$, then that compaction could be achieved in seven months.

Furthermore, strength typically decreases with increasing size scales. This suggests that comet nucleus material on the metre--scale and above might be weaker 
than the strength measured in the laboratory by \citet{guttleretal09} on much smaller scales. Indeed, by studying collapsed overhangs on 67P, \citet{groussinetal15} 
found that the compressive strength of cometary material was merely 0.03--$0.15\,\mathrm{kPa}$ on 5--$30\,\mathrm{m}$ scales. 

Compression of the type suggested here was not observed in the KOSI experiments 3--7 that included $\mathrm{CO_2}$ \citep{lammerzahl95}. However, the 
diffusivity was rather high \citep[$\sim 0.1\,\mathrm{m^2\,s^{-1}}$;][]{benkhoffandspohn91b} in these experiments, because of relatively coarse grains and similarly--sized 
pore spaces \citep[0.01--$1\,\mathrm{mm}$;][]{lammerzahl95}, resulting in modest $\mathrm{CO_2}$ partial pressures \citep[peaking at $13\,\mathrm{Pa}$;][]{hsiungandroessler89}. 
This is insufficient to cause compaction. We reach $\mathrm{kPa}$--level pressures by having $10^{-5}$--$10^{-4}\,\mathrm{m^2\,s^{-1}}$ diffusivities,  as expected for porous but 
homogeneous aggregates of $\mathrm{\mu m}$--sized monomers. 

The $\mathrm{CO_2}$ sublimation front can therefore be thought of as a slowly propagating wave that compacts the material before it, leaving behind a mixture of refractories and 
water ice that is compressed with respect to the deep interior. The compaction process might be self--reinforcing to a certain level. By decreasing the porosity, which lowers the diffusivity but increases heat conductivity, the 
$\mathrm{CO_2}$ sublimation front needs to develop ever increasing vapour pressures to elevate the vapour mass flux rate to the point where most heat conducted to the front is being 
consumed by net sublimation. This effect decreases the porosity further, which facilitates additional compaction, and so on. The material that is observable on the surface of comet nuclei 
could therefore very well have experienced substantial compression, structural alteration, and mechanical modification before it became exposed. Such processing could affect the way 
near--surface material fractures, and the size--frequency distribution function of coma material. 

\citet{davidssonetal22} found that the post--perihelion $\mathrm{CO_2}$ production rate of 67P required a substantial near--perihelion reduction of 
diffusivity (at the depth where $\mathrm{CO_2}$ vapour is being produced) on the southern hemisphere, by a factor 10--250. They proposed that this reduction 
of diffusivity was due to the movement of the $\mathrm{CO_2}$ sublimation front into a deeper layer that it previously had compressed during strong near--perihelion activity.  
Therefore,  sub--surface compression due to $\mathrm{CO_2}$ activity appears consistent with ROSINA data. 

This type of $\mathrm{CO_2}$--driven near--surface compaction might explain some puzzling observations of comets. Comparisons between the densities of the top few meters of comets 
inferred from radar observations, and those of the bulk nuclei derived from non--gravitational force modelling, led to suggestions of a thin compacted surface layer on comets even before \emph{Rosetta} 
\citep{davidssonetal09, kamounetal14}. The CONSERT experiment of \emph{Rosetta/Philae} showed that the upper few hundred metres of the small lobe of 67P  has an average dielectric 
constant (at $90\,\mathrm{MHz}$) of $\varepsilon'=1.27$, and an inferred porosity of $\psi=0.75$--$0.85$ \citep{kofmanetal15}. Further analysis of this data set revealed a radial gradient of the dielectric 
constant, taking values of $\varepsilon'=1.7$ near the surface that transitioned to $\varepsilon'=1.3$ at a depth of $150\,\mathrm{m}$ \citep{ciarlettietal15} or of 
`tens/hundreds of metres' \citep{ciarlettietal18}. This suggested an increase of porosity and/or decrease of the dust--to--ice mass ratio with depth \citep{ciarlettietal15}. 
Measurements by \emph{Philae}/SESAME--PP showed that $\varepsilon'=2.45\pm 0.2$ (at 409--$758\,\mathrm{Hz}$) within the top metre at Abydos, consistent with a porosity of $\psi<0.5$ or 
$\psi<0.75$ for dust analogues of carbonaceous or ordinary chondrites, respectively \citep{lethuillieretal16}. The work by \citet{brouetetal16} on SESAME--PP data suggests the upper limit on 
porosity may even be as low as 0.18--0.55. Comparing with ground--based Arecibo radar observations of 
67P \citep[$\varepsilon'=1.9$--2.1 at $2.38\,\mathrm{GHz}$ in the top $\sim 2.5\,\mathrm{m}$;][]{kamounetal14}, it was suggested by \citet{lethuillieretal16} that there 
`may also be a gradient in porosity in the first meters of the cometary mantle'. Temperature measurements by \emph{Philae}/MUPUS--TM at Abydos are consistent with a thermal 
inertia of $85\pm 35\,\mathrm{MKS}$ and a porosity of $\psi=0.3$--$0.65$ near the surface \citep{spohnetal15}. 

We propose that compaction of material passed by the $\mathrm{CO_2}$ sublimation front, caused by the $\mathrm{CO_2}$ vapour pressure gradient below the front, is responsible 
for the observed increase of porosity with depth (the $\mathrm{H_2O}$ vapour pressure is orders of magnitude weaker, and is not capable of such compression of the solids). We have demonstrated 
that the $\mathrm{CO_2}$ pressure difference between the front and the interior is sufficiently large to reduce the porosity from $\psi=0.75$--$0.85$ at large depth, 
to $\sim 0.6$ in near--surface regions that previously have been passed by the $\mathrm{CO_2}$ sublimation front, on time scales that are short compared to the orbital period. 
Such a porosity gradient is consistent with the porosity values at depth 
and near the surface inferred from the \emph{Rosetta} and Arecibo observations described above. \citet{davidssonetal22} show that the $\mathrm{CO_2}$ sublimation 
front on average is located $\sim 4\,\mathrm{m}$ below the surface on the northern 
hemisphere, and $\sim 2\,\mathrm{m}$ below the surface on the southern hemisphere of 67P, which roughly would correspond to the thickness of the compressed layer. 
We further propose that this type of compaction is responsible for the formation of consolidated material on 67P observed by OSIRIS \citep[e.g.][]{thomasetal15a, elmaarryetal15}. 
It is visually recognisable by the `rock--like' appearance caused by its brittleness that makes it prone to fracturing and forming angular faceted shapes \citep[e.g.][]{elmaarryetal15b, augeretal18}. 
Sheet--like overhangs with thicknesses of $5$--$30\,\mathrm{m}$ and lateral extensions of $10$--$100\,\mathrm{m}$ at the time of eventual collapse \citep{groussinetal15} 
serve as illustrations that the compacted near--surface layers of consolidated material are thin and stronger than the material that has been eroded from underneath them.

\section{Conclusions} \label{sec_conclusions}

We present \emph{Rosetta}/OSIRIS observations that document the gradual growth of a $\sim 0.5\,\mathrm{m}$ deep pit in the Hapi region on 
Comet 67P, from 2014 December 31 until 2015 March 17, when it reached lateral dimensions of $140\,\mathrm{m}\times 100\,\mathrm{m}$. 
We use thermophysical models and a radiative transfer equation solver, in order to analyse \emph{Rosetta}/MIRO microwave observations of the 
region in October and November, 2014, prior to pit formation. Our main conclusions are the following:

\begin{enumerate}
\item MIRO measurements are consistent with the existence of a solid--state greenhouse effect on 67P, which is active in 
coarse airfall material with high macro porosity.
\item MIRO has provided an estimated inferred depth of the $\mathrm{CO_2}$ sublimation front on 67P, 
being $\sim 0.5\,\mathrm{m}$ at the particular pit--forming region in Hapi.
\item In October 2014, the pit--forming region of Hapi was characterised by a low thermal inertia ($\sim 30\pm 10\,\mathrm{MKS}$), a thin ($\sim 2\,\mathrm{cm}$) dust 
mantle, a large diffusivity (indicative of pores and channels in the $1$--$10\,\mathrm{cm}$ range), a measurable solid--state greenhouse effect, the presence of 
shallow ($\sim 0.5\,\mathrm{m}$) $\mathrm{CO_2}$ ice, modest microwave extinction coefficients, and no evidence of multiple--scattering. 
\item The properties of the near--surface material of comet nuclei may change drastically and rapidly. Between October and November 2014, the pit--forming region 
at Hapi experienced a rapid removal of water ice, and a collapse of the dust mantle that resulted in significant compaction. These events measurably changed the 
thermal inertia (from $30\,\mathrm{MKS}$ to $110\,\mathrm{MKS}$, falling toward $65\,\mathrm{MKS}$ at the surface), the dust mantle thickness (from $\sim 2\,\mathrm{cm}$ to $\sim 21\,\mathrm{cm}$), reduced the 
diffusivity by four orders of magnitude (indicative of pores and channels in the $1$--$10\,\mathrm{\mu m}$ range), increased the microwave extinction coefficients by a factor 3--10, 
and introduced multiple--scattering at sub--millimetre wavelengths. 
\item Contextual thermophysical simulations with \textsc{nimbus} show that $\mathrm{CO_2}$ can be maintained within the $0.2$--$0.5\,\mathrm{m}$ region below 
the surface due to seasonally recurring pit--formation events, combined with low thermal inertia and low diffusivity near aphelion.
\item We constrain the conditions under which the $\mathrm{CO_2}$ vapour pressure becomes sufficiently high to overcome the tensile 
strength of porous $1\,\mathrm{\mu m}$--grain dust mantle analogues, and to do so at a location in the orbit that coincides with the observed start of pit formation: a high 
thermal inertia of $\sim 100\,\mathrm{MKS}$, a low diffusivity consistent with homogeneous assemblages of $\mathrm{\mu m}$--sized grains (with a non--zero but low tortuosity), and 
$\mathrm{CO_2}$ ice located to within $0.5\,\mathrm{m}$ of the surface.
\item Accordingly, we propose that pit formation and escarpment propagation in smooth terrain on comet nuclei may be primarily due to rapid ejection of 
dust into the coma driven by superficial $\mathrm{CO_2}$ ice sublimation. 
\item We find that the $\mathrm{CO_2}$ vapour pressure at the sublimation front may become sufficiently strong to cause an inward compression of comet material. 
Accordingly, we propose that the moving $\mathrm{CO_2}$ front compresses material before it, and leaves behind a compacted mixture of dust and water ice that is exposed at the surface of comets. 
\item Accordingly, we propose that the compaction of material in the upper few meters in 67P with respect to the more porous material at depth, inferred from observations 
by CONSERT and SESAME on \emph{Rosetta/Philae}, as well as by the Arecibo radar, is caused by processing during   $\mathrm{CO_2}$ sublimation.
\item We propose that the consolidated terrain observed on the surface of 67P (where it is not covered by smooth terrain) is the observable surface 
expression of material that has been partially or primarily compacted by previous $\mathrm{CO_2}$ sublimation.
\item We recommend enhanced efforts to investigate other pit--forming events observed by OSIRIS and MIRO, vis--\'{a}--vis near--surface supervolatile deposits 
and visually observable surface expressions of recent pit formation and escarpment movements. This is particularly important in the context of cryogenic comet 
sample--return missions, for which deep excavation in the hunt for cold ice drastically drives up costs and technical complexity. We recommend the usage of 
microwave instruments on orbiting reconnaissance spacecraft in order to locate suitable sampling sites.
\end{enumerate}

\section*{Acknowledgements} 

\emph{We dedicate this paper to the memory of our dear friend and colleague, Dr. Claudia J. Alexander (1959-2015), Project Scientist of the US portion of the Rosetta mission, 
whose dedication to science and the search for knowledge remains a lasting inspiration to all of us.} Parts of this research were carried out at the Jet Propulsion Laboratory, 
California Institute of Technology, under a contract with the National Aeronautics and Space Administration. PJG acknowledges financial support from the State Agency for Research 
of the Spanish Ministerio de Ciencia, Innovac\'{i}on y Universidades through project PGC2018--099425--B--I00 and through the `Center of Excellence Severo Ochoa' award to the 
Instituto de Astrof\'{i}sica de Andaluc\'{i}a (SEV--2017--0709). M.~R.~ELM. is partly supported by the internal grant (8474000336--KU--SPSC). 
The MIRO instrument was developed by an international collaboration led by NASA and the Jet Propulsion Laboratory, 
California Institute of Technology, with contributions from France, Germany, and Taiwan. OSIRIS was built by a consortium led by the Max--Planck--Institut f\"{u}r Sonnensystemforschung, G\"{o}ttingen, Germany, in collaboration 
with CISAS, University of Padova, Italy, the Laboratoire d'Astrophysique de Marseille, France, the Instituto de Astrof\'{i}sica de Andaluc\'{i}a, 
CSIC, Granada, Spain, the Scientific Support Office of the European Space Agency, Noordwijk, The Netherlands, the Instituto Nacional 
de T\'{e}cnica Aeroespacial, Madrid, Spain, the Universidad Polit\'{e}chnica de Madrid, Spain, the Department of Physics and Astronomy 
of Uppsala University, Sweden, and the Institut f\"{u}r Datentechnik und Kommunikationsnetze der Technischen Universit\"{a}t Braunschweig, 
Germany. The support of the national funding agencies of Germany (DLR), France (CNES), Italy (ASI), Spain (MEC), Sweden (SNSB), and the 
ESA Technical Directorate is gratefully acknowledged. We thank the \emph{Rosetta} Science Ground Segment at ESAC, the \emph{Rosetta} Mission Operations 
Centre at ESOC and the \emph{Rosetta} Project at ESTEC for their outstanding work enabling the science return of the \emph{Rosetta} Mission.\\

\noindent
\emph{COPYRIGHT}.  \textcopyright\,2022. All rights reserved.

\section*{Data Availability}

The data underlying this article will be shared on reasonable request to the corresponding author.

\bibliography{MN-22-1702-MJ.R1.bbl}

\vspace{0.5cm}
$^{1}$Jet Propulsion Laboratory, California Institute of Technology,  M/S 183--401, 4800 Oak Grove Drive, Pasadena, CA 91109, USA\\
$^{2}$University of Massachusetts, Department of Astronomy, LGRT--B 847 710 North Pleasant Street, Amherst, MA 01003--9305, USA\\
$^{3}$LESIA, Universit\'e Paris Cit\'e, Observatoire de Paris, Universit\'e PSL, CNRS,  Sorbonne Universit\'e, 5 place Jules Janssen, 92195 Meudon, France\\
$^{4}$Institut Universitaire de France (IUF), 1 rue Descartes, 75231 PARIS CEDEX 05, France\\
$^{5}$Independent researcher, Berlin, Germany\\
$^{6}$Instituto de Astrof\'{i}sica de Andaluc\'{i}a (CSIC), Glorieta de la Astronom\'{i}a s/n. 18080 Granada, Spain\\
$^{7}$Jet Propulsion Laboratory, California Institute of Technology,  M/S 183--601, 4800 Oak Grove Drive, Pasadena, CA 91109, USA\\
$^{8}$Jet Propulsion Laboratory, California Institute of Technology,  M/S 321--655, 4800 Oak Grove Drive, Pasadena, CA 91109, USA\\
$^{9}$200 W Highland Dr., Unit 104, Seattle, WA 98119, USA\\
$^{10}$Jet Propulsion Laboratory, California Institute of Technology,  183--301, 4800 Oak Grove Drive, Pasadena, CA 91109, USA\\
$^{11}$Institut f\"{u}r Geophysik und extraterrestrische Physik (IGeP), Technische Universit\"{a}t Braunschweig, Mendelssohnstr. 3, 38106 Braunschweig, Germany\\
$^{12}$Deutsches Zentrum f\"{u}r Luft-- und Raumfahrt (DLR), Institut f\"{u}r Planetenforschung, Asteroiden und Kometen, Rutherfordstr. 2, 12489 Berlin, Germany\\
$^{13}$Max Planck Institute for Solar System Research, Justus--von--Liebig--Weg 3, 37077 G{\"o}ttingen, Germany\\
$^{14}$European Space Agency (ESA), European Space Astronomy Centre (ESAC), Camino Bajo del Castillo s/n, 28692 Villanueva de la Ca\~{n}ada, Madrid, Spain\\
$^{15}$Centrum Bada\'{n} Kosmicznych Polskiej Akademii Nauk, Bartycka 18A, PL--00716 Warszawa, Poland\\
$^{16}$ Department of Physics and Astronomy, Uppsala University, Box 516, SE--75120 Uppsala, Sweden\\
$^{17}$Jet Propulsion Laboratory, California Institute of Technology,  M/S 168--200, 4800 Oak Grove Drive, Pasadena, CA 91109, USA\\
$^{18}$CNR--IFN Padova, Via Trasea 7, 35131 Padova, Italy\\
$^{19}$Aix Marseille Univ, CNRS, CNES, LAM, Marseille, France\\
$^{20}$German Aerospace Center (DLR), Institute of Optical Sensor Systems, Rutherfordstr. 2, 12489 Berlin, Germany\\
$^{21}$Space Research and Planetology Division, Physikalisches Inst., University of Bern, Sidlerstrasse 5, CH--3012 Bern, Switzerland\\
$^{22}$Istituto di Astrofisica e Planetologia Spaziali -- IAPS/INAF, Via del Fosso del Cavaliere 100, 00133 Roma, Italy\\
$^{23}$Space and Planetary Science Center, and Department of Earth Sciences, Khalifa University, P O Box 127788, Abu Dhabi, UAE\\
$^{24}$Department of Physics and Astronomy, University of Padova, vicolo Osservatorio 3, 35122 Padova, Italy\\
$^{25}$INAF--Astronomical Observatory of Padova, Vic. Osservatorio 5, 35122 Padova, Italy\\

\bsp	
\label{lastpage}
\end{document}